\documentclass[amsmath,amssymb,
 aps, physrev,reprint,onecolumn, superscriptaddress]{revtex4-2}
\usepackage{esdiff}
\usepackage{amsmath}
\usepackage{amsfonts}
\usepackage[percent]{overpic} 
\usepackage{graphicx}
\graphicspath{{Images}}
\usepackage{subcaption}
\usepackage{hyperref}
\usepackage{xcolor}
\usepackage{booktabs}
\usepackage{tabularray}
\usepackage[capitalise]{cleveref}
\Crefname{section}{Section}{Sections}
\crefname{figure}{Figure}{Figures}

\newcommand{\force}{p}
\newcommand{\quadrupole}{q}
\newcommand{\source}{s}
\newcommand{\forceavg}{\overline{p}}
\newcommand{\quadavg}{\overline{q}}
\newcommand{\sourceavg}{\overline{s}}
\newcommand{\effsource}{B_s}
\newcommand{\effforce}{B_p}
\newcommand{\effquad}{B_q}
\newcommand{\effswim}{U}
\newcommand{\avg}{\overline}
\newcommand{\vect}{\mathbf}

\begin{document}

\title{Far-field approximations for multi-timescale microswimmers near a boundary }
\author{Sara Drummond-Curtis}
\email{Email: sara.drummond-curtis.24@ucl.ac.uk}
\affiliation{ Department of Mathematics, University College London, London, WC1E 6BT, UK}
\author{ Mohit P. Dalwadi}
\affiliation{ Mathematical Institute, University of Oxford, Oxford, OX2 6GG, UK}
\affiliation{ Department of Mathematics, University College London, London, WC1E 6BT, UK}
\author{Benjamin J. Walker}
\affiliation{ Department of Mathematics, University College London, London, WC1E 6BT, UK}

\date{\today}

\begin{abstract}
Hydrodynamic interactions with boundaries can significantly affect the trajectories of microscale swimmers. In simple swimmer models, a common assumption is that swimmer shape remains constant, essentially averaging over the rapid oscillations in geometry and associated fluid flows that often are the source of propulsion. Previous work in minimal force-dipole models has shown how the inclusion of time-dependent swimmer changes can lead to a fundamentally wider class of behaviours than for their classic (implicitly averaged) counterparts. However, since force dipole models correspond to the leading-order term in the far-field description of the swimmer-induced flow, they break down as the swimmer approaches a boundary and predictions can become qualitatively inaccurate. Here, we extend the minimal force-dipole model by incorporating higher order flow singularities, systematically accounting for rapid oscillations in shape and singularity strength through a multiscale analysis. We demonstrate that the inclusion of time-dependence into these higher order models significantly expands the reachable parameter space, in particular by increasing its dimensionality. In these extended dynamics, we observe three distinct behaviours: crashing, escaping and hovering. Notably, hovering states are absent from the dynamics predicted by the simplest models, but are observed in more complex models.

\end{abstract}
\maketitle
\section{Introduction }\label{sec: intro}
The dynamics of microscale swimming near a boundary are key to understanding a wide range of biological and physical phenomena. Motile bacteria often accumulate near surfaces, where they can form  biofilms that enable antibiotic-resistant infections  to establish and persist in both natural and industrial settings   \citep{klapper_mathematical_2010,guttenplan_regulation_2013}.  Similarly, sperm navigate confined environments such as the female reproductive tract during fertilisation \citep{suarez_sperm_2006,elgeti_hydrodynamics_2010}. Moreover, surfaces will almost always be present in  experiments, for example, the glass of a microscope slide or the walls of a  microfluidic device. Therefore, there is a need to understand how these boundaries impact a swimmer's dynamics and their emergent behaviours.

The interactions between an active swimmer and a boundary can lead to fundamentally different swimming behaviour to that in a bulk fluid.  First reported by \citet{rothschild_non-random_1963} with bull sperm, many microswimmers are hydrodynamically attracted to surfaces. In a broad range of systems, swimmers such as bacteria are seen experimentally to maintain a stable `hovering' distance above the surface  \citep{frymier_three-dimensional_1995,bianchi_3d_2019}.  For instance, \textit{Escherichia coli}  have been observed to swim stably in circles above boundaries \citep{berg_chemotaxis_1990}, in contrast to the largely linear runs that punctuate tumbling motion in bulk swimming. More generally, the nature of the boundary can influence behaviour; \textit{E. coli} have been observed to swim in a clockwise direction above a no-slip surface and counter-clockwise above a stress-free surface \citep{frymier_three-dimensional_1995,berke_hydrodynamic_2008,di_leonardo_swimming_2011}. Although direct contact with surfaces impacts the scattering dynamics of a swimmer \citep{kantsler_ciliary_2013,bianchi_holographic_2017}, hydrodynamic interactions have been noted to be important in determining trajectories \citep{spagnolie_hydrodynamics_2012,elgeti_microswimmers_2016} and the motility of bacteria affects the adhesion rate of a bacterial suspension in shear flow \cite{yeo_shear-induced_2025}.
Overall, this complexity in observed behaviours has led to the development of mathematical models of varying intricacy. 

For a microswimmer near a boundary, many mathematical models are able to predict a range of observed behaviours \cite{lauga_swimming_2006,berke_hydrodynamic_2008,or_dynamics_2009,shum_modelling_2010,giacche_hydrodynamic_2010,di_leonardo_swimming_2011,manabe_shape_nodate}.
The simplest models consider the swimmer-induced flow field several body lengths away from the swimmer, which can often be described by an expansion of fundamental singularities of Stokes flow.  Due to the force-free nature of microscale swimming, the leading-order term of this far-field expansion is a force dipole \citep{lauga_fluid_2020}. Even with such a minimal level of detail, by only considering hydrodynamic effects, the force dipole model  agrees favourably with observed swimming behaviours \cite{berke_hydrodynamic_2008, drescher_direct_2010,drescher_fluid_2011}. For instance, as seen experimentally, hydrodynamic `pushers' such as \textit{E. coli} or spermatozoa are predicted to be attracted to the boundary, whereas `pullers' like \textit{Chlamydomonas} are repelled \citep{berke_hydrodynamic_2008}. Models that incorporate rotlet dipoles (higher-order terms in the singularity representation) successfully predict the clockwise and counter-clockwise circular swimming that has been seen experimentally \cite{lauga_fluid_2020}. This behaviour, however, cannot be predicted by the simplest force dipole model due to its left-right symmetry, highlighting that the leading-order singularity is not always able to predict the dynamics of the swimmer \citep{lauga_fluid_2020}.  

Close to a boundary, the far-field approximation fails to be a formal asymptotic series, and the neglect of higher-order terms is no longer formally justified. Therefore, the validity of predictions made by the simplest force dipole models is unclear. As partial resolutions of this challenge, more detailed mathematical models tend to specialise and, in doing so, more faithfully represent the geometry and hydrodynamic properties of individual swimmers.  For example, resistive-force theory accounts for hydrodynamic resistance along a swimmer's body through a leading-order local relation by exploiting slenderness, whereas more complicated boundary element methods and variations thereof often incorporate detailed descriptions of swimmer geometry. Such methods can increase the fidelity of the simulations, leading to predictions of entrapment, stable hovering, and circular motion for swimmers near a boundary \citep{lauga_swimming_2006,giacche_hydrodynamic_2010,di_leonardo_swimming_2011,ishimoto_squirmer_2013,smith_human_2009}, behaviours that are beyond the reach of the simplest models.

In more complex models, geometric characteristics of swimmers have been shown to be important in accurately predicting dynamics near a boundary. For instance,  \citet{or_dynamics_2009} found that breaking symmetry of shape is essential for stable hovering of multi-sphere swimmers near a boundary. For bacteria, \citet{shum_modelling_2010} reported that attraction or repulsion from a no-slip boundary can depend purely on geometrical features, such as the elongation of the cell body \cite{htet_hydrodynamic_2024}. For ciliates, \citet{manabe_shape_nodate} found that fore-aft asymmetry and constriction dominate the existence of stable hovering states. The geometrical effects of both the finite size of the cell body and the fore-aft asymmetry are therefore important in accurately predicting swimmer dynamics near a boundary. Although these more sophisticated models show that geometric characteristics are crucial, the assumption of a specific swimmer type limits their ability to provide insight into more universal swimming behaviours. Alternatively, the simplicity of minimal models often allows more generalisable insight to emerge from the analysis. Therefore, we seek a balance between the accuracy of specialised models and the tractability of the far-field expansion. 

In order to induce swimming, organisms often rapidly change their body shape by, for example, beating their flagella. The significant computational expense of including any rapid oscillations in more complex models that already accurately describe the geometry of the swimmer means that a general understanding of the role of the rapid oscillations is lacking. Recently, the method of multiple scales has been used to understand the effect of fast oscillations of swimmers over longer timescales. By systematically including the change in body shape of a swimmer in Poiseuille flow with the method of multiple scales, \citet{walker_emergent_2022} were able to match the predictions of the model considered by \citet{omori_rheotaxis_2022} to experimental data. The method of multiple scales has also been used to derive emergent angular dynamics of a swimmer undergoing rapid yawing and rotation in both a two and three-dimensional shear flow. Whilst the emergent dynamics are equivalent to those for a passive particle in shear flow, the shape of this effective passive particle is not necessarily the same as the underlying active swimmer  \citep{walker_effects_2022,gaffney_canonical_2022,dalwadi_generalised_2024,dalwadi_generalised_2024_b,dalwadi_rapidly_2025}. For a swimmer near a boundary, \citet{walker_systematic_2023} considered the rapid variation of the singularity strength and body shape in the force dipole model. The subsequent, systematically averaged, equations  revealed new pitching steady states at which the swimmer was no longer aligned with the boundary. These changes to the angular dynamics suggest that accounting for the rapid motion can lead to qualitatively different predictions in the long-time swimmer trajectories. Hence, owing to the potential significance of both boundary and multi-timescale effects, our primary aim is to consider the effect of including time-dependent parameters in a simple but not quite minimal model: the linear combination of the leading and next-order terms from the far-field approximation. 

In this study, seeking a balance between accuracy and tractability, we analyse the impacts of rapid oscillatory effects on higher order far-field models of microscale swimming. In particular, we include additional terms from the far-field expansion that can be interpreted as accounting for various geometric features of the swimmer \citep{spagnolie_hydrodynamics_2012}. By systematically accounting for the multi-timescale effects, the models predict a range of long-time behaviours such as stable hovering, escaping or crashing into the boundary, whereas in \textit{a priori}-averaged models (those which assume a constant swimmer shape), predictions of boundary collision often dominate. In \cref{sec: model overview}, we formulate these models by considering terms beyond leading order in a far-field expansion of microswimmer dynamics. Using the method of multiple scales \citep{holmespertubation,bender_advanced_1999}, we systematically average over the fast oscillation timescale and derive effective governing equations for the dynamics of the swimmer. In \cref{sec: FD+SD,sec: FD+Q}, we  analyse the effective autonomous dynamical system for the combination of a force dipole with either a source dipole or a quadrupole, respectively, above a stress-free surface.  We then compare the predictions to those of the \textit{a priori}-averaged model. In \cref{sec: no slip}, we consider how changing the boundary condition on the surface from stress-free to no-slip changes the dynamics, before extending the model to include a non-zero self-generated swimming speed in \cref{sec: nonzero swim}. In \cref{sec: FD+SD+Q analysis}, we consider emergent behaviours of the combination of a force dipole with both a source dipole and quadrupole. Finally, in \cref{sec: conclusion}, we discuss the implications and limitations of the systematically averaged effective equations in describing the dynamics of a swimmer near a boundary. 

\section{Model setup}\label{sec: model overview}
 
 Length, time and velocity scales typically associated with microscale swimming mean that the fluid flow around a microswimmer is well described by Stokes equations \citep{lauga_fluid_2020}. In the far field, the fluid flow generated by a swimmer can be formally written as a multipole expansion, consisting of fundamental singular solutions to Stokes equations \citep{chwang_hydromechanics_1975,kim2013microhydrodynamics}. For a point force acting at the origin, singularities are associated with the Green's function for Stokes flow, which in the direction of a unit vector $\vect{n}$ and suitably non-dimensionalised is 
 \begin{equation}
     \mathbf{G}(\vect{x},\vect{n}) = \frac{1}{r}\left( \vect{n} +\frac{\vect{n\cdot x}}{r^2}\vect{x}\right).
 \end{equation}
 Here, $\vect{x}$ is the position relative to the origin and $r = |\vect{x}|$. Other singular solutions can be derived from a point source, with flow field given by
 \begin{equation}
     \vect{u}(\vect{x}) = \frac{\vect{x}}{r^3}.
 \end{equation}
We assume the swimmer is axisymmetric about some unit vector $\vect{d}$ with centre at $\mathbf{x}_0$ and take $\vect{n} = \vect{d}$. As these swimmers are both force and torque-free, the swimmer-generated flow in the far field has the form
 \begin{equation}
    \vect{u}(\vect{x}) = \underbrace{ \vect{\force}(\vect{x}-\vect{x}_0)}_{\text{force dipole}}+\underbrace{\vect{\source}(\vect{x}-\vect{x}_0)}_{\text{source dipole}}+\underbrace{\vect{\quadrupole}(\vect{x}-\vect{x}_0)}_{\text{quadrupole}}+\underbrace{\vect{r}(\vect{x}-\vect{x}_0)}_{\text{rotlet dipole}}+O(|\vect{x}-\vect{x}_0|^{-4}).
\end{equation}
where the terms on the right-hand side correspond to a linear combination of the force dipole, source dipole, quadrupole and rotlet dipole, respectively. At leading order, the force dipole, $\mathbf{\force}$, accounts for the flow field generated by the swimmer. The flow field is further altered by the next-order singularities: the source dipole, $\mathbf{\source}$, captures effects from the finite size of the cell body; the quadrupole, $\mathbf{\quadrupole}$, accounts for fore-aft asymmetry; and the rotlet dipole, $\mathbf{r}$, approximates rotation and counter-rotation within the swimmer, such as that between the flagella and cell body in many bacterial species.

\begin{figure}[hbtp]
    \centering
    \includegraphics[width=0.3\linewidth]{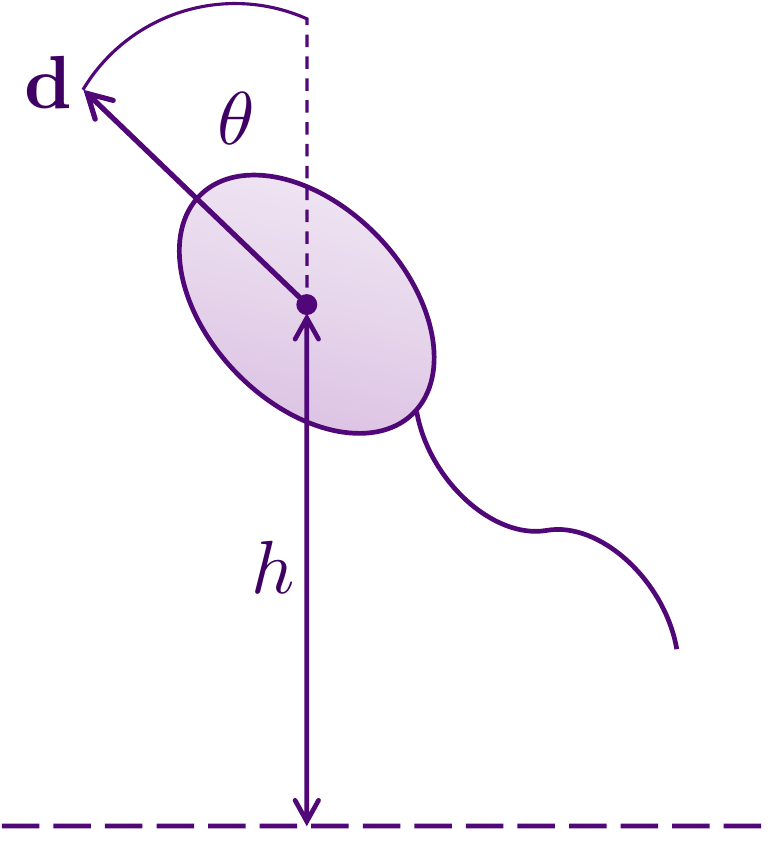}
    \caption{Schematic of a microswimmer above a fixed boundary. We denote the distance between the centre of the swimmer body and the surface by $h$. The swimmer's director, $\mathbf{d}$, makes an angle of $\theta$ with respect to the normal of the surface. The swimmer faces directly away from the surface when $\theta = 0$, parallel to the surface when $\theta = \pi/2$, and towards the surface when $\theta = \pi$.}
    \label{fig: microswimmer above boundary schematic}
\end{figure}

We consider a microswimmer moving in a half-space bounded by a fixed planar surface. We consider the dimensionless problem, achieved by scaling velocities with the velocity attained by the swimmer in free space, $U$, and lengths with respect to ten times the swimmer length, $10L$. We denote the distance between the centre of the swimmer and the surface by $h$ and the angle between the body-fixed director, $\bf d$, and the normal to the surface by $\theta$, as shown in  \cref{fig: microswimmer above boundary schematic}. Therefore, the smallest distance between the swimmer and the boundary is $h = 0.1$.   We will focus on angular and attraction dynamics, so restrict the motion to the $(x,h)$ plane. As the rotlet dipole acts normal to this plane, we omit the potential effects of torque dipoles on the resulting dynamics. The remaining singularities have the form
\begin{align}
   \vect{\force}(\vect{x}) = \frac{\force}{r^2}\left(\frac{3(\vect{d}\cdot\vect{x})^3}{r^3}\vect{x}-\frac{\vect{x}}{r}\right),\quad  \vect{\source}(\vect{x}) =  \frac{\source}{r^3}\left(- \vect{d} + \frac{3 (\vect{d\cdot x})}{r^2}\vect{x}\right), \;\\ \vect{\quadrupole}(\vect{x}) = \frac{\quadrupole}{r^3}\left(\vect{d}-\frac{3}{r^3}(\vect{d}\cdot\vect{x})\vect{x}+(\vect{d}\cdot\vect{x})^2\vect{d}+\frac{15(\vect{d}\cdot\vect{x})^3}{r^4}\vect{x}\right),
\end{align}
where $\force,\,\source,\,$  and $\quadrupole$ are the corresponding dimensionless singularity strengths. 

The interaction of the boundary with the swimmer-generated flow field induces a new flow field that, in turn, acts to change the dynamics of the swimmer. By imposing no-shear and no-stress boundary conditions on the surface, \citet{spagnolie_hydrodynamics_2012} compute the resulting swimming dynamics near a stress-free surface 
\begin{subequations}
\label{eq:FD+SD+Q unaveraged system freeslip}
\begin{align}
\diff{\theta}{t} =&
 \frac{3\force\sin 2\theta}{32h^3}\left[2+B(1+\cos 2\theta)\right] - \frac{3B \source \sin\theta}{32h^4}\left(3+\cos2 \theta\right)  \label{eq:FD+SD+Q unaveraged system theta} \\
 &-\frac{3\quadrupole \sin \theta}{32 h^4}\left[3+5\cos 2\theta +\frac{B}{4}(3\cos 4\theta + 8 \cos 2\theta -11)\right], \nonumber \\
\diff{h}{t} =&   \frac{\force}{8h^2}\left(1+3\cos 2\theta\right)  -\frac{\source\cos\theta}{4h^3}+ \frac{\quadrupole \cos \theta}{4h^3}(1-3\cos2\theta)+u\cos\theta. \label{eq:FD+SD+Qunaveraged system h} 
\end{align}
\end{subequations}
Here, $u$ is the swimmer-induced velocity and we have shifted Spagnolie and Lauga's definition of $\theta$ by $\pi/2$. $B$ is the Bretherton parameter, which encodes information about the swimmer's instantaneous shape \cite{bretherton_motion_1962}. For spheroids, such as those studied by \citet{jeffery_motion_1922}, $B = {(1-e^2)}/{(1+e^2})$ where $e$ is the aspect ratio of the spheroid. In particular, $B = 0$ for spheres and $ B \in(-1,0)$, $B\ \in (0,1)$ represent oblate and prolate spheroids, respectively. We refer to the autonomous system \cref{eq:FD+SD+Q unaveraged system freeslip}  as the \textit{a priori}-averaged system.

Periodic changes in shape, such as those that occur due to a beating flagellum, can cause $\force$, $\source,$ $\quadrupole,$ and $B$ to vary rapidly in time \cite{guasto_oscillatory_2010,ishimoto_coarse-graining_2017, schoeller_flagellar_2018}. However, it is often convenient to treat these as constant in time \cite{spagnolie_hydrodynamics_2012,lauga_fluid_2020}, effectively averaging out the fast time dependence of these quantities before considering their impact on the dynamics. In this study, we instead assume $\force = \force(\omega t)$, $\source = \source(\omega t)$, $\quadrupole = \quadrupole(\omega t)$ and $B = B (\omega t)$ are periodic in $\omega t$ with a period of $2\pi$ and, for convenience, that the average of $\force$, $\source$ and $\quadrupole$ are non-zero. Here,  $\omega \gg 1$ is the frequency of  oscillation, so that the variation occurs  on a timescale much faster than that associated with the motion of interest for the swimmer. The equations \cref{eq:FD+SD+Q unaveraged system freeslip} are now non-autonomous and read
\begin{subequations}
\label{eq:FD+SD+Q time dep system freeslip}
\begin{align}
\diff{\theta}{t} =&
 \frac{3\force(\omega t)\sin 2\theta}{32h^3}\left[2+B(\omega t)(1+\cos 2\theta)\right] - \frac{3B(\omega t) \source(\omega t) \sin\theta}{32h^4}\left(3+\cos2 \theta\right)  \nonumber \\
 &-\frac{3\quadrupole(\omega t) \sin \theta}{32 h^4}\left[3+5\cos 2\theta +\frac{B(\omega t)}{4}(3\cos 4\theta + 8 \cos 2\theta -11)\right], \label{eq:FD+SD+Q time dep system theta} \\
\diff{h}{t} =&   \frac{\force(\omega t)}{8h^2}\left(1+3\cos 2\theta\right)  -\frac{\source(\omega t)\cos\theta}{4h^3}+ \frac{\quadrupole(\omega t) \cos \theta}{4h^3}(1-3\cos2\theta)+u\cos\theta. \label{eq:FD+SD+Q time dep system h} 
\end{align}
\end{subequations}

For a swimmer near a no-slip boundary, the equivalent equations can be adapted from the autonomous analogues given in \citet{spagnolie_hydrodynamics_2012}, and are 
\begin{subequations}
\label{eq:FD+SD+Q time dep system noslip}
\begin{align}
\diff{\theta}{t} =&
 \frac{3\force(\omega t)\sin 2\theta}{64h^3}\left[4+B(\omega t)(3+\cos2\theta)\right] - \frac{3\source(\omega t)\sin\theta}{32h^4}\left[4+3B(\omega t)(3+\cos2\theta)\right] \nonumber \\
&- \frac{3\quadrupole(\omega t)\sin\theta}{16h^4}\left[1+3\cos 2\theta  +\frac{B(\omega t)}{16}(3\cos4\theta +12\cos2\theta -79)\right], \label{eq:FD+SD+Q time dep theta noslip}\\
\diff{h}{t} =&   \frac{3\force(\omega t)}{16h^2}\left(1+3\cos2\theta \right)  -\frac{\source(\omega t)\cos\theta}{h^3}+\frac{\quadrupole(\omega t)\cos\theta}{8h^3}\left(5-9\cos 2\theta\right)+u\cos\theta. \label{eq:FD+SD+Q time dep noslip h} 
\end{align}
\end{subequations}
Compared to \cref{eq:FD+SD+Q time dep system freeslip}, the equations for a stress-free surface, \cref{eq:FD+SD+Q time dep system noslip} primarily differs in the particular constants in the system.

\subsection{Multiscale analysis}
We now exploit the separation of timescales between the fast shape change and slow overall swimming speed to systematically average the equations using the method of multiple scales \cite{holmespertubation,bender_advanced_1999}. As is standard in this method, we introduce the fast timescale $T = \omega t$ so that $\force = \force(T)$, $\source = \source(T)$ and $B = B(T)$ and treat $t$ and $T$ as independent so that $\theta = \theta(T,t)$ and $h = h(T,t)$. We expand the time derivative 
\begin{equation}\label{eq: time deriv expansion}
    \diff{}{ t  } \mapsto \diffp{}{t} + \omega \diffp{}{T},
\end{equation}
transforming the system of ordinary differential equations for a stress-free surface \cref{eq:FD+SD+Q time dep system freeslip} into a system of partial differential equations 
\begin{subequations}
\label{eq:FD+SD+Q expanded deriv system freeslip}
\begin{align}
\diffp{\theta}{t} + \omega \diffp{\theta}{T}  =&
 \frac{3\force(T)\sin 2\theta}{32h^3}\left[2+B(T)(1+\cos 2\theta)\right] - \frac{3B(T) \source(T) \sin\theta}{32h^4}\left(3+\cos2 \theta\right)  \nonumber \\
 &-\frac{3\quadrupole(T) \sin \theta}{32 h^4}\left[3+5\cos 2\theta +\frac{B(T)}{4}(3\cos 4\theta + 8 \cos 2\theta -11)\right], \label{eq:FD+SD+Q expanded deriv system theta} \\
\diffp{h}{t} + \omega \diffp{h}{T} =&   \frac{\force(T)}{8h^2}\left(1+3\cos 2\theta\right)  -\frac{\source(T)\cos\theta}{4h^3}+ \frac{\quadrupole(T) \cos \theta}{4h^3}(1-3\cos2\theta)+u\cos\theta. \label{eq:FD+SD+Q expanded deriv system h} 
\end{align}
\end{subequations}
The extra degrees of freedom generated by this are removed later by imposing periodicity in the fast scale $T$. We seek asymptotic solutions to the system of the form
\begin{equation}\label{eq: theta and h expansion}
\theta \sim \theta_0(t,T) +\frac{1}{\omega}\theta_1(t,T) + \dotsc, \quad h \sim h_0(t,T)+\frac{1}{\omega}h_1(t,T) \dotsc.
\end{equation}
After substituting the expansions \cref{eq: theta and h expansion} into the system of PDEs \cref{eq:FD+SD+Q expanded deriv system freeslip},  we obtain to leading order $(O(\omega))$

\begin{equation}\label{eq: FD+SD leading order}
    \diffp{\theta_0}{T} = 0, \quad \diffp{h_0}{T} = 0,
\end{equation}
so that $\theta_0 = \theta_0(t)$ and $h_0 = h_0(t)$ are both independent of the fast timescale. Our goal is to determine how these functions depend on $t$. To achieve this, we must consider the next order ($O(1)$) terms in \cref{eq:FD+SD+Q expanded deriv system freeslip}:
\begin{subequations}
\label{eq: FD+SD+Q order 1}
\begin{align}
    \diff{\theta_0}{t} +\diffp{\theta_1}{T} =& \frac{3\force(T)\sin 2\theta_0}{32h_0^3}\left[2+B(T)(1+\cos 2\theta_0)\right] - \frac{3B( T) \source ( T)\sin\theta_0}{32h_0^4}\left(3+\cos2 \theta_0\right) \nonumber\\
& -\frac{3\quadrupole(T) \sin \theta_0}{32 h_0^4}\left[3+5\cos 2\theta_0 +\frac{B(T)}{4}(3\cos 4\theta_0 + 8 \cos 2\theta_0 -11)\right], \\
    \diff{h_0}{t}+\diffp{h_1}{T} =& \frac{\force( T)}{8h_0^2}\left(1+3\cos 2\theta_0\right)  -\frac{\source( T)\cos\theta_0}{4h_0^3}
     + \frac{\quadrupole(T) \cos \theta_0}{4h_0^3}(1-3\cos2\theta_0) +u\cos\theta_0. 
\end{align}
\end{subequations}
To obtain the averaged equations that we are seeking, we integrate \cref{eq: FD+SD+Q order 1} over a period of the fast timescale $T$, imposing the method of multiple scales constraint that $\theta$ and $h$ are both periodic in $T$. This procedure yields
\begin{subequations}
\label{eq: FD+SD+Q averaged equations freeslip}
\begin{align}
    \diff{\theta_0}{t}  =& \frac{3\sin 2\theta_0}{32h_0^3}\left[2\forceavg+\avg{B\force}(1+\cos 2\theta_0)\right] - \frac{3\avg{B\source} \sin\theta_0}{32h_0^4}\left(3+\cos2 \theta_0\right), \nonumber\\
    &-\frac{3\sin \theta_0}{32 h_0^4}\left[\quadavg(3+5\cos 2\theta_0) +\frac{\avg{B\quadrupole}}{4}(3\cos 4\theta_0 + 8 \cos 2\theta_0 -11)\right], \label{eq: FD+SD+Q avg freeslip theta} \\
    \diff{h_0}{t} =& \frac{\forceavg}{8h_0^2}\left(1+3\cos 2\theta_0\right)  -\frac{\sourceavg\cos\theta_0}{4h_0^3}+ \frac{\quadavg \cos \theta_0}{4h_0^3}(1-3\cos2\theta_0)+u\cos\theta_0,
\end{align}
\end{subequations}
where we introduce the notation $\overline{\cdot}$ to represent the average over one period of oscillation of the fast timescale
\begin{equation}
    \overline{a} = \frac{1}{2\pi}\int_0^{2\pi}a \,\text{d}T.
\end{equation}

As a result of this systematic averaging,  six averages appear in the resulting autonomous dynamical system:  $\forceavg$, $\sourceavg$, $\quadavg$, $\avg{B\force}$, $\avg{B\source}$, $\avg{B\quadrupole}$ and the swimming speed, $u$; in contrast, five flow and shape parameters appeared in the original, \textit{a priori}-averaged problem \cref{eq:FD+SD+Q unaveraged system freeslip}: $\force,\, \source,\, \quadrupole, \, B,\, u$. It is convenient to write the averaged equations \cref{eq: FD+SD+Q averaged equations freeslip} as
\begin{subequations}
\label{eq: FD+SD+Q averaged freeslip final}
\begin{align}
    \diff{\theta}{t}  =& \frac{3 \forceavg \sin 2\theta}{32h^3}\left[2+\effforce(1+\cos 2\theta)\right] - \frac{3\effsource \sourceavg \sin\theta}{32h^4}\left(3+\cos2 \theta\right) \nonumber\\
   & -\frac{3\quadavg\sin \theta}{32 h^4}\left[3+5\cos 2\theta  +\frac{\effquad}{4}(3\cos 4\theta + 8 \cos 2\theta -11)\right], \label{eq: FD+SD+Q avg freeslip theta 0 dropped}\\
    \diff{h}{t} =& \frac{\forceavg}{8h^2}\left(1+3\cos 2\theta\right)  -\frac{\sourceavg\cos\theta}{4h^3}+ \frac{\quadavg \cos \theta}{4h^3}(1-3\cos2\theta)+u\cos\theta \label{eq: FD+SD+Q avg freeslip h 0 dropped},
\end{align}
\end{subequations}
where we have introduced three effective shape parameters
\begin{equation}
\effforce = \frac{\overline{B\force}}{\forceavg}, \quad \effsource  = \frac{\overline{B\source}}{\sourceavg }, \quad \effquad = \frac{\overline{B\quadrupole}}{\quadavg}.
\end{equation}
corresponding to the interaction between the different singularities and the shape of the swimmer. Whilst the Bretherton parameter $B$ is constrained between $-1$ and $1$ for a spheroidal swimmer, the three new effective shape parameters are unconstrained in general. Furthermore, the systematically averaged system \cref{eq: FD+SD+Q averaged freeslip final} only reduces to the original, \textit{a priori}-averaged system \cref{eq:FD+SD+Q unaveraged system freeslip} if $\effforce = \effsource =\effquad = B$, which is not true in general. In fact, it is enough to have oscillations in the body shape and one of the singularity strengths to see the emergence of an extra parameter, which fundamentally alters the system. Hence, rapid oscillations in shape and generated flow fields can qualitatively alter the effective dynamical system at leading order.

For a no-slip boundary, the structure of the equations and techniques needed for the multiscale analysis do not differ significantly to those used for a stress-free surface. We therefore proceed as before to find that, to leading order, the systematically averaged system now reads
\begin{subequations}
\label{eq:FD+SD+Q averaged system noslip}
\begin{align}
\diff{\theta}{t} =&
 \frac{3\forceavg \sin 2\theta}{64h^3}\left[4+\effforce (3+\cos2\theta)\right] - \frac{3\sourceavg\sin\theta}{32h^4}\left[4+3\effsource(3+\cos2\theta)\right] \nonumber \\
& -\frac{3\quadavg\sin\theta}{16h^4}\left[1+3\cos 2\theta  +\frac{\effquad}{16}(3\cos4\theta +12\cos2\theta -79)\right], \label{eq:FD+SD+Q averaged theta noslip}\\
\diff{h}{t} =&   \frac{3\forceavg}{16h^2}\left(1+3\cos2\theta \right)  -\frac{\sourceavg\cos\theta}{h^3}+\frac{\quadavg\cos\theta}{8h^3}\left(5-9\cos 2\theta\right)+u\cos\theta. \label{eq:FD+SD+Q averaged noslip h} 
\end{align}
\end{subequations}
Again, the same three effective shape parameters have emerged ($\effforce = \overline{B\force}/\forceavg$, $\effsource = \overline{B\source}/\sourceavg$, $\effquad = \overline{B\quadrupole}/\quadavg$), increasing the number of parameters from five in the \textit{a priori}-averaged system (\cref{eq:FD+SD+Q time dep system noslip} with $\force(\omega t) = \forceavg$, $\quadrupole(\omega t) = \quadavg$, $\source(\omega t) = \sourceavg$ and $B(\omega t) = \avg{B}$) to seven in the systematically averaged system \cref{eq:FD+SD+Q averaged system noslip}. To recover the \textit{a priori}-averaged system, we again require $\effforce = \effsource =\effquad = B$.

\subsection{Classifying emergent behaviours} \label{subsec: model setup emerge}
 In order to understand the individual effect of each additional singularity, we first consider the addition of the source dipole and quadrupole separately for both the stress-free and no-slip boundary conditions. We then investigate any combined effects by considering all three singularities at once for a stress-free boundary. We initially consider the case in which there is no net self propulsion, so that $u = 0$. This applies to, for instance, a reciprocal `swimmer' \cite{lauga_fluid_2020}.  We later consider how a non-zero swimming speed changes the dynamics.  A full overview of  which effects are included in each section is detailed in \cref{tab:inclusion exclusion}.
\setlength{\tabcolsep}{10 pt}
\begin{table}[htbp]
    \centering
   \begin{tabular}{lcccccc}
    \toprule
     &\multicolumn{3}{c}{Singularity} && \multicolumn{2}{c}{Boundary condition} \\
   \cmidrule{2-4} \cmidrule{6-7}
   
    & Force dipole & Source dipole & Quadrupole &  Speed & Stress-free & No-slip \\
    \midrule
    \cref{sec: FD+SD} & $\checkmark$ & $\checkmark$ & $\times$ & $\times$ & $\checkmark$ &  \\
    \cref{sec: FD+Q} & $\checkmark$ & $\times$ & $\checkmark$ & $\times$ & $\checkmark$ &  \\
    \cref{subsec: FD+SD no slip} & $\checkmark$ & $\checkmark$ & $\times$ & $\times$ &  & $\checkmark$ \\
    \cref{subsec: FD+Q no slip} & $\checkmark$ & $\times$ & $\checkmark$ & $\times$ &  & $\checkmark$ \\
    \cref{subsec: FD+SD+U} & $\checkmark$ & $\checkmark$ & $\times$ & $\checkmark$ & $\checkmark$ &  \\
    \cref{subsec: FD+Q+U} & $\checkmark$ & $\times$ & $\checkmark$ & $\checkmark$ & $\checkmark$ &  \\
    \cref{sec: FD+SD+Q analysis} & $\checkmark$ & $\checkmark$ & $\checkmark$ & $\times$ & $\checkmark$ &  \\
    \bottomrule
    \end{tabular}
 
    \caption{Summary of the different combinations of  parameters and boundary conditions considered in this study.}
    \label{tab:inclusion exclusion}
\end{table}

In each section, we explore the emergent behaviours predicted by the appropriate systematically averaged model and compare these to the predictions of the \textit{a priori}-averaged model. We compute the numerical solution for the systematically derived swimmer dynamics and classify the long-time behaviours in relation to the corresponding $(\theta, h)$ phase portrait. To classify the resulting trajectory, we first determine if a linearly stable state exists for $\theta \in (-\pi,\pi)$ and $h>0$ in the $(\theta,h)$ plane. If it does, and the trajectory ends within a $0.1$-ball of any steady state (recalling that the dimensionless swimmer lengthscale is 0.1), then we label this behaviour as `hovering'. If the swimmer is not near a steady state, but is within two swimmer lengths of the wall, then we say the swimmer crashes into the boundary. This threshold separation reflects the assumption that the flow field is only valid several swimmer lengths away from the boundary and avoids numerical issues associated with the singularity in the dynamics that is present at $h=0$. However, increasing this threshold to, for example, three body lengths does not qualitatively change the results. If the swimmer is neither hovering nor crashed, we classify it as having escaped from the boundary. The impacts of finite-time simulation on classification were mitigated by verifying that classifications were unchanged after doubling simulation time.

\section{The effect of a source dipole} \label{sec: FD+SD}
\subsection{Governing equations }\label{subsec: FD+SD gov eqn}
We first consider a swimmer near a stress-free boundary and focus on the addition of a source dipole, which reflects the effects of finite swimmer size. Therefore, we analyse the following  (systematically averaged) dynamical system
\begin{subequations}
\label{eq: FD+SD averaged freeslip final}
\begin{align}
    \diff{\theta}{t}  =& \frac{3 \forceavg \sin 2\theta}{32h^3}\left[2+\effforce(1+\cos 2\theta)\right] - \frac{3\effsource \sourceavg \sin\theta}{32h^4}\left(3+\cos2 \theta\right), \label{eq: FD+SD avg freeslip theta 0 dropped}\\
    \diff{h}{t} =& \frac{\forceavg}{8h^2}\left(1+3\cos 2\theta\right)  -\frac{\sourceavg\cos\theta}{4h^3} \label{eq: FD+SD avg freeslip h 0 dropped}.
\end{align}
\end{subequations}
For this to be quantitatively equivalent to the original \textit{a priori}-averaged system \cref{eq:FD+SD+Q unaveraged system freeslip}, we require $\force = \forceavg,\, \source =\sourceavg $, and $\effforce = \effsource = B$.  In order to simplify our exploration below, we note that the system \cref{eq: FD+SD averaged freeslip final} is invariant under the transformation $\sourceavg\mapsto -\sourceavg$, $\theta \mapsto \theta + \pi$. As in \textit{a priori}-averaged models, the dynamics depend strongly on whether or not the swimmer is, on average, a pusher or puller. Here, we define $\forceavg>0$ to be an effective pusher and $\forceavg<0$ to be an effective puller. 

\subsection{Emergent behaviours}\label{subsec: FD+SD emergent behaviour}

\subsubsection{Existence of trivial steady states}\label{subsubsec: FD+SD trivial}
We note  $\mathrm{d} \theta / \mathrm{d} t = 0$ for $\theta = n\pi$, $n \in \mathbb Z $ in \cref{eq: FD+SD avg freeslip theta 0 dropped}, independent of the value of $h$ or the system parameters.  We therefore refer to the steady states with $\theta = 0,\, \pi$ as \emph{trivial}, corresponding to the swimmer facing away from and towards the boundary, respectively. Without loss of generality, we only consider $\sourceavg<0$ as we can map to $\sourceavg>0$ by shifting $\theta$ by $\pi$. Substituting $\theta = 0$ into \cref{eq: FD+SD avg freeslip h 0 dropped}, the trivial steady states are therefore 
\begin{equation}
    \theta = 0, \; h = \frac{\sourceavg}{2\forceavg} \quad \text{and} \quad \theta =  \pi,\; h = -\frac{\sourceavg}{2\forceavg}.
\end{equation}
Calculating the linear stability, the eigenvalues are
\begin{equation}\label{eq: FD+SD trivial eigen}
\lambda_0 \in\left\{  \frac{3{\forceavg}^4}{\sourceavg^3}\left(\effforce+1-2\effsource\right), \frac{4\forceavg^4}{\sourceavg^3}\right\} \quad\text{and} \quad \lambda_{\pi} \in \left\{-\frac{3 {\forceavg}^4}{\sourceavg^3}\left(\effforce+1-2\effsource\right), -\frac{4\forceavg^4}{\sourceavg^3}\right\},
\end{equation}
for the steady states corresponding to $\theta = 0$ and $\theta =  \pi$ respectively.

Since $h>0$ in the domain, $\theta = 0$ is a relevant steady state when $\forceavg<0$ and $\theta =  \pi$ when $\forceavg>0$. For effective pushers, $\forceavg>0$, the trivial steady state $\theta = \pi$ is always linearly unstable as $- 4\forceavg^4/\sourceavg^3>0$.  For effective pullers $\forceavg<0$,  the trivial steady state ($\theta = \pi$) is linearly stable if and only if $\effforce+1-2\effsource >0$. Projecting  onto the \textit{a priori}-averaged system, we set $\effforce = \effsource = B$ and note that the trivial steady state is linearly stable when $B<1$.  Therefore, by including a source dipole, for \textit{a priori}-averaged pullers we always have a stable steady state at which the swimmer hovers above the boundary at a finite height. Without the additional singularity, no such steady state exists in the \textit{a priori}-averaged force dipole model and the swimmer is predicted to crash into the boundary. These conclusions continue to hold for $\sourceavg>0$ with the relevant steady states now $\theta = \pi$ for effective pushers, $\forceavg>0$, and $\theta = 0$ for effective pullers, $\forceavg<0$.

\subsubsection{Existence of non-trivial steady states}
To find the remaining steady states, which we refer to as \textit{non-trivial} steady states, we note that seeking $\text{d}{h}/{\text{d}t} = 0$ in \cref{eq: FD+SD avg freeslip h 0 dropped} yields the steady height as a function of $\theta$ 
\begin{equation} \label{eq: FD+SD h steady state}
    h = \frac{2\sourceavg \cos\theta}{\forceavg(1+3\cos 2\theta)}.
\end{equation}
 By substituting \cref{eq: FD+SD h steady state} into \cref{eq: FD+SD avg freeslip theta 0 dropped} and factoring out ${3 \sourceavg \sin \theta}/{32h^3}$, we find that the non-trivial $\theta$ steady states are solutions to
\begin{equation}\label{eq: FD+SD poly theta}
    \frac{4\cos^2\theta}{1+3\cos 2\theta}\left(2+\effforce (1+\cos 2\theta)\right) - \effsource (3+\cos 2\theta) = 0.
\end{equation}
We make the further substitution $x = \cos^2 \theta$ in \cref{eq: FD+SD poly theta} to deduce that the solutions satisfy
\begin{equation}\label{eq: FD+SD poly x}
    \frac{2x}{3x-1}(1+\effforce x) -\effsource (1+x) = 0,
\end{equation}
which reduces to the quadratic
\begin{equation}(2\effforce-3\effsource)x^2+(2-2\effsource)x+\effsource =0,
\end{equation}
in the case $B_p \neq -3$. Therefore, the non-trivial steady states are given by 
\begin{equation}
 x = 
    \begin{cases}
        -\frac{\effsource}{2+\effsource}, &\text{if}\; \effforce = -3\\
        \frac{\effsource}{2\effsource-2}, & \text{if}\; \effforce = 3\effsource/2\\
        \frac{\effsource-1\pm \sqrt{(\effsource-1)^2-\effsource(2\effforce-3\effsource)}}{2\effforce-3\effsource}, &\text{otherwise.}
    \end{cases}
\end{equation}
Importantly, although we have four parameters in the effective dynamical system, the existence of non-trivial steady states depends only on the effective shape parameters $\effforce,\,\effsource$. The effective strengths of the force and source dipole, $\forceavg,\, \sourceavg$, affect the stability of these states as the parameters change sign. We continue to assume, without loss of generality, that $\sourceavg<0$. This implies that the possible dynamics of the swimmer depend only on the two effective shape parameters, $\effforce$, $\effsource$, and the sign of the effective strength of the force dipole, $\forceavg$.

\begin{figure}[hbtp]
\centering
\begin{overpic}[width=\textwidth, grid = false]{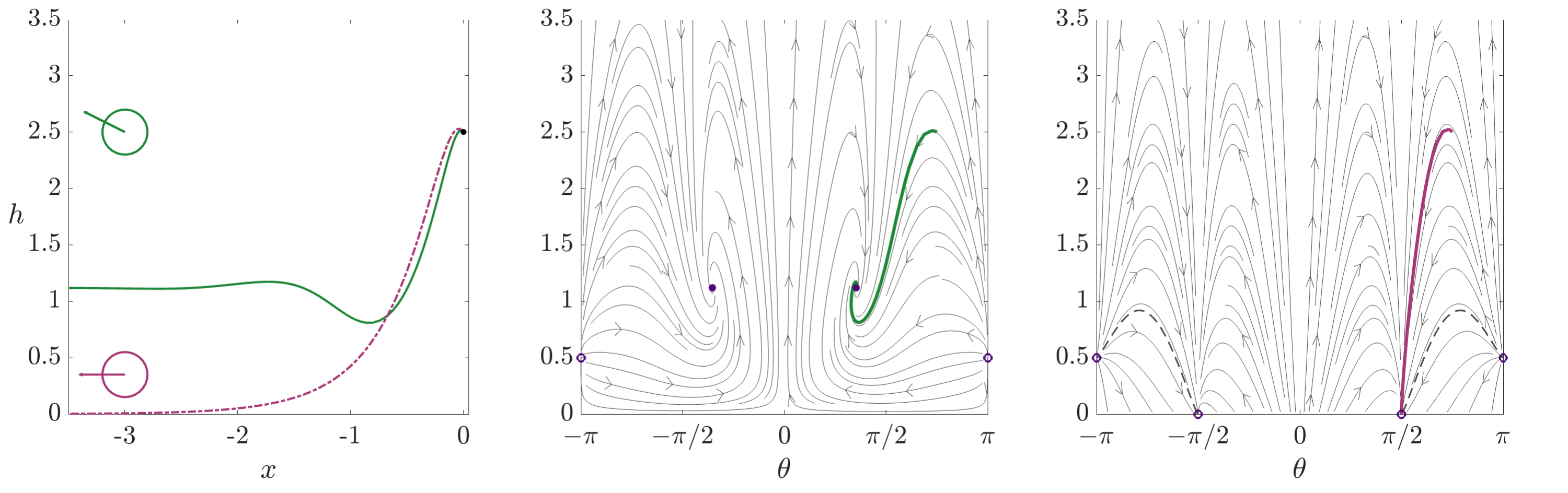}
\put(-1.8, 30){(a)} \put(31,30){(b)} \put(64,30){(c)}
\end{overpic}

\caption{Hovering dynamics of a single swimmer modelled as a force dipole and source dipole near a stress-free surface. (a) Trajectories in the $(x,h)$ plane and illustrative long-time configurations. The systematically averaged system (green) approaches a stable height, whereas the \textit{a priori}-averaged system (pink, dashed-dot) crashes into the boundary despite the same initial condition. (b,c) The $(\theta,h)$ phase portraits corresponding to the dynamics of (a) for (b) the systematically averaged system and (c) the \textit{a priori}-averaged system, with the sample trajectories of (a) shown in the phase planes.  In (b), we observe the local stability of the hovering state in the systematically averaged system. In (c), the dashed lines correspond to the unstable manifold of the saddle at $(\theta,h) = (\pi,0.5)$. We take $p(T) = 1+4\sin T$, $s(T) =-1-4\sin T $, $B(T) = (\sin T)/2$ so that $ \forceavg = 1$, $\sourceavg = -1$, $\effforce = 1$, $\effsource = -1$, $\overline{B} = 0$. }
\label{fig: stable behaviour FD+SD freeslip}

\end{figure}

\begin{figure}[hbtp]
\centering
\begin{overpic}[width = \textwidth, grid = false]{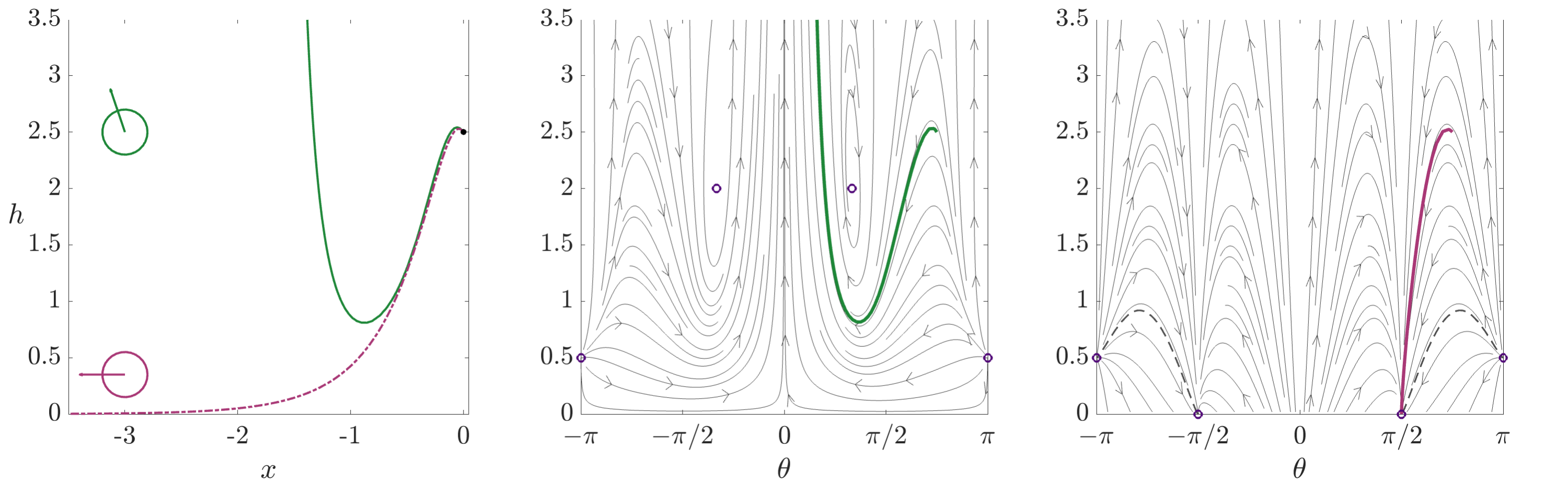}
\put(-1.8, 30){(a)} \put(31,30){(b)} \put(64,30){(c)}
\end{overpic}

\caption{Escape dynamics of a single swimmer modelled as a force dipole and source dipole near a stress-free surface. (a) Trajectories in the $(x,h)$ plane and illustrative long-time configurations. The systematically averaged system (green) escapes from the surface, whereas the \textit{a priori}-averaged system (pink, dashed-dot) crashes into the boundary despite the same initial condition. (b,c) The $(\theta,h)$ phase portraits corresponding to the dynamics of (a) for (b) the systematically averaged system and (c) \textit{a priori}-averaged system, with the sample trajectories shown in the phase planes. In (c), the dashed lines correspond to the unstable manifold of the saddle at $(\theta,h) = (\pi,0.5)$. We take $p(T) = 1-6\sin T$, $s(T) =-1-4\sin T $, $B(T) = (\sin T)/2$ so that $ \forceavg = 1$, $\sourceavg = -1$, $\effforce = -1.5$, $\effsource = -1$, $\overline{B} = 0$.}
\label{fig: escape trajectories FD+SD freeslip}

\end{figure}

Unlike the trivial steady states, the effect of the non-trivial steady states on the dynamics of the swimmer are more complex since the effective body shape parameters, $\effforce,\, \effsource$, are unconstrained.  We therefore calculate the fixed points for a range of effective body shape parameters and numerically evaluate the corresponding Jacobian to find the associated linear stability. Again, for certain body shapes we find stable states at which the swimmer is tilted towards or away from the boundary. In these configurations, the hydrodynamic interaction with the boundary causes the swimmer to translate horizontally along the boundary whilst remaining at a fixed height, as in \cref{fig: stable behaviour FD+SD freeslip}. We also find occurrences where there are no stable steady states and the swimmer escapes from the boundary regardless of the initial condition, as shown in \cref{fig: escape trajectories FD+SD freeslip}. Additionally, as in the (classic) force dipole model, there are also scenarios in which the swimmer crashes into the boundary.

\begin{figure}[hbtp]
\centering
\begin{subfigure}[hbtp]{0.35\textwidth}
  \centering
\begin{overpic}[width = 0.985\textwidth, grid = false]{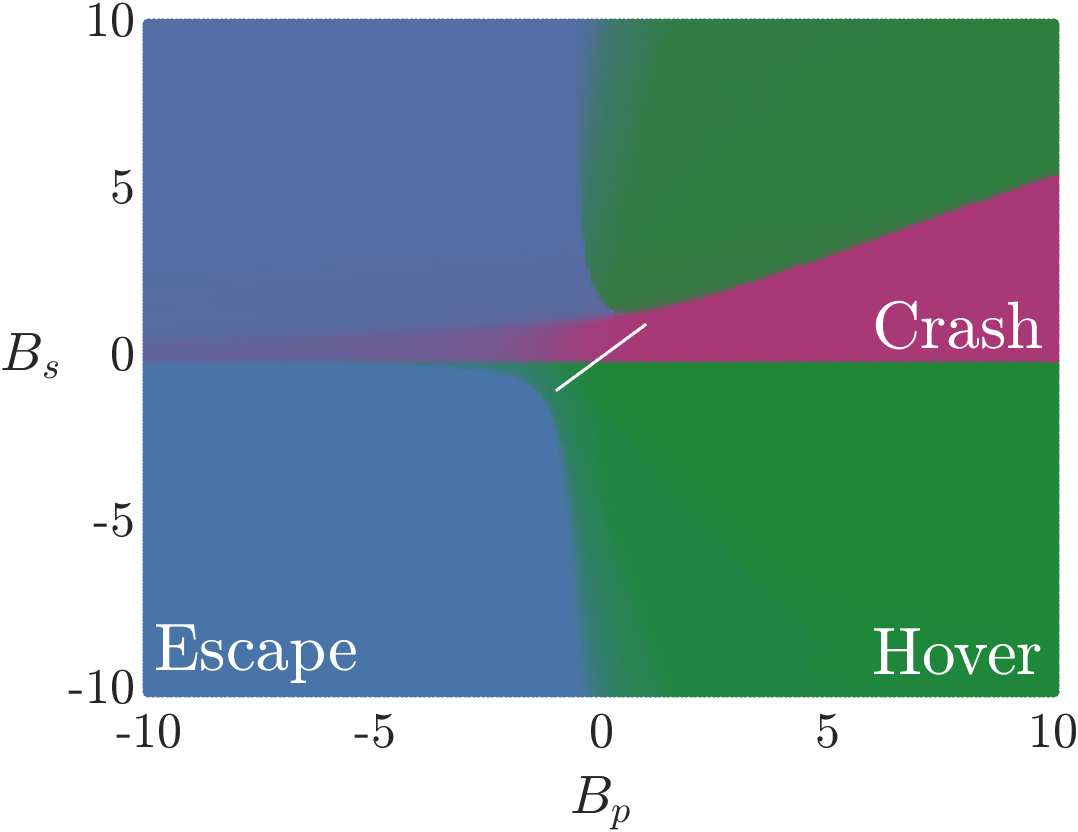}
  \put(-1,75){(a)} \put(103,75){(b)} \put(40,78){Pusher $\avg{p}>0$} 
\end{overpic}
\end{subfigure}
\hfill
\begin{subfigure}[hbtp]{0.35\textwidth}
  \centering
  \begin{overpic}[width=0.935\textwidth, grid = false]{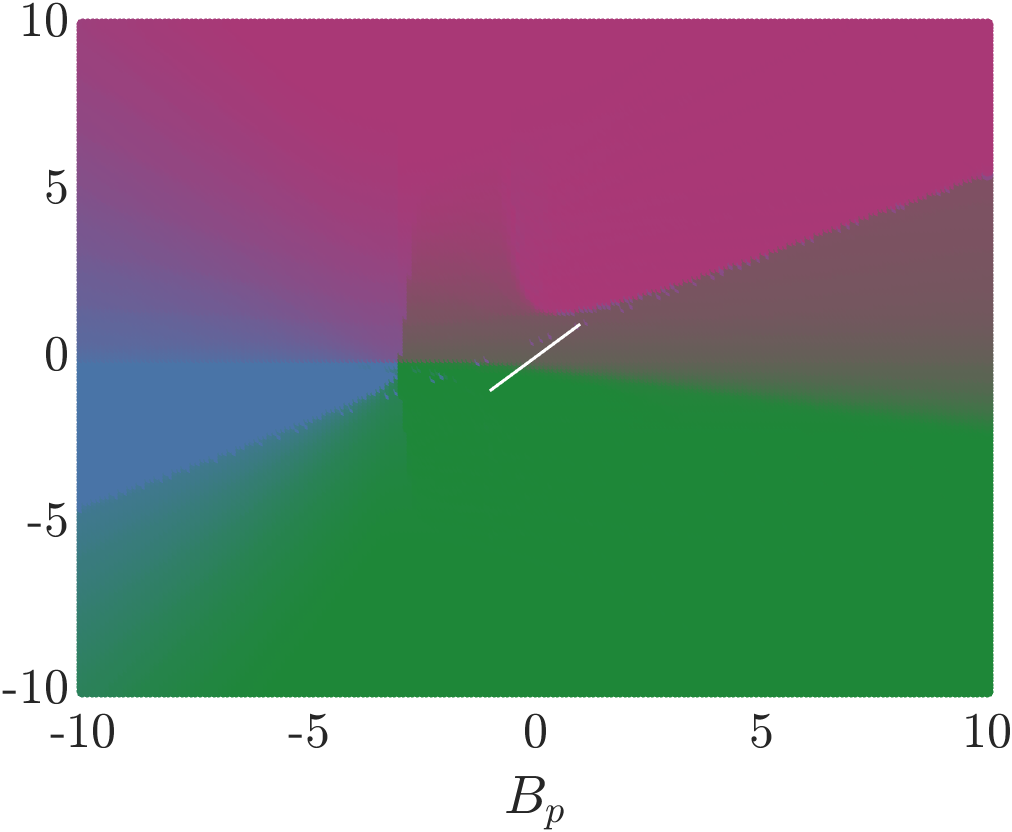}
  \put(40,83){Puller $\avg{p}<0$}
  \end{overpic}
\end{subfigure}
\hfill
\begin{subfigure}[hbtp]{0.25\textwidth} 
\centering
\includegraphics[width = \textwidth]{{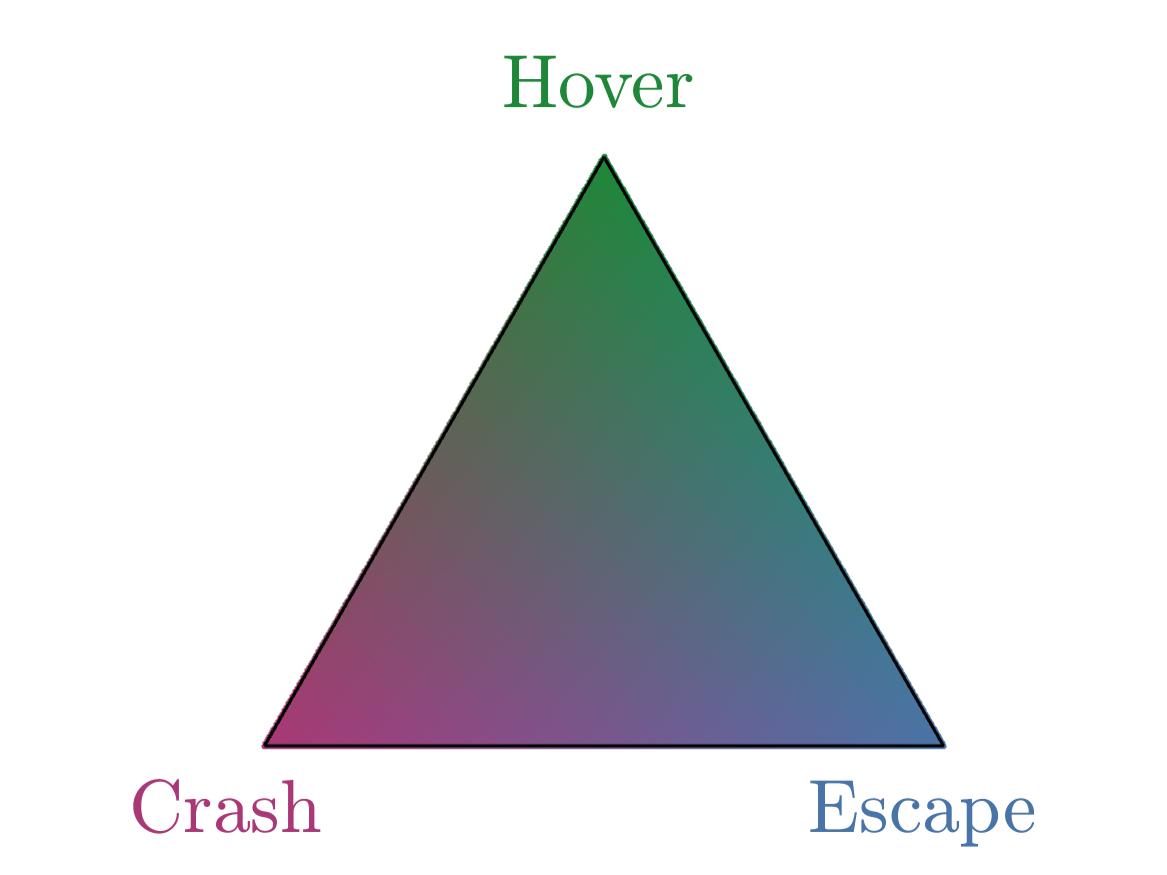}}
    \label{fig: main behaviour legend}
\end{subfigure}
\caption{The proportion of the main behaviours (hovering, escaping or crashing)  for a single swimmer modelled as a force dipole and source dipole above a stress-free surface. The possible dynamics depend on the effective shape parameters ($\effforce$, $\effsource$) and the sign of the effective force dipole strength $\forceavg$. In panel (a), the swimmer is an effective pusher, $\forceavg>0$; in panel (b), the swimmer is an effective puller $\forceavg<0$. The possible dynamics of the \textit{a priori}-averaged system correspond to the white line segments in (a), (b) which generally do not reflect the behaviour proportions observed in the wider $(\effforce,\effsource)$ space of the systematically averaged system.  Each colour denotes a unique proportion of behaviours with all possibilities shown in the colourmap. The corners correspond to the extreme cases in which all trajectories result in a single state, irrespective of the sampled initial condition. We take $\sourceavg = -1$, $\forceavg = \pm 1$ and vary $\effforce$ and $\effsource$ between $-10$ and $10$ beyond which the trends continue.  }
\label{fig: main behaviour FD+SD freeslip}

\end{figure}

In order to determine how changing the effective body shape affects the behaviour of the swimmer,  we repeat the parameter sweep over $\effforce$ and $\effsource$ for effective pushers and pullers whilst keeping $\sourceavg = -1$ in order to determine the steady states and their associated stability. Additionally, for each pair $(\effforce, \effsource)$ we uniformly scatter 144 initial conditions for $(\theta, h) \in [0.1,\pi-0.1]\times [0.1,5]$ to sample behaviour throughout the phase space.  We consider an upper limit on $h$ as we are interested in the dynamics of a swimmer near a boundary, and have verified that our conclusions are not sensitive to changes in this upper limit.  We then  numerically solve the ODE system \cref{eq: FD+SD averaged freeslip final} using \texttt{ode15s} in MATLAB \cite{MATLAB}. We run each simulation for time $t = 5\times10^4$, which allows sufficient time for the swimmer to reach the fixed point or escape from the surface, as verified by also running the simulation for time $t = 10^5$. We classify the swimmer as hovering, crashing or escaping as detailed in \cref{subsec: model setup emerge}. We then calculate the proportion of trajectories for each behaviour before assigning it a unique colour based on the linear combination of the three proportions for each $(\effforce,\effsource)$.

\begin{figure}[hbtp]
\centering
\begin{overpic}[width = \textwidth, grid = false]{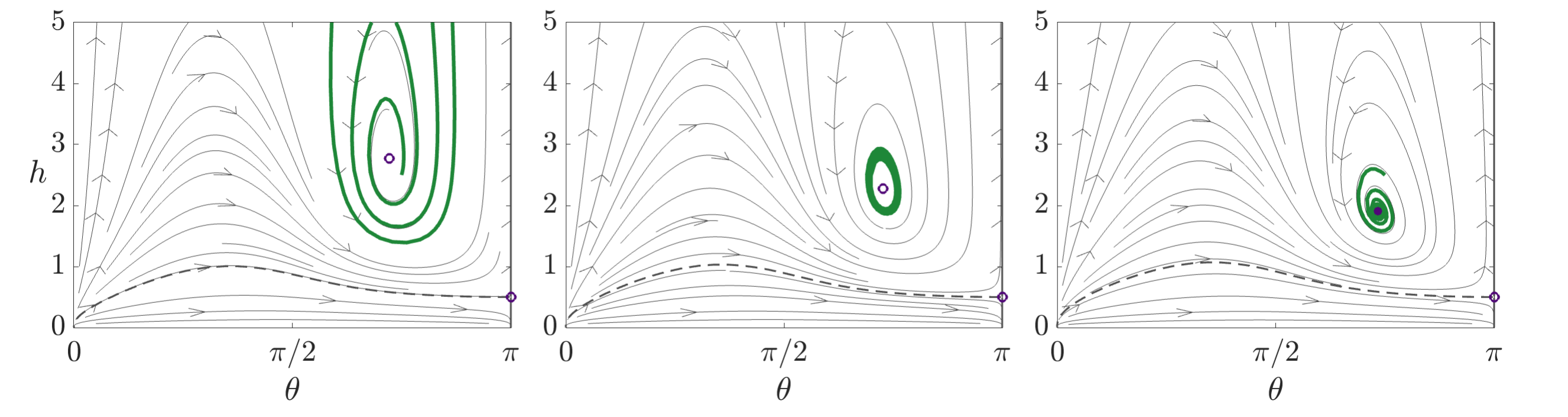}
\put(0,26){(a)} \put(31.5,26){(b)} \put(63.4,26){(c)}
\end{overpic}
\caption{ Phase portraits in $(\theta, h)$ space showing the transition from (a) a linearly unstable steady state to (c) a linearly stable steady state via (b) a non-linear centre, for a swimmer modelled as a force dipole and source dipole. An example  trajectory is shown on each phase portrait demonstrating the stability of the fixed point. We take $\forceavg =1$, $\sourceavg = -1$, $\effsource = 2$ and  $\effforce = -0.5,\, -0.1, \, 0.3$ in (a), (b) and (c), respectively. }
\label{fig:FD+SD spiral}
\end{figure}

Regardless of whether the swimmer is a pusher or a puller, the analysis in  \cref{fig: main behaviour FD+SD freeslip} shows the swimmer either follows  a trajectory to a stable fixed point (green) or moves off into the bulk (blue) for a large proportion of the parameter space. The sharp boundaries between different prevalent behaviours suggest that only small regions of parameter space give rise to varied outcomes. Some of these bifurcations in swimmer behaviour have straightforward interpretations. For example, the sharp transition at $\effsource = 0$ in \cref{fig: main behaviour FD+SD freeslip}(a)  corresponds to the point at which the solution for the non-trivial steady states \cref{eq: FD+SD poly theta} is $0$, coinciding with a trivial steady state. Others, such as the transitions from escape to hover in  \cref{fig: main behaviour FD+SD freeslip}, only become apparent from the linear stability analysis. Highlighting an interesting behaviour, we note that as the predominant behaviour in the $(\effforce, \effsource)$ phase space switches between escaping and hovering, the stability of one of the fixed points in the $(\theta, h)$ plane switches from unstable to stable and a non-linear centre emerges at the transition. This bifurcation occurs for all sharp transitions between escaping and hovering in \cref{fig: main behaviour FD+SD freeslip} and is shown in \cref{fig:FD+SD spiral} for a constant $\effsource$ and increasing $\effforce$.

In summary, by accounting for the rapid oscillations of a swimmer's body shape as well as considering the effect of a higher order singularity close to the boundary, we have seen two new possible behaviours compared to the \textit{a priori}-averaged force dipole model. The possible behaviours of the \textit{a priori}-averaged system, even with the addition of a source dipole, are limited by the restriction $\effforce = \effsource = B \in (-1,1 )$. This restricted parameter space corresponds to the white line segment in each of the main behaviour plots \cref{fig: main behaviour FD+SD freeslip}. Furthermore, the behaviour predicted by the \textit{a priori}-averaged system is not necessarily representative of the equivalent systematically averaged predictions. For instance, in \cref{fig: stable behaviour FD+SD freeslip,fig: escape trajectories FD+SD freeslip} the systematically averaged system predicts  hovering and escaping trajectories, respectively, whereas the \textit{a priori}-averaged system predicts that the swimmer will crash in both cases. Hence, the multi-timescale nature of microscale swimming can fundamentally change the predicted dynamics of a swimmer but, through systematic multiscale analysis, these effects can be accounted for.

\section{The effect of a quadrupole }\label{sec: FD+Q}
\subsection{Governing equations}\label{subsec: gov eqn}
We now consider the addition of a quadrupole, which reflects effects of any fore-aft asymmetry, for example between a cell body and flagella in some bacteria. We continue to neglect the bulk swimming speed, $u = 0$, and set $\source(\omega t) = 0$ for all $t$. The equations \cref{eq: FD+SD+Q averaged freeslip final} are therefore
\begin{subequations}
\label{eq:FD+Q  averaged system final}
\begin{align}
\diff{\theta}{t} =&
 \frac{3\forceavg\sin 2\theta}{32h^3}\left[2+\effforce(1+\cos 2\theta)\right] -\frac{3\quadavg \sin \theta}{32 h^4}\left[3+5\cos 2\theta +\frac{\effquad}{4}(3\cos 4\theta + 8 \cos 2\theta -11)\right], \label{eq: FD+Q averaged system theta final}\\
\diff{h}{t} =&   \frac{\forceavg}{8h^2}\left(1+3\cos 2\theta\right) + \frac{\quadavg \cos \theta}{4h^3}(1-3\cos2\theta). \label{eq: FD+Q averaged system h final} 
\end{align}
\end{subequations}
Now, the systematically averaged system is essentially equivalent to  the \textit{a priori}-averaged system \eqref{eq:FD+SD+Q unaveraged system freeslip}, where $\force$, $\quadrupole$ and $B$ are constant, when $\force = \forceavg$, $\quadrupole = \quadavg$, $\effforce = \effquad = B\in (-1,1)$. As the governing equations \eqref{eq:FD+Q  averaged system final} are similar to those for the addition of a source dipole \eqref{eq: FD+SD averaged freeslip final}, we expect to follow a similar analysis. In particular, we note that the invariance of the system under the transformation $\quadavg \mapsto -\quadavg$, $\theta \mapsto \theta+\pi$ will simplify the following exploration and the dynamics to depend strongly on the sign of the effective force dipole strength $\forceavg$. We retain the definition of $\forceavg>0$ as an effective pusher and $\forceavg<0$ an effective puller. 

\subsection{Emergent behaviours}\label{subsec: FD+Q emergent behaviour}
\subsubsection{Existence of trivial steady states}\label{subsubsec: FD+Q trivial}
We again find that $\text{d}\theta/\text{d}t = 0$ for $\theta = n\pi$, $n \in \mathbb Z $ in \cref{eq: FD+Q averaged system theta final} for all values of $h$ and system parameters, and continue to refer to $\theta = 0,$ $ \pi$ as trivial. We assume $\quadavg>0$ as we can map to $\quadavg <0$ by shifting by $\theta$ by $\pi$. Substituting $\theta = 0,\; \pm \pi$ into \cref{eq: FD+Q averaged system h final} the trivial steady states are 
\begin{equation}
    \theta = 0, \; h = \frac{\quadavg}{\forceavg} \quad \text{and}\quad \theta =  \pi,\; h = -\frac{\quadavg}{\forceavg}.
\end{equation}
We find that the eigenvalues for the corresponding linear stability are
\begin{equation}
\lambda_0 \in \left\{  \frac{3{\forceavg}^4}{8\quadavg^3}\left(\effforce-1\right), \frac{\forceavg^4}{2\quadavg^3}\right\} \quad\text{and} \quad \lambda_{\pm \pi} = \{-\frac{3 {\forceavg}^4}{8\quadavg^3}\left(\effforce-1\right), -\frac{\forceavg^4}{2\quadavg^3}\},
\end{equation}
for the steady states corresponding to $\theta = 0$ and $\theta = \pi$, respectively.

As $h>0$ in the domain, there is a relevant trivial steady state at $\theta = 0$ for effective pushers $\forceavg>0$; however, it is always linearly unstable. For effective pullers, $\forceavg<0$, there is a relevant steady state at $\theta = \pi$ which is linearly stable when $\effforce>1$. As we require $B \in (-1, 1)$, the trivial steady states can never be linearly stable for \textit{a priori}-averaged swimmers. This directly contrasts the  addition of a source dipole, which ensures the existence of a linearly stable steady state for effective pullers in the \textit{a priori}-averaged system. 

\subsubsection{Existence of non-trivial steady states}
To find the remaining, non-trivial, steady states we proceed as before and seek solutions to $\text{d}{h}/{\text{d}t} = 0$ in \cref{eq: FD+Q averaged system h final}. The steady height as a function of $\theta $ at fixed points is
\begin{equation}\label{eq: height FD+Q steady}
h = -\frac{2\quadavg\cos\theta}{\forceavg(1+3\cos\theta)}\left(1-3\cos2\theta\right).
\end{equation}
We substitute \cref{eq: height FD+Q steady} into \cref{eq: FD+Q averaged system theta final} and factor out $3\quadavg\sin\theta/32h^4$ (the roots of which correspond to the trivial steady states characterised in \cref{subsubsec: FD+Q trivial}). Setting $x = \cos^2\theta$, we find that the remaining steady states satisfy 
\begin{equation}\label{eq: FD+Q roots}
4 x \frac{2-3 x}{3 x-1}(1+\effforce x) -1+5 x+\effquad(3 x^2-x-2)=0,
\end{equation}
which corresponds to solving the cubic equation
\begin{equation}\label{FD+Q reduced system: cubic}
(9\effquad-12\effforce)x^3 + (3+8\effforce-6\effquad)x^2-5\effquad x+1+2\effquad=0
\end{equation}
in the case $\effforce \neq -3$, and the quadratic equation
\begin{equation}\label{eq: FD+Q reduced quad}
    (3\effquad +12)x^2 - (\effquad+3)x-2\effquad-1 = 0
\end{equation}
otherwise. While explicit expressions for the roots of \cref{FD+Q reduced system: cubic,eq: FD+Q reduced quad} are possible, they are cumbersome and do not provide significant insight. However, as \cref{FD+Q reduced system: cubic,eq: FD+Q reduced quad} are in terms of  $\effforce,\, \effquad$, the existence of non-trivial steady states will only depend on these effective body shape parameters. The singularity strengths, $\forceavg,$ $ \quadavg$, only affect the linear stability when they change sign, allowing the possible  dynamics of the swimmer to be determined from the effective shape parameters $\effforce,\, \effquad$ and the sign of the effective dipole strength $\forceavg$ (since we continue to assume, without loss of generality, that $\quadavg>0$). 

\begin{figure}[hbtp]
\centering
\begin{overpic}[width = \textwidth, grid = false]{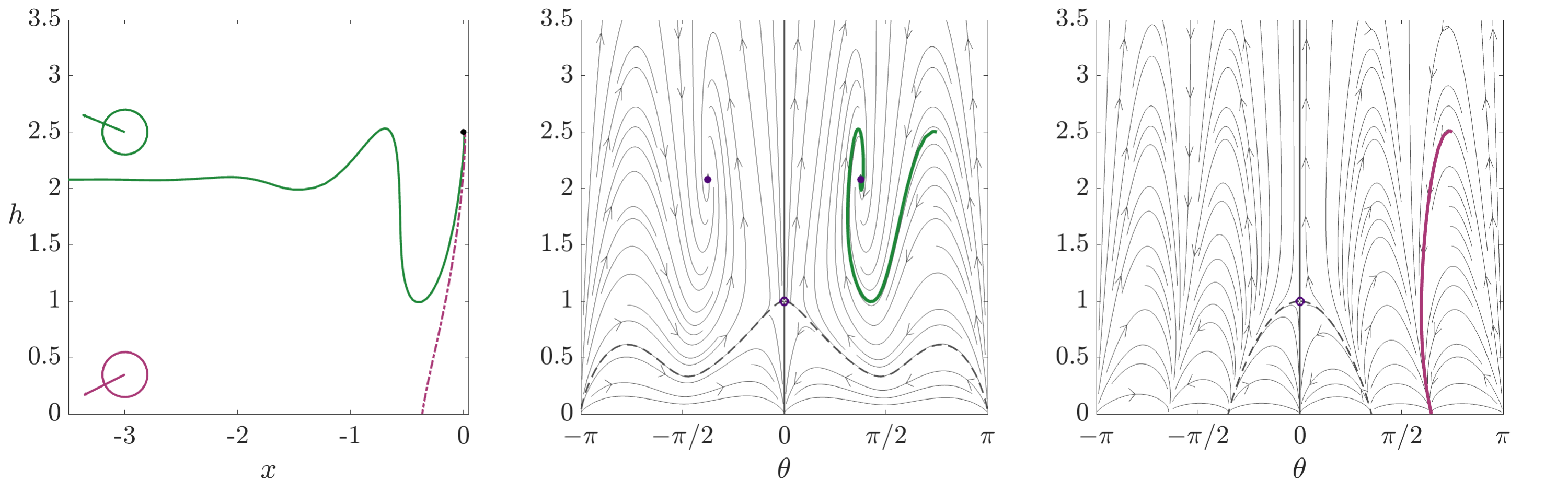}
\put(-1, 30){(a)} \put(31,30){(b)} \put(64,30){(c)}
\end{overpic}

\caption{Hovering dynamics of a single swimmer modelled as a force dipole and quadrupole near a  stress-free surface. (a) Trajectories in the $(x,h)$ plane and illustrative long-time configurations. The systematically averaged system (green) approaches a stable height, whereas the \textit{a priori}-averaged system (pink, dashed-dot) crashes into the boundary despite the same initial condition. (b,c) The $(\theta,h)$, phase portraits corresponding to the dynamics of (a) for (b) the systematically averaged system and (c) the \textit{a priori}-averaged system, with the sample trajectories of (a) shown in the phase planes. In (b), we note the local stability of the hovering state in the systematically averaged system. In (c), the dashed lines correspond to the unstable manifold for the saddle at $(\theta, h) = (\pi,0.5)$. We take $p(T) = 1+4\sin T$, $q(T) =1-4\sin T $, $B(T) = (\sin T)/2$ so that $ \forceavg = 1$, $\quadavg = 1$, $\effforce = 1$, $\effquad = -1$, $\overline{B} = 0$. }
\label{fig: stable behaviour FD+Q freeslip}
\end{figure}

\begin{figure}[hbtp]
\centering
\begin{overpic}[width = \textwidth, grid = false]{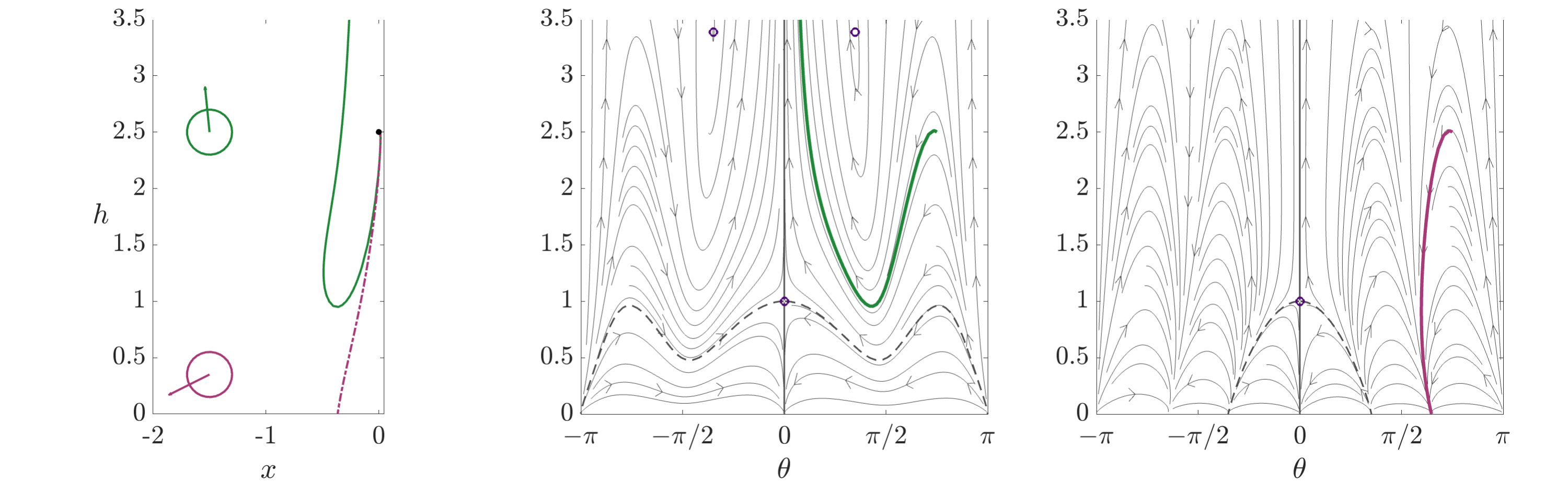}
\put(4, 30){(a)} \put(31,30){(b)} \put(64,30){(c)}
\end{overpic}

\caption{Escape dynamics of a single swimmer modelled as a force dipole and quadrupole near a  stress-free surface. (a) Trajectories in the $(x,h)$ plane and illustrative long-time configurations. The systematically averaged system (green) escapes from the boundary, whereas the \textit{a priori}-averaged system (pink, dashed-dot) crashes into the boundary despite the same initial condition. (b,c) The $(\theta,h)$, phase portraits corresponding to the dynamics of (a) for (b) the systematically averaged system and (c) the \textit{a priori}-averaged system, with the sample trajectories of (a) shown in the phase planes.  In (c), the dashed lines correspond to the unstable manifold for the saddle at $(\theta, h) = (\pi,0.5)$. We take $p(T) = 1-6\sin T$, $q(T) = 1 -4\sin T $, $B(T) = \sin T/2$ so that $ \forceavg = 1$, $\quadavg = 1$, $\effforce = -1.5$, $\effquad = -1$, $\overline{B} = 0$. }
\label{fig: escape behaviour FD+Q freeslip}

\end{figure}

For different combinations of body shape parameters, we solve for the steady states numerically with the \texttt{roots} function in MATLAB \cite{MATLAB}. We then calculate the corresponding Jacobian to determine the linear stability. As in \cref{sec: FD+SD}, we find stable steady states where the swimmer is tilted towards or away from the boundary at a fixed height. At these steady states, the swimmer hovers above the boundary and translates horizontally along it as in \cref{fig: stable behaviour FD+Q freeslip}. There are also combinations of body shape parameters at which there are no stable steady states and instead the swimmer escapes from the boundary regardless of the initial angle such as in \cref{fig: escape behaviour FD+Q freeslip} as well as examples where the swimmer crashes into the boundary as in the \textit{a priori}-averaged force dipole model.

\begin{figure}[hbtp]
\centering
\begin{subfigure}[hbtp]{0.35\textwidth}
  \centering
  \begin{overpic}[width=\textwidth, grid = false]{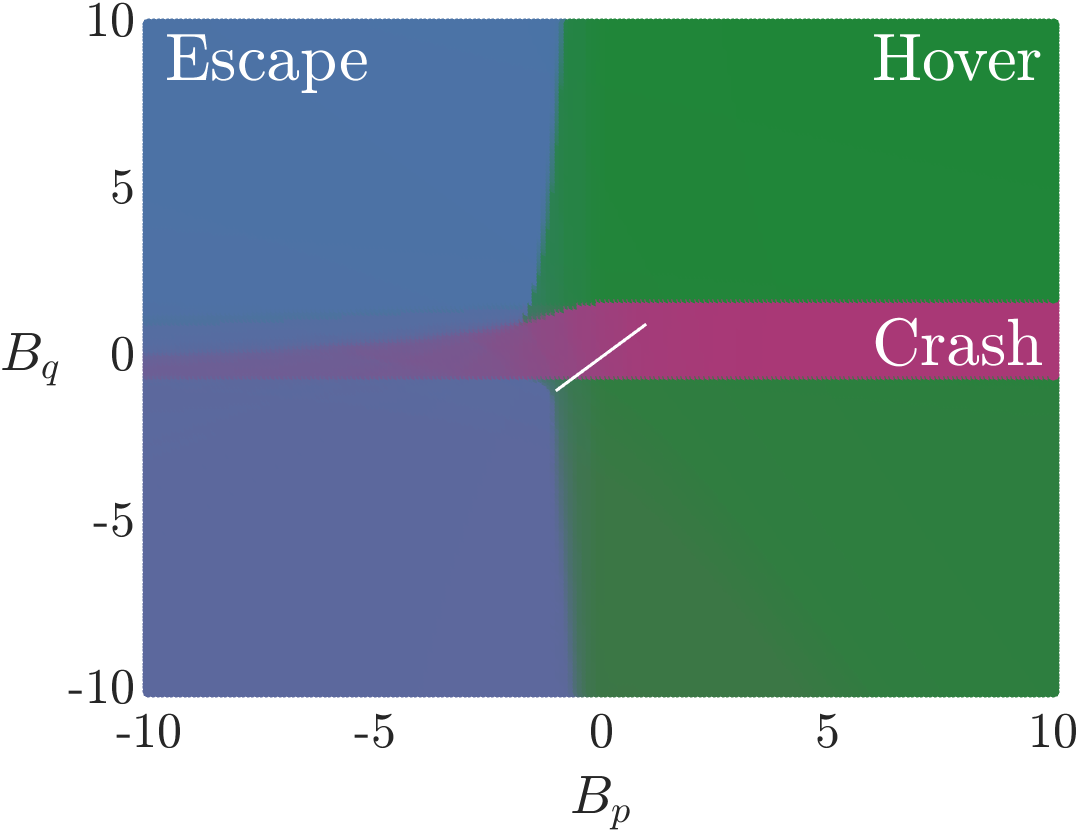}
  \put(0,75){(a)} \put(106,75){(b)}\put(40,78){Pusher $\avg{p}>0$}
  \end{overpic}

  \label{fig: FD+Q main dipole pos}
\end{subfigure}
\hfill
\begin{subfigure}[hbtp]{0.35\textwidth}
  \centering
  \begin{overpic}[width=0.935\textwidth, grid = false]{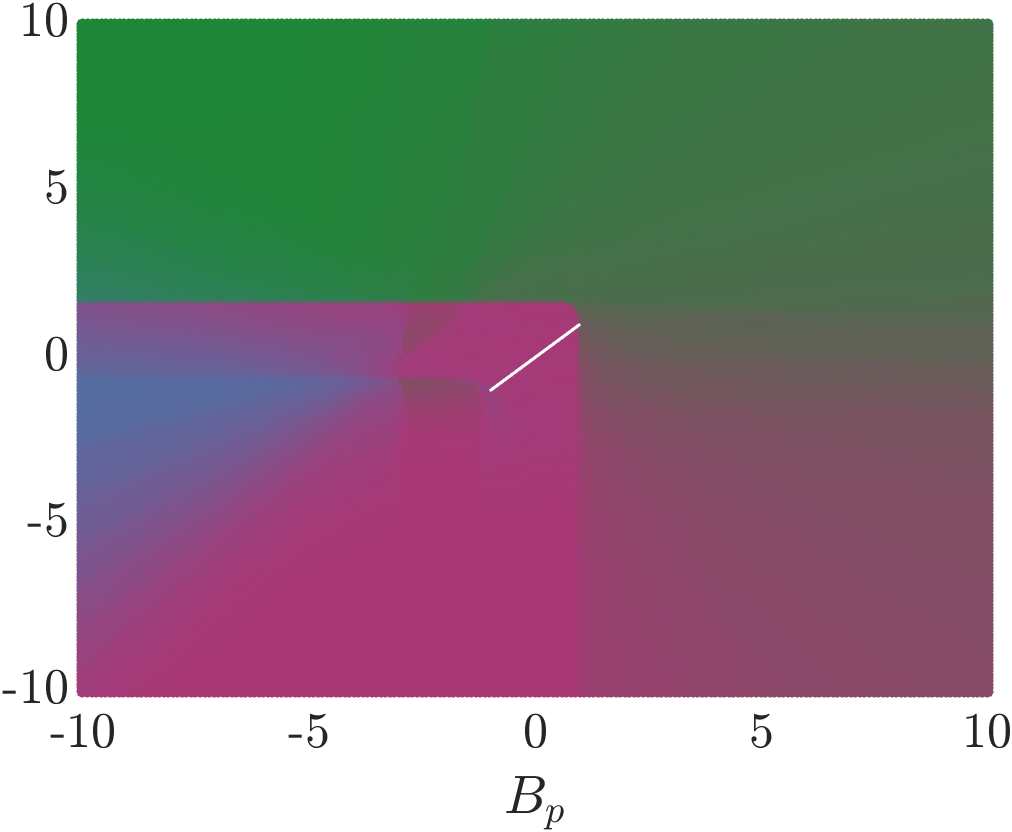}
  \put(40,83){Puller $\avg{p}<0$}
  \end{overpic}
  \label{fig: FD+Q main dipole neg}
\end{subfigure}
\begin{subfigure}[hbtp]{0.25\textwidth} 
\centering
\includegraphics[width = \textwidth]{{Images/triangle_updated_legend.pdf}}

\end{subfigure}
\caption{The proportion of the main behaviours (hovering, escaping or crashing)  for a single swimmer modelled as a force dipole and quadrupole above a stress-free surface. In panel (a), the swimmer is an effective pusher, $\forceavg>0$; in panel (b), the swimmer is an effective puller $\forceavg<0$. The possible dynamics of the \textit{a priori}-averaged system correspond to the white line segments in (a), (b) which typically do not reflect the proportions observed in the wider $(\effforce,\effsource)$ space of the systematically averaged system.  Each colour denotes a unique proportion of behaviours with all possibilities shown in the colourmap. We take $\quadavg = 1$, $\forceavg = \pm 1$ and vary $\effforce$ and $\effquad$ between $-10$ and $10$ beyond which the trends continue. } 
\label{fig: main behaviour FD+Q freeslip}

\end{figure}

\begin{figure}[hbtp]
\centering
\begin{overpic}[width = \textwidth, grid = false]{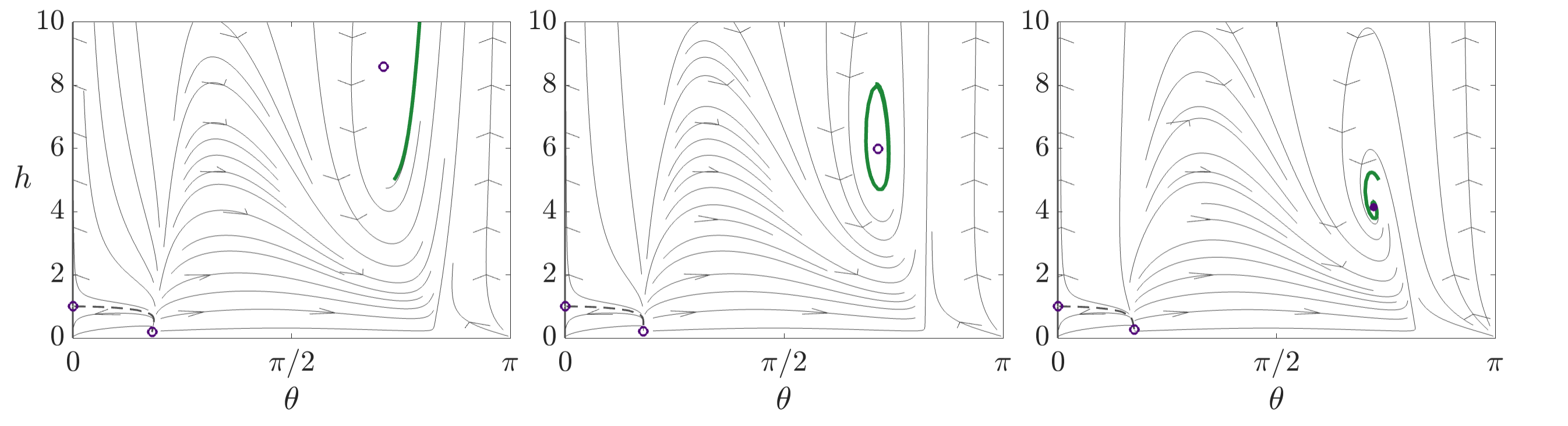}
\put(-0.5,26.5){(a)} \put(31,26.5){(b)} \put(62.5,26.5){(c)}
\end{overpic}
\caption{ Phase portraits in $(\theta, h)$ space showing the transition from (a) a linearly unstable steady state, to (c) a linearly stable steady state, with a (b) non-linear centre, for a swimmer modelled as a force dipole and quadrupole. An example  trajectory is shown on each phase portrait demonstrating the linear stability of the fixed point. We take $\forceavg =1$, $\quadavg = 1$, $\effquad = 2.2$ and  $\effforce = -1.8,\, -1.43, \, -1$ in (a), (b) and (c), respectively. }
\label{fig:FD+Q spiral}
\end{figure}
Therefore, we have found that the effect of the quadrupole investigated in this section generates the same three emergent behaviours as the source dipole case considered in \cref{sec: FD+SD}. We proceed as in \cref{subsec: FD+SD emergent behaviour} and repeat the above parameter sweep over the effective body shape parameters ($\effforce$ and $\effquad$) now also determining the proportion of each behaviour (hovering, escaping or crashing). As in \cref{sec: FD+SD}, \cref{fig: main behaviour FD+Q freeslip} now shows that for both pushers and pullers, crashing (pink) is rare for a large proportion of the parameter space. However, contrasting the source dipole case, we also observe from \cref{fig: main behaviour FD+Q freeslip}(a), for effective pushers there appears to be a finite effective quadrupole shape, $\effquad$, above which the swimmer is never predicted to crash into the boundary regardless of $\effforce$. Whilst this bifurcation is unique to the addition of a quadrupole, other bifurcations are similar to those found in the source dipole case. In particular, we again highlight the bifurcation that occurs for all transitions between escaping and hovering in  \cref{fig: main behaviour FD+Q freeslip}(a). As for the source dipole, \cref{fig:FD+Q spiral} shows the emergence of a non-linear centre at the transition in the stability of steady states in the $(\theta,h)$ plane.

Overall, for the addition of either a source dipole or a quadrupole, fundamentally different behaviours can occur in the systematically averaged system compared to the \textit{a priori}-averaged system. For the \textit{a priori}-averaged force dipole and quadrupole model, the possible dynamics correspond to the white line segment in the expanded parameter space of \cref{fig: main behaviour FD+Q freeslip}. Similar to the source dipole case, we note that the behaviours corresponding to this restricted line segment are not necessarily illustrative of the possible dynamics in the wider, systematically averaged system. Therefore, by  accounting for multi-timescale effects, fundamentally different behaviours can arise in the systematically averaged system compared to the \textit{a priori}-averaged system. 

\section{Behaviour near a no-slip interface}\label{sec: no slip}
Until this point, we have only considered the dynamics of a stress-free boundary condition. However, a no-slip boundary condition is often appropriate, for example, experimentally on a microscope slide. Due to the prevalence of this no-slip boundary condition, we briefly modify our earlier analysis to account for this, and report the effects on long-term behaviours. We continue to assume that $u = 0$ and examine the effect of the source dipole and quadrupole terms individually. 
\subsection{Emergent behaviours from a source dipole}\label{subsec: FD+SD no slip}
For the linear combination of a force dipole and source dipole, the systematically averaged system \cref{eq:FD+SD+Q averaged system noslip} is now
\begin{subequations}
\label{eq:FD+SD avg system noslip}
\begin{align}
\diff{\theta}{t} =&
 \frac{3\forceavg\sin 2\theta}{64h^3}\left[4+\effforce(3+\cos2\theta)\right] - \frac{3\sourceavg\sin\theta}{32h^4}\left[4+3\effsource(3+\cos2\theta)\right], \label{eq:FD+SD avg theta noslip} \\
\diff{h}{t} =&   \frac{3\forceavg}{16h^2}\left(1+3\cos2\theta \right)  -\frac{\sourceavg\cos\theta}{h^3}. \label{eq:FD+SD avg h noslip} 
\end{align}
\end{subequations}
 We again have that if $\forceavg = \force,\, \sourceavg = \source$ and $\effforce = \effsource = B\in (-1, 1)$, the systematically averaged system projects onto the (original) \textit{a priori}-averaged system. Compared to the equations for a stress-free boundary condition \cref{eq: FD+SD averaged freeslip final}, the equations for a no-slip \cref{eq:FD+SD avg system noslip} differ in an additional $\sourceavg$ term in the $\theta$-evolution equation \eqref{eq:FD+SD avg theta noslip} and the constants only. We therefore expect similar conclusions to those in \cref{sec: FD+SD} to hold.  In particular, we note that the system \cref{eq:FD+SD avg system noslip} is invariant under the transformation, $\sourceavg \mapsto -\sourceavg,$ $\theta \mapsto - \theta$. Therefore, as in the stress-free case, we expect the possible behaviours of the swimmer to be determined from the effective shape parameters $\effforce,\, \effsource$, and the sign of the effective dipole strength, $\forceavg$. We therefore continue to assume that $\sourceavg<0$, noting that cases with $\sourceavg>0$ can be mapped to $\sourceavg<0$ by shifting $\theta$ by $\pi$.

The trivial roots, $\theta = n\pi$, $n \in \mathbb Z$ are  steady states of the angular evolution equation \cref{eq:FD+SD avg theta noslip}. We therefore define the trivial steady states for a no-slip interface \cref{eq:FD+SD avg system noslip} to be
\begin{align}
 \theta = 0, \; h = \frac{4\sourceavg}{3\forceavg} \quad\text{ and }\quad \theta =  \pi,\; h = -\frac{4\sourceavg}{3\forceavg}.
\end{align}

For $h>0$ in the domain, similar analysis to the free-slip case \cref{subsubsec: FD+SD trivial,subsubsec: FD+Q trivial} shows that for effective pushers, $\forceavg>0$, again there are no linearly stable trivial steady states. Now, for effective pullers, $\forceavg<0$, a state trivial steady state exists if $4\effforce-9\effsource+1>0$. This corresponds to $B<1/5$ in the \textit{a priori}-averaged system.

To find the remaining, non-trivial, steady states, we consider the $h$ evolution equation \cref{eq:FD+SD avg h noslip} and note that the steady height as a function of theta is given by
\begin{equation}\label{eq: FD+SD steady height no slip}
  h = \frac{8 \sourceavg\cos \theta}{3\forceavg(3\cos^2\theta -1)}.   
\end{equation}
Substituting \cref{eq: FD+SD steady height no slip} into \cref{eq:FD+SD avg theta noslip}, we factor out $3\sourceavg\sin\theta/(32h^4)$, roots of which coincide with the trivial steady states.  After  substituting $x = \cos^2\theta$, the steady states are solutions to the quadratic equation
\begin{equation}   
(8\effforce-27\effsource)x^2 +(8\effforce-18\effsource-2)x+9\effsource+6 = 0.
\end{equation}

We therefore have that the existence of steady states only depends on the effective body shape parameters, $\effforce$, $\effsource$.  Therefore, regardless of the boundary condition, the possible dynamics of the swimmer depend on whether the swimmer is an effective pusher, $\forceavg>0$, or puller, $\forceavg<0$, as well as the effective shape parameters, $\effforce$, $\effsource$. An exploratory numerical sweep of the steady states and their corresponding linear stability suggests that the main behaviours are hovering, escaping and crashing, as before. We therefore perform a thorough parameter and behaviour sweep as in \cref{subsec: FD+SD emergent behaviour} to determine the most prevalent behaviours for different effective body shapes, $\effforce$, $\effsource$.
\begin{figure}[hbtp]
\centering
\begin{subfigure}[hbtp]{0.35\textwidth}
  \centering
  \begin{overpic}[width = \textwidth, grid = false]{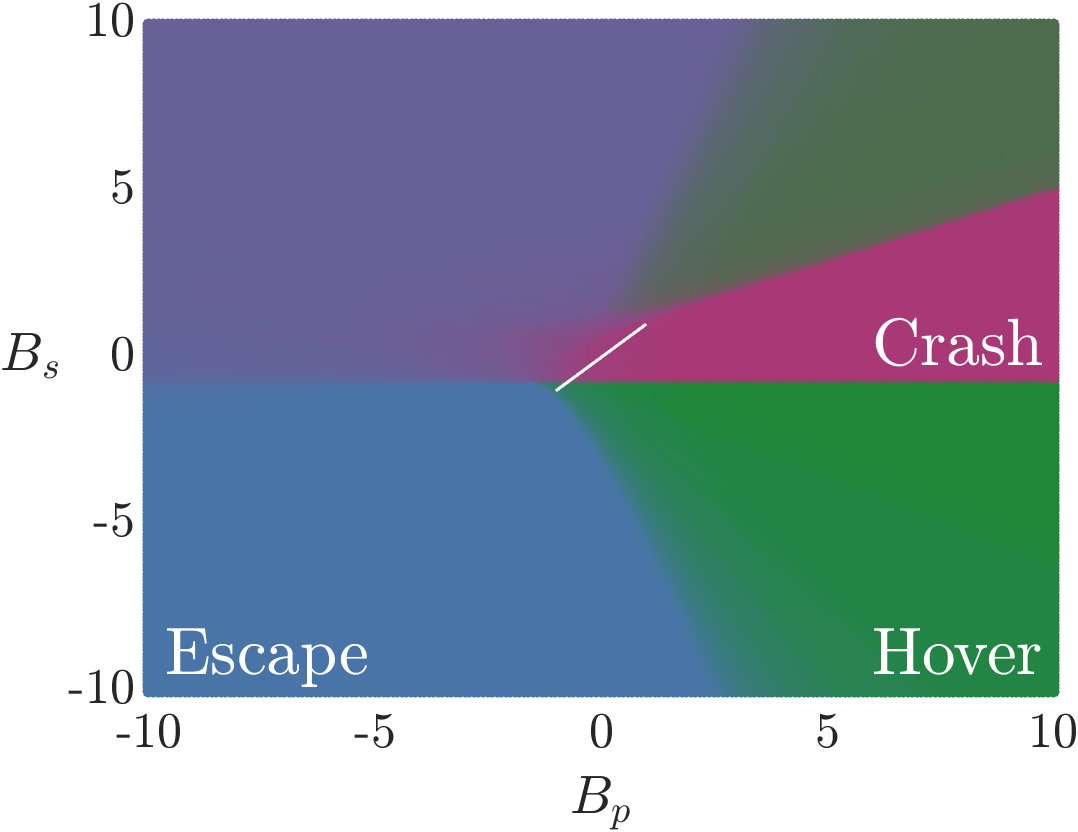}
  \put(-0.5,75){(a)} \put(108,75){(b)} \put(40,78){Pusher $\avg{p}>0$} 
\end{overpic}

\end{subfigure}
\hfill
\begin{subfigure}[hbtp]{0.35\textwidth}
  \centering
  \begin{overpic}[width=0.935\textwidth, grid = false]{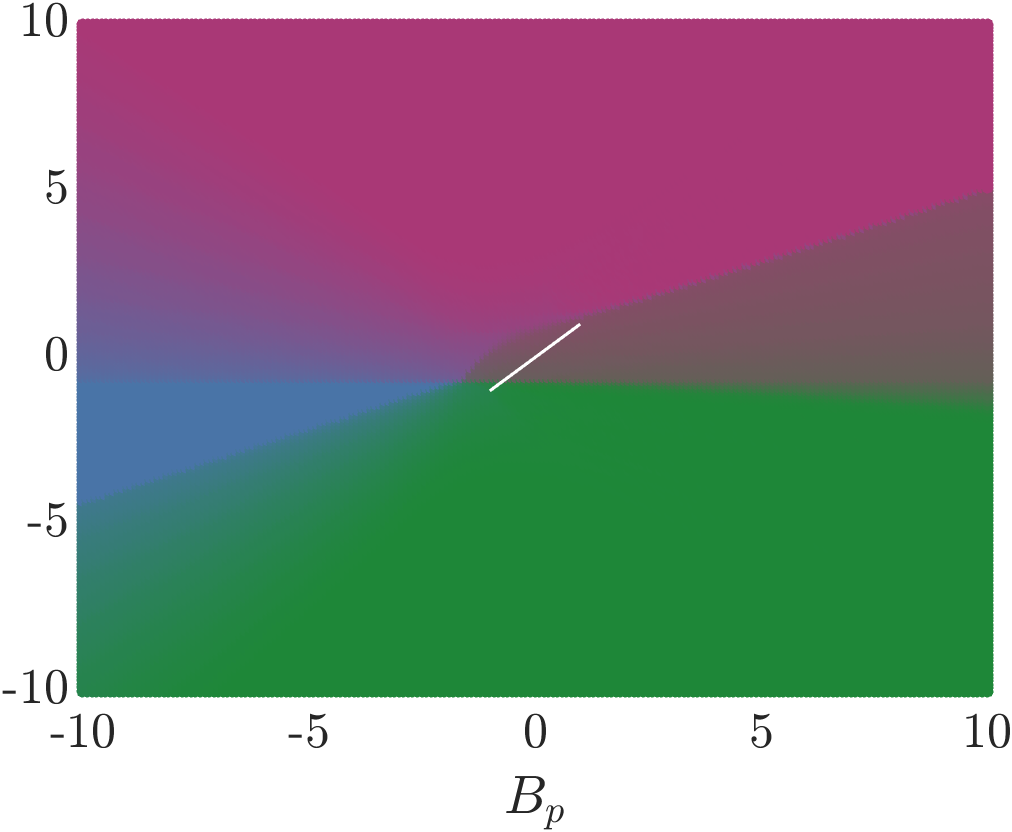}
\put(40,83){Puller $\avg{p}<0$}
\end{overpic}
\end{subfigure}
\begin{subfigure}[hbtp]{0.25\textwidth} 
\centering
\includegraphics[width = \textwidth]{{Images/triangle_updated_legend.pdf}}
\end{subfigure}
\caption{The proportion of the main behaviours (hovering, escaping or crashing)  for a single swimmer modelled as a force dipole and source dipole above a no-slip surface. In panel (a), the swimmer is an effective pusher, $\forceavg>0$; in panel (b), the swimmer is an effective puller $\forceavg<0$. The possible dynamics of the \textit{a priori}-averaged system correspond to the white line segments in (a), (b) which do not generally reflect the proportions observed in the wider $(\effforce,\effsource)$ space of the systematically averaged system. We take $\sourceavg = -1$, have $\forceavg = \pm 1$ and vary $\effforce$, and $\effsource$ between $-10$ and $10$ beyond which the trends continue. }
\label{fig: FD+SD main behaviour noslip}

\end{figure}

Despite changing the interface, the possible behaviours  for a no-slip boundary condition (\cref{fig: FD+SD main behaviour noslip}) is qualitatively similar to those observed in the stress-free case (\cref{fig: main behaviour FD+SD freeslip}). However, there are now more instances of ambiguous regions such as the purple region in \cref{fig: FD+SD main behaviour noslip}(a), where the swimmer may either crash into or escape from the boundary depending on the initial height and orientation. Compared to the \textit{a priori}-averaged system (indicated by the white line segments in \cref{fig: FD+SD main behaviour noslip}), by systematically including multi-timescale effects, qualitatively different behaviours can arise regardless of the boundary condition. 

\subsection{Emergent behaviours from a quadrupole}\label{subsec: FD+Q no slip}
For a swimmer assumed to be the linear combination of a force dipole and a quadrupole, we find \cref{eq:FD+SD+Q averaged system noslip} becomes
\begin{subequations}
\label{eq:FD+Q avg system noslip}
\begin{align}
\diff{\theta}{t} =&
 \frac{3\forceavg\sin 2\theta}{64h^3}\left[4+\effforce(3+\cos2\theta)\right] - \frac{3\quadavg\sin\theta}{16h^4}\left[1+3\cos 2\theta  +\frac{\effquad}{16}(3\cos4\theta +12\cos2\theta -79)\right], \label{eq:FD+Q avg theta noslip} \\
\diff{h}{t} =&   \frac{3\forceavg}{16h^2}\left(1+3\cos2\theta \right)  +\frac{\quadavg\cos\theta}{8h^3}\left(5-9\cos 2\theta\right)+u\cos\theta. \label{eq:FD+Q avg h noslip} 
\end{align}
\end{subequations}
The  equations for  stress-free \cref{eq:FD+Q  averaged system final} and no-slip \cref{eq:FD+Q avg system noslip} only differ in the values of the constants. We therefore proceed as in \cref{subsec: FD+Q emergent behaviour} and again find that the number of steady states continues to depend only on the effective body shape parameters, $\effforce,\, \effquad$.

After exploring the steady states and their linear stability numerically, we find that the most prevalent behaviours are hovering, escaping and crashing, as before. We therefore perform a parameter and behaviour sweep as detailed in \cref{subsec: FD+SD emergent behaviour} to determine the most prevalent behaviour for different combinations of effective body shape parameters, $\effforce$,  $\effquad$.

\begin{figure}[hbtp]
\centering
\begin{subfigure}[hbtp]{0.35\textwidth}
  \centering
    \begin{overpic}[width = \textwidth, grid = false]{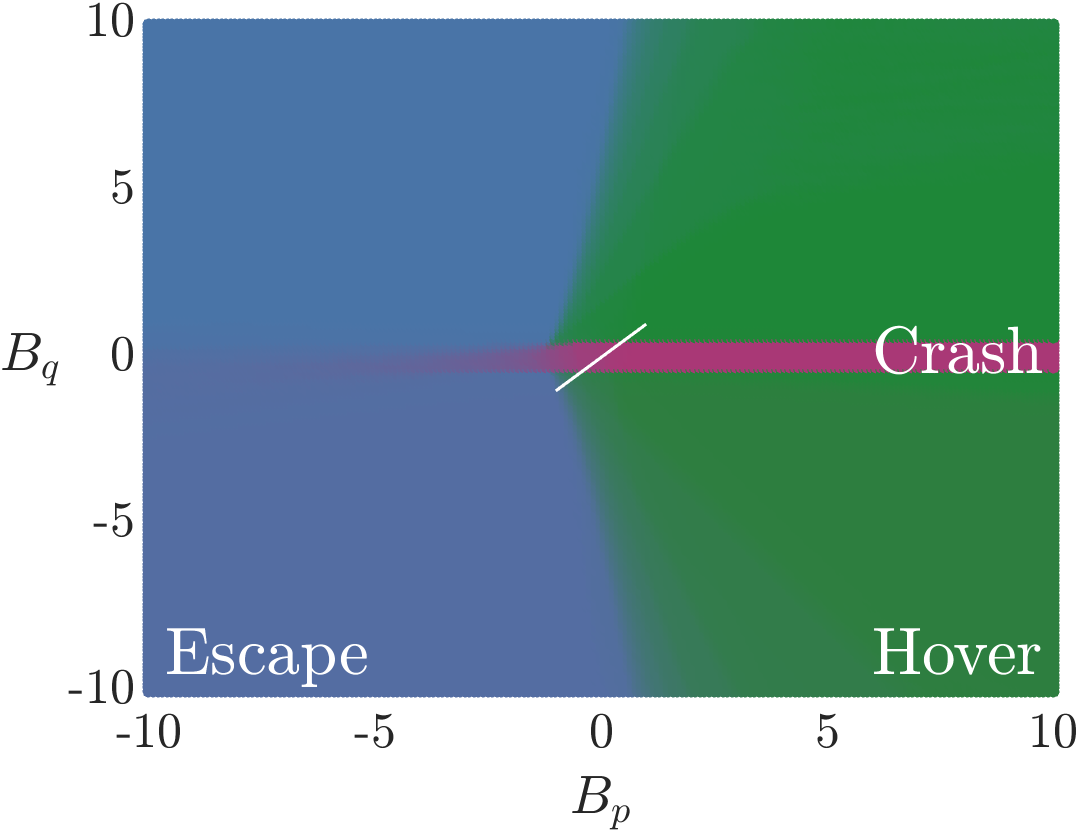}
  \put(-0.5,75){(a)} \put(108,75){(b)} \put(40,78){Pusher $\avg{p}>0$} 
\end{overpic}
\end{subfigure}
\hfill
\begin{subfigure}[hbtp]{0.35\textwidth}
  \centering
  \begin{overpic}[width=0.935\textwidth, grid = false]{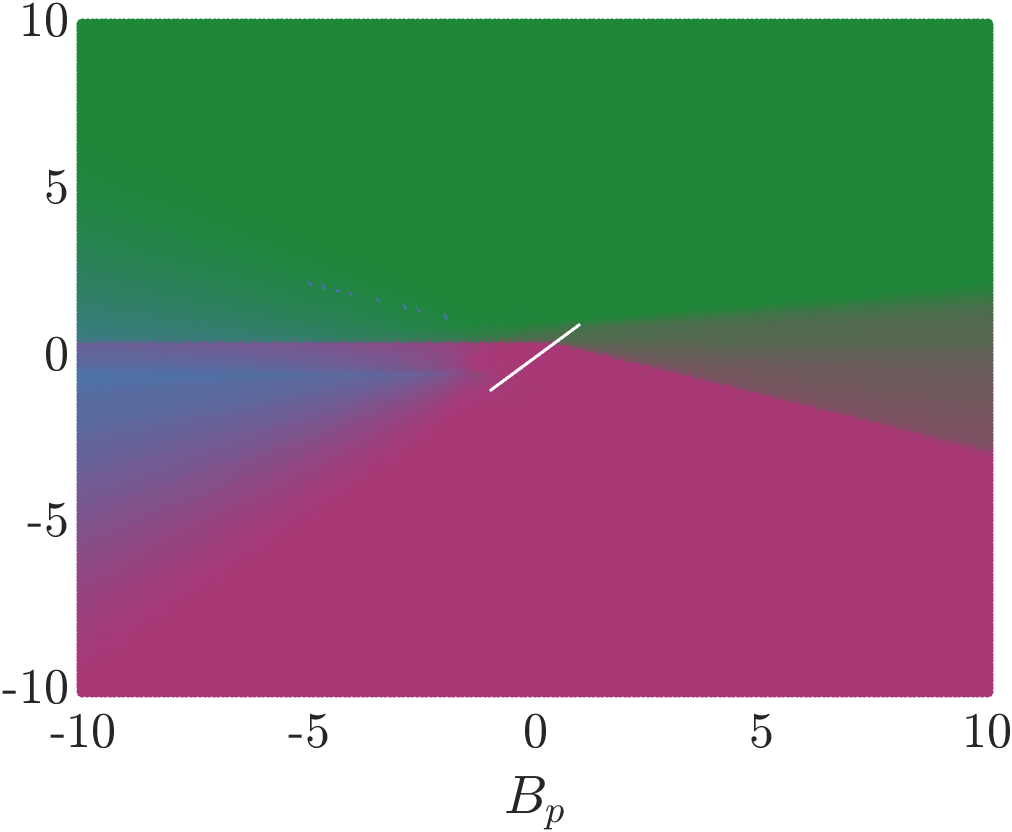}
\put(40,83){Puller $\avg{p}<0$}
\end{overpic}
\end{subfigure}
\begin{subfigure}[hbtp]{0.25\textwidth} 
\centering
\includegraphics[width = \textwidth]{{Images/triangle_updated_legend.pdf}}
\end{subfigure}
\caption{The proportion of the main behaviours (hovering, escaping or crashing)  for a single swimmer modelled as a force dipole and quadrupole above a no-slip surface. In panel (a), the swimmer is an effective pusher, $\forceavg>0$; in panel (b), the swimmer is an effective puller $\forceavg<0$. The possible dynamics of the \textit{a priori}-averaged system correspond to the white line segments in (a), (b) which generally do not reflect the proportions observed in the wider $(\effforce,\effsource)$ space of the systematically averaged system. We take $\quadavg = 1$, have $\forceavg = \pm 1$ and vary $\effforce$, and $\effquad$ between $-10$ and $10$ beyond which the trends continue.}
\label{fig: FD+Q main behaviour noslip}

\end{figure}

Again the predictions for the systematically averaged system for a stress-free (\cref{fig: main behaviour FD+Q freeslip}) and no-slip surface (\cref{fig: FD+Q main behaviour noslip}) are qualitatively similar. For both boundary conditions, by systematically including multi-timescale effects, fundamentally different behaviours are able to occur compared to the \textit{a priori}-averaged system, indicated by the white line segment in \cref{fig: FD+Q main behaviour noslip}. In particular, regardless of the boundary condition considered, crashing becomes much less frequent for effective pullers, $\forceavg>0$ (Figures \ref{fig: main behaviour FD+Q freeslip}(a) and \ref{fig: FD+Q main behaviour noslip}(a)).

\section{Including a non-zero swimming speed}\label{sec: nonzero swim}
Until now, our analysis has only considered $u = 0$ and, hence, the swimmer trajectories are only due to hydrodynamic interactions with the boundary.  We investigate how a non-negligible swimming speed changes the predicted dynamics of the systematically averaged systems \cref{eq: FD+SD averaged freeslip final,eq:FD+Q  averaged system final} in the case of a stress-free boundary. For no-slip boundaries, the predicted dynamics are qualitatively similar to stress-free predictions, as was also the case for negligible swimming speed discussed in \cref{sec: no slip}. 

\subsection{Emergent behaviours from a source dipole}\label{subsec: FD+SD+U}
For the linear combination of a force dipole and source dipole, the $h$ evolution equation \cref{eq: FD+SD avg freeslip h 0 dropped} is now
\begin{equation}\label{eq: FD+SD h eqn with swim}
   \diff{h}{t} = \frac{\forceavg}{8h^2}\left(1+3\cos 2\theta\right)  -\frac{\sourceavg\cos\theta}{4h^3} +u\cos\theta, 
\end{equation}
reintroducing the swimming speed from \cref{eq: FD+SD+Q averaged freeslip final}. Combining \cref{eq: FD+SD h eqn with swim} with the $\theta$ evolution equation \cref{eq: FD+SD avg freeslip theta 0 dropped}, we now have that the system is invariant under the transformation $\sourceavg \mapsto -\sourceavg,\, u \mapsto -u,\, \theta \mapsto \theta +\pi$ and therefore assume that $\sourceavg<0$.

\subsubsection{Trivial steady states}\label{subsubsec: FD+SD+u trivial}
As the $\theta$ equation \cref{eq: FD+SD avg freeslip theta 0 dropped} does not depend on the swimming speed, we have the same trivial steady states $\theta=0$, $\theta =  \pi$ as \cref{subsec: FD+SD emergent behaviour}. Now, however, the corresponding heights are solutions to cubic polynomials:
\begin{subequations}\label{eq: FD+SD+u steady h cubic}
\begin{align}
 & \quad4uh^3+2\forceavg h-\sourceavg = 0,  \\  & \quad
    -4uh^3+2\forceavg h +\sourceavg = 0,
\end{align}
\end{subequations}
 for $\theta = 0$ and $\theta = \pi$, respectively. For $h>0$ in the domain, we find three cases: no solution to either equation, a solution to both of the equations, or two solutions from one equation. In particular, the scenario of two solutions from one equation occurs when the turning point of the equation \cref{eq: FD+SD+u steady h cubic} crosses the $h$-axis. By considering the derivative of each cubic equation, zeros of which are the location of the turning point, we find that this saddle-node bifurcation occurs for the relevant  $\theta = 0$ steady state when $\forceavg <0$ and $u {\sourceavg^2}/{\forceavg^3} = -8/27 $. For the relevant steady state corresponding to   $\theta =  \pi$, the bifurcation occurs for effective pushers $\forceavg>0$ and $u {\sourceavg^2}/{\forceavg^3} = 8/27 $. We write $\effswim = u {\sourceavg^2}/{\forceavg^3} $ and refer to  $\effswim$ as the effective swimming speed of the swimmer.

For general singularity strengths, we can determine the heights and calculate the corresponding linear stabilities numerically. We note that only the signs of the singularity strengths affect the existence of steady states and, therefore, we choose one $\forceavg>0$ and one $\forceavg<0$ and vary the swimming speed.  We determine the linear stability of the trivial steady states by numerically evaluating the corresponding Jacobian in a wider parameter sweep for the non-trivial steady states as well.

\subsubsection{Non-trivial steady states}

For the remaining, non-trivial, steady states, we seek the steady values of $h$ as a function of $\theta$. However, the $h$ equation \cref{eq: FD+SD h eqn with swim}  corresponds to solving   the cubic equation
\begin{equation}\label{eq: FD+SD+swim h cubic}
   8 u\cos\theta h^3+\forceavg(1+3\cos2\theta)h-2\sourceavg\cos\theta =0.
\end{equation}
Instead of seeking solutions directly, we factor out $3\sin\theta/32h^4$ from the $\theta$ equation \cref{eq: FD+SD avg freeslip theta 0 dropped} and write 
\begin{equation}
    h = \frac{\effsource \sourceavg}{2\forceavg\cos\theta }\frac{3+\cos 2\theta}{2+\effforce(1+\cos2\theta)},
\end{equation}
which we substitute into \cref{eq: FD+SD+swim h cubic}. After factoring out $\effsource\sourceavg/(8\cos^2\theta)$ and substituting $x = \cos\theta$, we seek solutions to the polynomial
\begin{equation}\label{eq: FD+SD+u poly}
    x(1+x^2)(3x^2-1)(1+\effforce x^2)^2-\frac{2}{\effsource}x^3(1+\effforce x^2)^3 +\effswim\effsource^2 (1+x^2)^3 = 0.
\end{equation}
Here we have maintained the definition for effective swimming speed, $\effswim = u {\sourceavg^2 }/{\forceavg^3}$ from \cref{subsubsec: FD+SD+u trivial}. This allows us to write \cref{eq: FD+SD+u poly} in terms of three parameters only, $\effforce,\,\effsource,\, \effswim$. Therefore, the possible dynamics depend on the two effective body shape parameters, $\effforce,\, \effsource$, the effective swimming speed, $\effswim$,  and if the swimmer is an effective puller, $\forceavg>0$ or pusher, $\forceavg<0$. We note that when $\forceavg<0$, a negative $\effswim$ corresponds to a positive swimming speed, that is $u >0$.

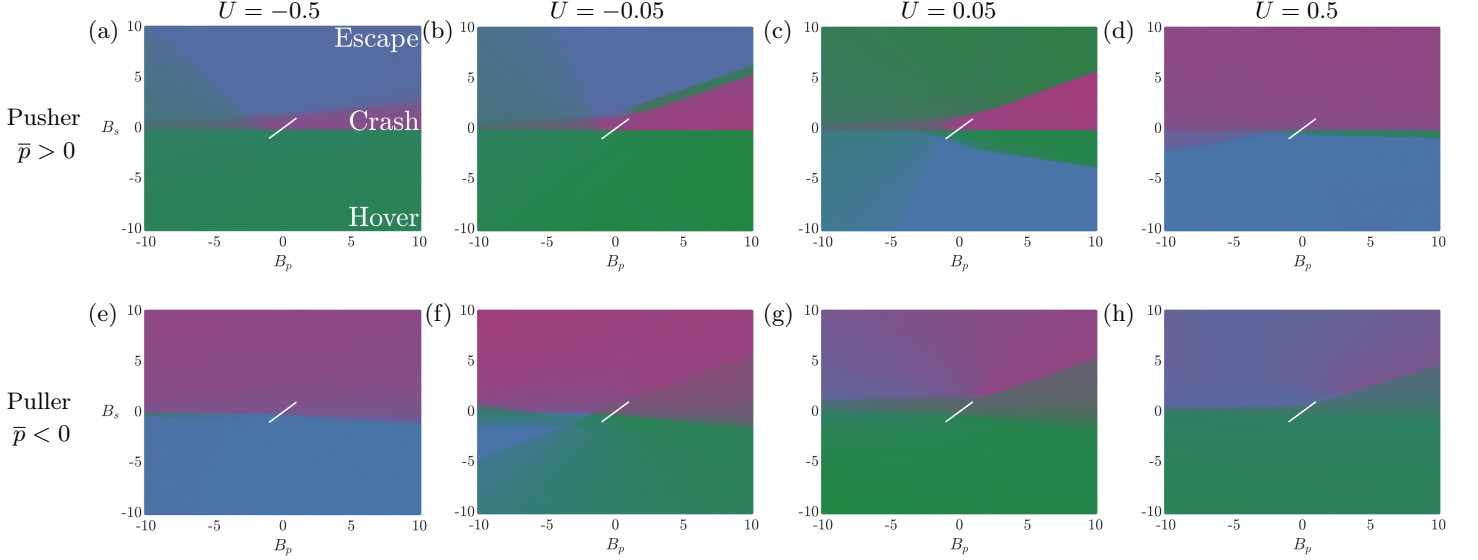
\begin{figure}[hbtp]

\vspace{0.1cm}
\centering
\begin{subfigure}[hbtp]{0.24\textwidth}
  \centering 
     \begin{overpic}[width = \textwidth, grid = false]{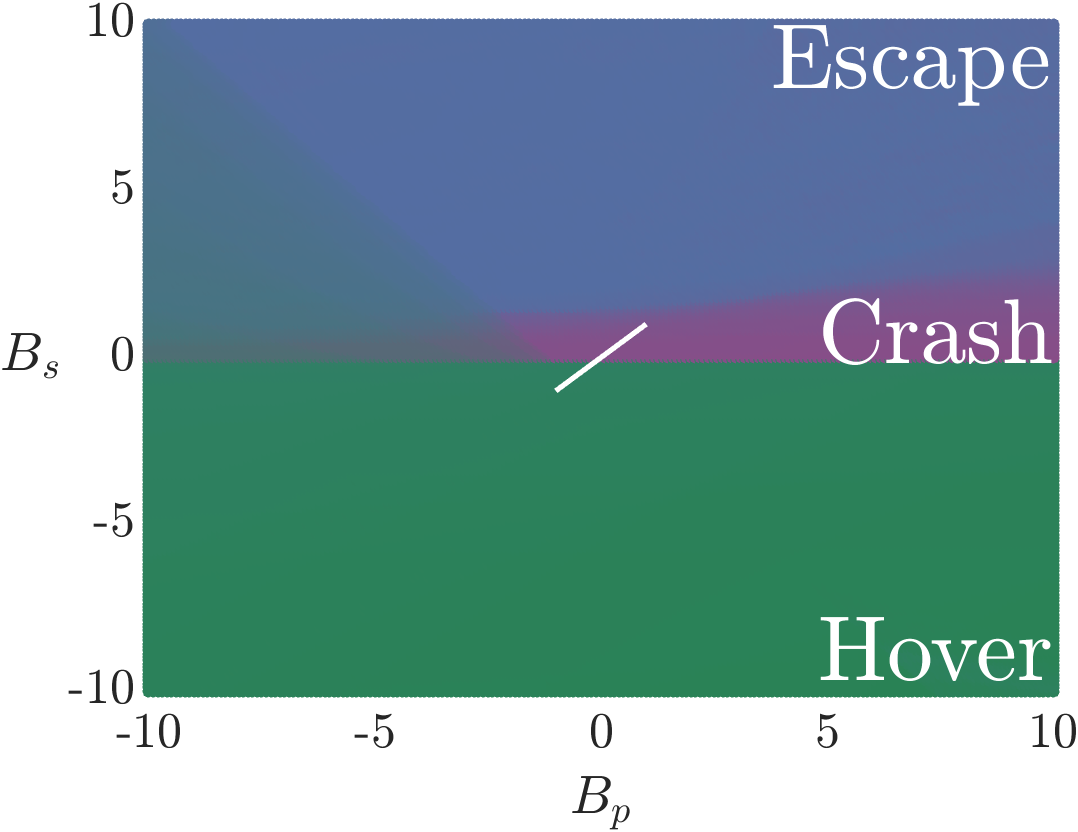}
  \put(-4,72){(a)}\put(203.5,72){(c)} \put(308,72){(d)} \put(-29,45){Pusher} \put(-28,35){ $\forceavg >0$}  \put(36,78){$U = -0.5$} \put(138,78){$U = -0.05$} \put(246,78){$U = 0.05$}
      \put(355,78){$U = 0.5$}
\end{overpic}
\end{subfigure}
\hfill
\begin{subfigure}[hbtp]{0.24\textwidth}
  \centering
  \begin{overpic}[width=0.935\textwidth, grid = false]{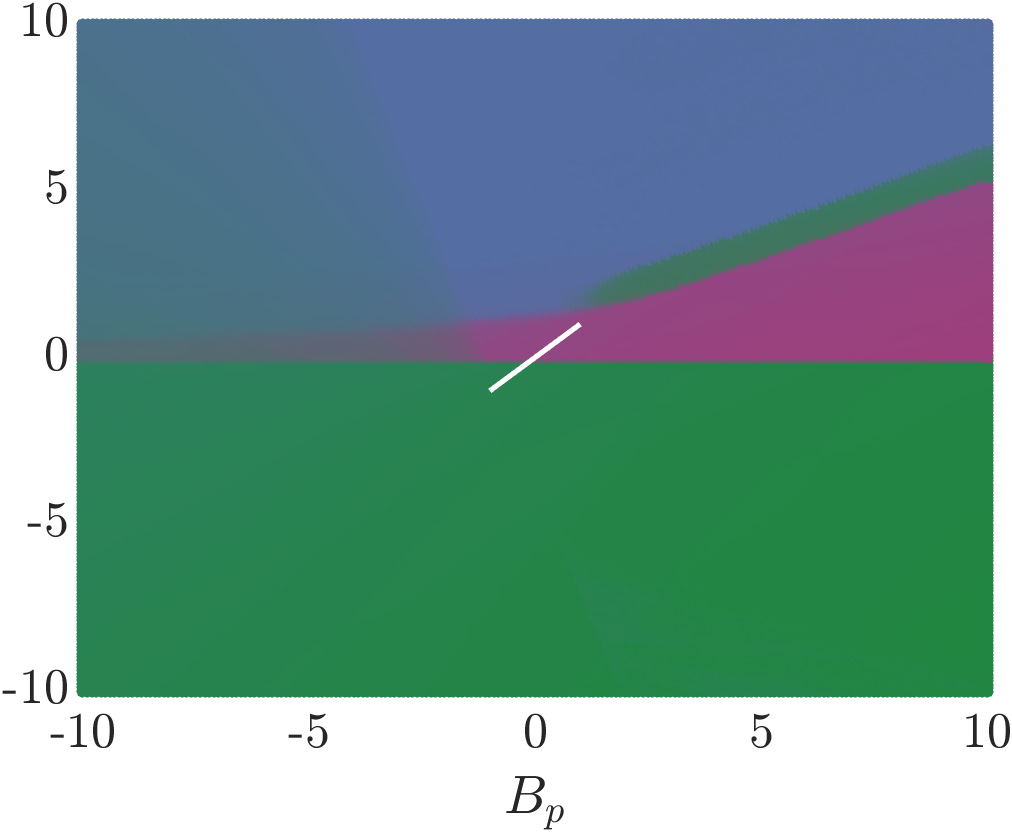} 
  \put(-11, 76.75){(b)} 
  \end{overpic}

\end{subfigure}
\hfill
\begin{subfigure}[hbtp]{0.24\textwidth}
  \centering
  \includegraphics[width=0.935\textwidth]{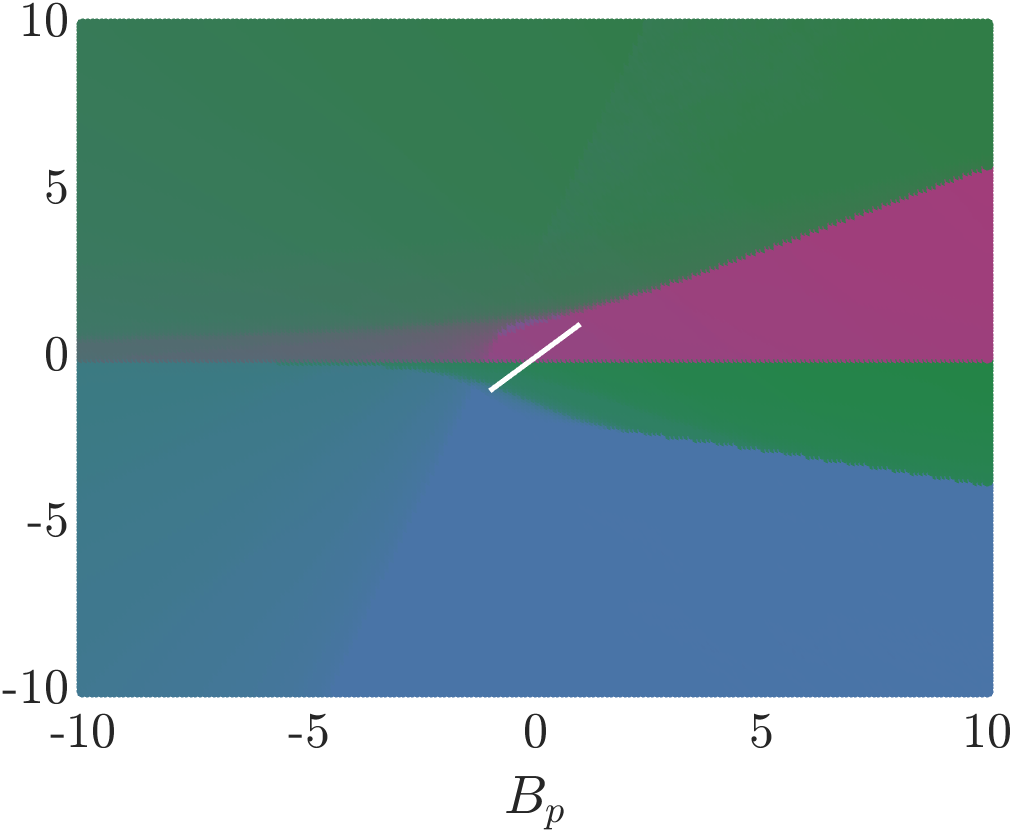}

\end{subfigure}
\hfill
\begin{subfigure}[hbtp]{0.24\textwidth}
  \centering
  \includegraphics[width=0.935\textwidth]{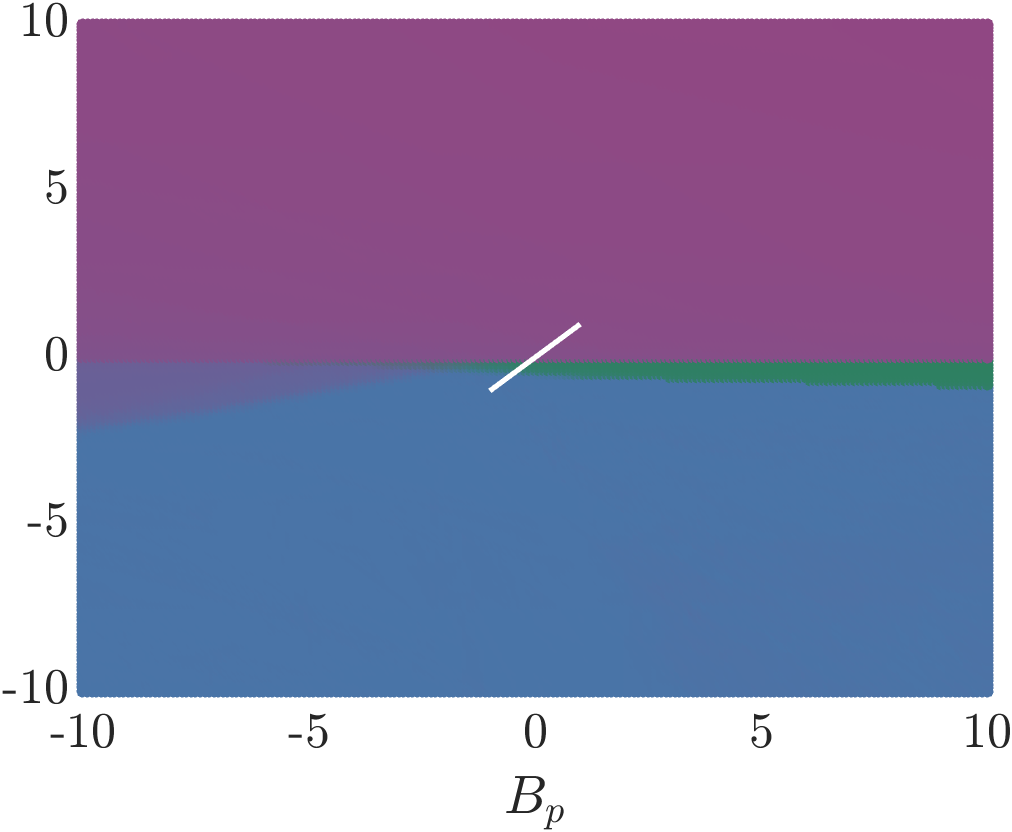}

\end{subfigure}

\vspace{0.4cm}

\begin{subfigure}[hbtp]{0.24\textwidth}
  \centering
  \begin{overpic}[width=\textwidth, grid = false]{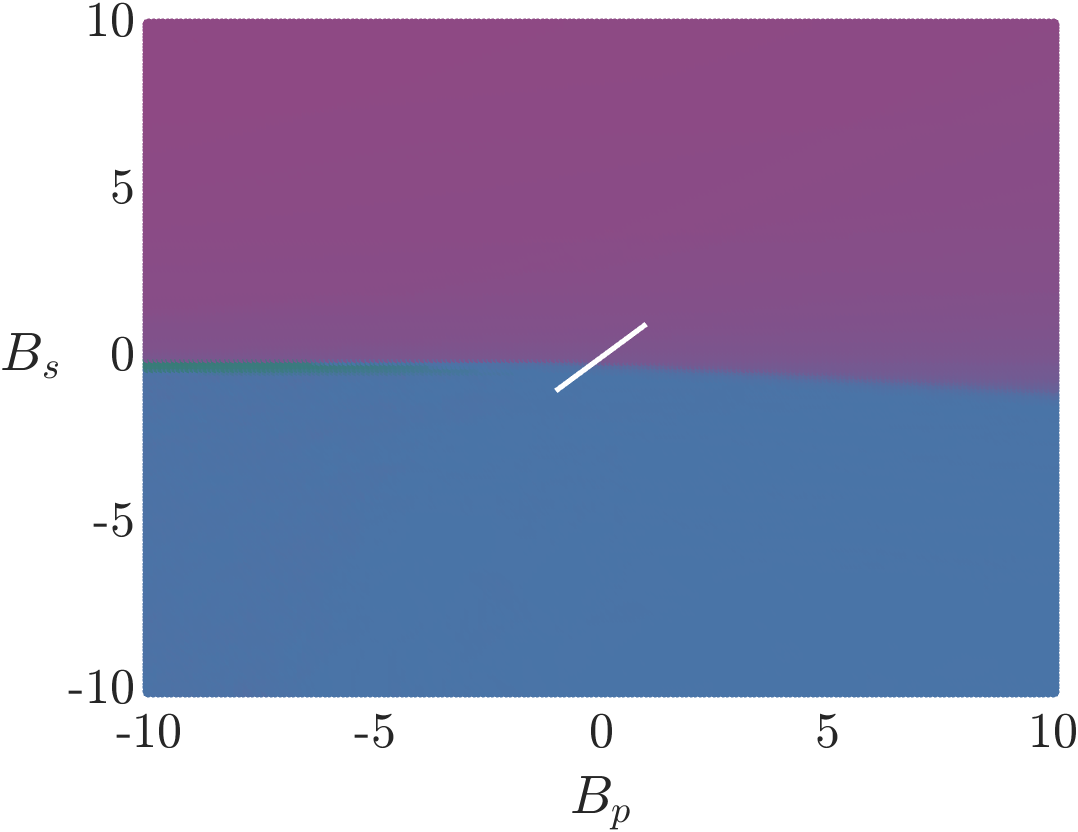}
      \put(-4,72){(e)} \put(203.5,72){(g)} \put(308,72){(h)} \put(-29,45){Puller} \put(-27,35){$\forceavg<0$}
  \end{overpic}
\end{subfigure}
\hfill
\begin{subfigure}[hbtp]{0.24\textwidth}
  \centering
   \begin{overpic}[width=0.935\textwidth, grid = false]{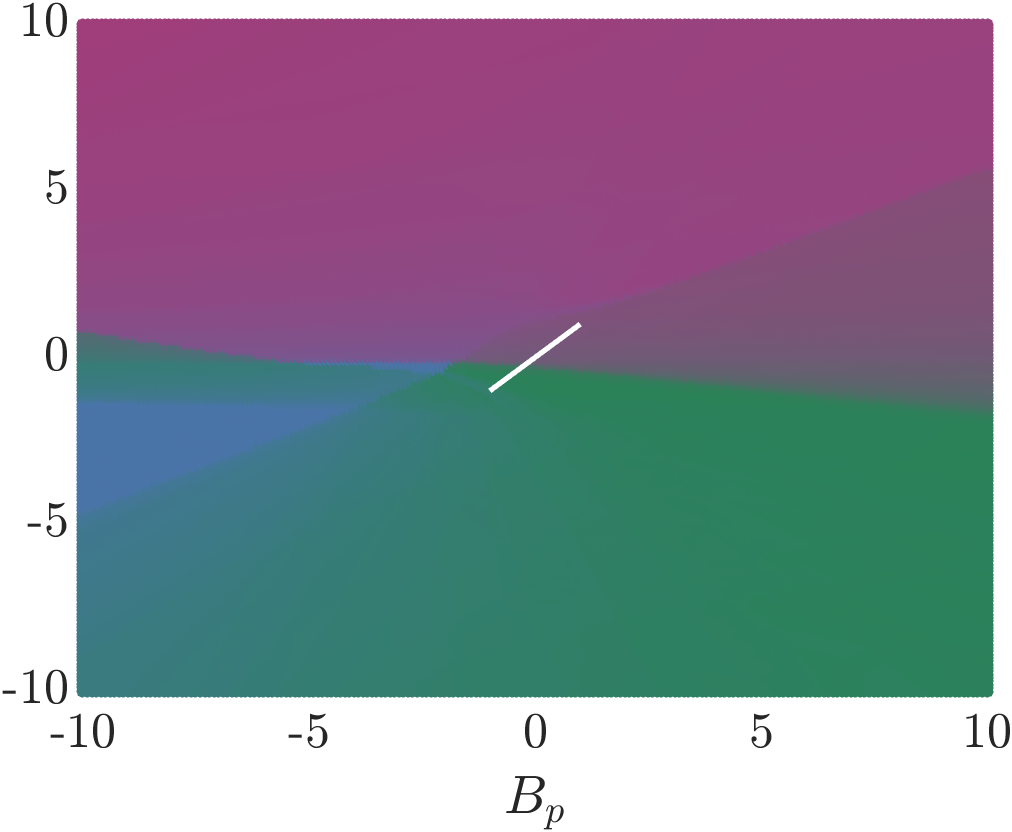}
  \put(-10, 76.75){(f)} 
  \end{overpic}

\end{subfigure}
\hfill
\begin{subfigure}[hbtp]{0.24\textwidth}
  \centering
  \includegraphics[width=0.935\textwidth]{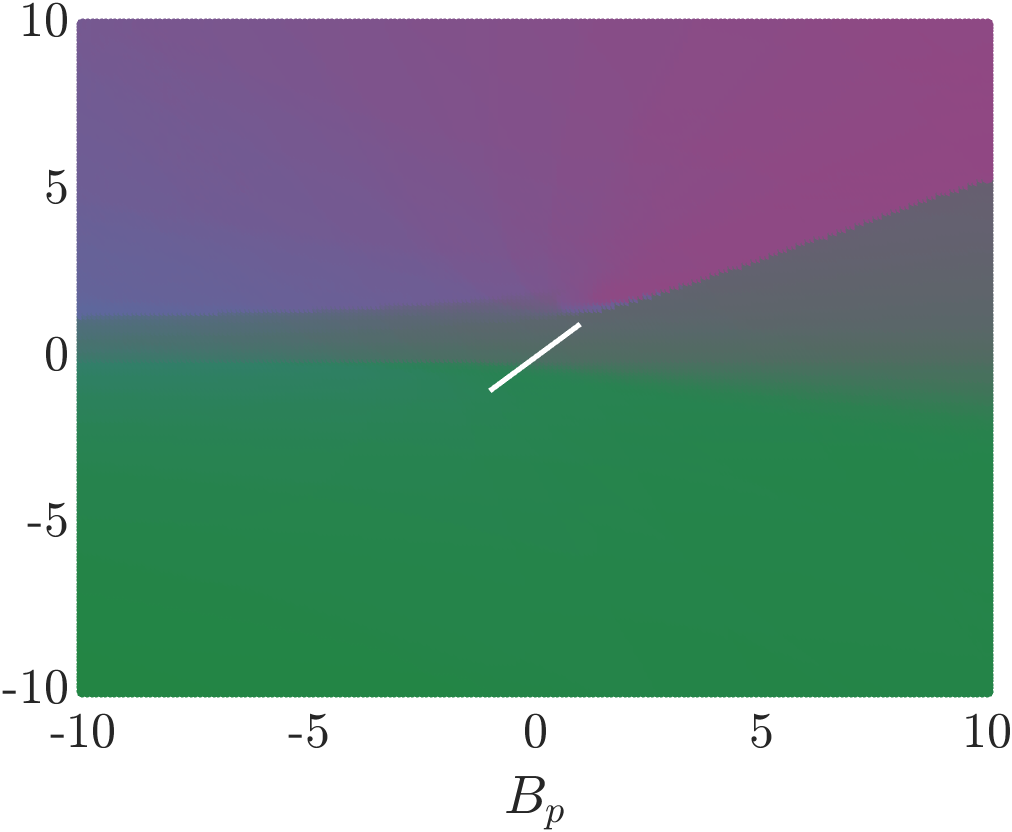}

\end{subfigure}
\hfill
\begin{subfigure}[hbtp]{0.24\textwidth}
  \centering
  \includegraphics[width=0.935\textwidth]{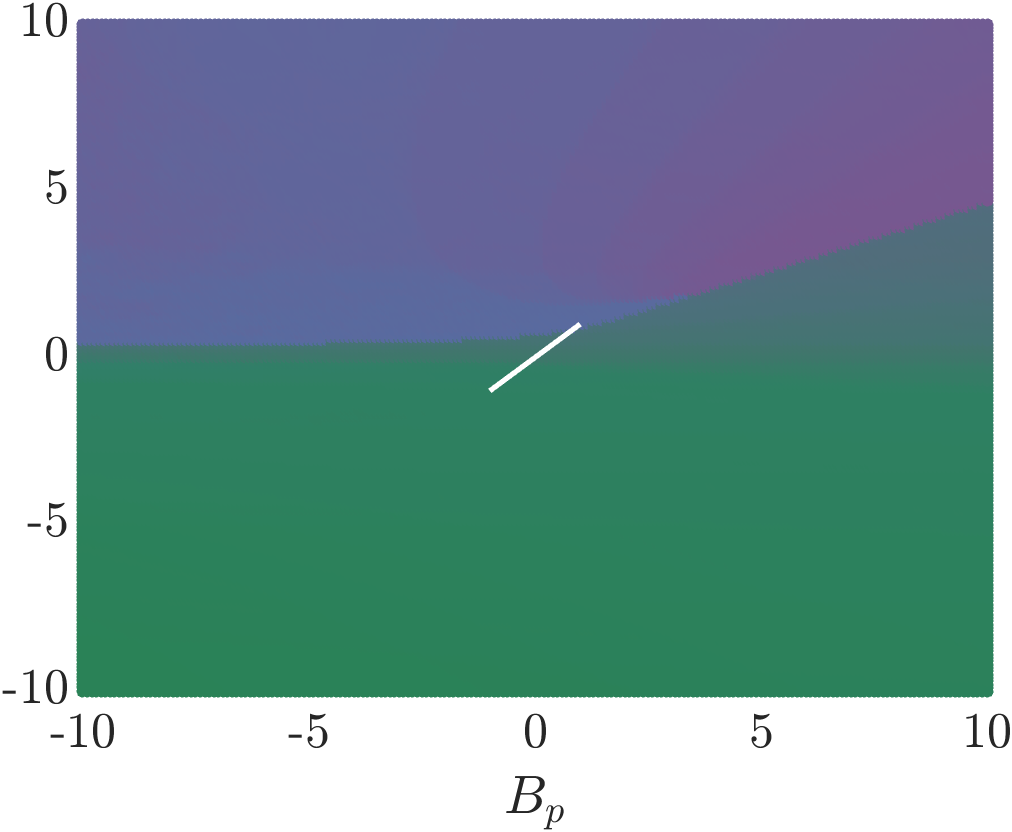}

\end{subfigure}
\caption{The effect of a changing effective swimming speed on the prevalent behaviours for a single swimmer modelled as a force dipole and source dipole above a stress-free surface. The possible dynamics depend on the effective shape parameters $(\effforce,\effsource)$, the effective swimming speed $\effswim$ and the sign of the effective force dipole strength $\forceavg$. In panels (a)-(d), the swimmer is an effective pusher, $\forceavg>0$; in panels (e)-(h), the swimmer is an effective puller $\forceavg<0$.  We choose $\effswim =$ (a),(e) $-0.5$, (b),(f) $-0.05$, (c),(g) $0.05$, and (d),(h) $0.5$ to show how the prevalent behaviour qualitatively changes as we vary the effective swimming speed. The possible dynamics of the \textit{a priori}-averaged system correspond to the white line segments which do not generally reflect the proportions observed in the wider $(\effforce,\effsource)$ space of the systematically averaged system.  We take $\sourceavg = -1$, have $\forceavg = \pm 1$ and vary $\effforce$, and $\effsource$ between $-10$ and $10$, beyond which the trends continue. }
\label{fig: FD+SD +swim}

\end{figure}

For different combinations of the parameters $\effforce,\,\effsource,\, \text{and}\;\effswim$ we calculate the linear stability of the steady states by numerically evaluating the Jacobian. We find the same three  behaviours as before: hovering, escaping and crashing. 

To classify the expected behaviour for propulsive swimmers, we repeat our numerical parameter and behaviour sweeps over the effective body shape parameters $\effforce,\,\effsource$, now for different effective swimming speeds, $\effswim$, as detailed in \cref{subsec: FD+SD emergent behaviour}. By considering the effective-body-shape parameter space for different values of the effective swimming speed, $\effswim$, we find four different regions of $\effswim$ within which we have qualitatively similar behaviour, as shown in \cref{fig: FD+SD +swim}. These regions are bounded by three critical effective swimming speeds. For effective pushers, $\forceavg>0$, the critical values correspond to $\effswim = 8/27$, $\effswim = 0$ and $\effswim \approx -0.1$. For effective pullers, $\forceavg<0$, the critical values correspond to $\effswim \approx 0.1$, $\effswim = 0$ and $\effswim = -8/27$. We note that the bifurcations at $\effswim = \pm 8/27$ correspond to the critical effective swimming speed found in \cref{subsubsec: FD+SD+u trivial} and the $\effswim = 0$ bifurcation is due to the non-trivial steady states coalescing with the trivial steady state $\theta = 0$. 

 The different possible dynamics are fundamentally different from the no-swimming case (\cref{fig: main behaviour FD+SD freeslip}). In particular, for sufficiently large swimming speeds ($\effswim>8/27$ for pushers and $\effswim< -8/27$ for pullers), stable hovering becomes rare across the majority of the parameter space. Additionally, the same effective body shape combination can predict qualitatively different outcomes: hovering, escaping or crashing, depending on the swimming speed regime. However, the limited variation in the \textit{a priori}-averaged system (indicated by white line segments in \cref{fig: FD+SD +swim}) compared to the wider parameter space highlights how the multi-timescale dynamics of microscale swimming can significantly alter the predictions of swimmer behaviour. 

\subsection{Emergent behaviours from a quadrupole}\label{subsec: FD+Q+U}
For a force dipole and quadrupole, we proceed as before and find the equations \cref{eq:FD+Q  averaged system final} are
\begin{subequations}\label{eq: FD+Q+u full system}
 \begin{align}
\diff{\theta}{t} =&
 \frac{3\forceavg\sin 2\theta}{32h^3}\left[2+\effforce(1+\cos 2\theta)\right] -  \frac{3\quadavg \sin \theta}{32 h^4}\left[3+5\cos 2\theta +\frac{\effquad}{4}(3\cos 4\theta + 8 \cos 2\theta -11)\right], \label{eq: FD+Q+ u averaged system theta final}\\
  \diff{h}{t} =&   \frac{\forceavg}{8h^2}\left(1+3\cos 2\theta\right) + \frac{\quadavg \cos \theta}{4h^3}(1-3\cos2\theta) + u\cos \theta. \label{eq: FD+Q with swim h eqn}
  \end{align}
\end{subequations}
Now the system \cref{eq: FD+Q+u full system} is invariant under the transformation $\quadavg \mapsto -\quadavg,\, \effswim \mapsto -\effswim,\, \theta \mapsto \theta +\pi$ and we assume that $\quadavg>0$.

For the addition of a quadrupole, by substituting the trivial $\theta$ steady states, $\theta = 0,\, \theta =  \pi$, into the $h$-evolution equation \cref{eq: FD+Q with swim h eqn}, we find the corresponding $h$ polynomials
\begin{align}
 uh^3+\forceavg h-\quadavg = 0, \\
 -2 u h^3+\forceavg h +\quadavg = 0,
\end{align}
for $\theta = 0$ and $\theta = \pi$, respectively. For $h> 0$ in the domain, we find the same three cases: no solution to either cubic polynomial, a solution to both, or two solutions to one. As in the source dipole case (\cref{subsubsec: FD+SD+u trivial}), we find that a saddle-node bifurcation occurs for the relevant $\theta = 0$ steady state when $\forceavg>0$ and $u\quadavg^2/\forceavg^3 = -4/54$.  For $\theta = \pi$, the relevant steady state now requires $\forceavg<0$ and $u\quadavg^2/\forceavg^3 = 4/54$. We now write $\effswim = u\quadavg^2/\forceavg^3$ and continue to refer to $\effswim$ as the effective swimming speed of the swimmer. We determine the trivial steady states and their associated linear stability numerically with the non-trivial steady states in a wider parameter sweep of effective body shape and swimming speed.

In order to find the remaining non-trivial steady states, we consider steady values of $h$ as a function of $\theta$. From the $\theta $ equation \cref{eq: FD+Q averaged system theta final}, we write $h$ as a function of $\theta$. Substituting this in the $h$-evolution equation \cref{eq: FD+Q with swim h eqn}, we determine that the non-trivial steady states are solutions to the polynomial:
\begin{multline}\label{eq: FD+Q+swim poly}
x(1+\effforce x^2)^2(5x^2-1+\effquad(3x^4-x^2-2))(3x^2-1)+\\
4x^3(1+\effforce x^2)^3(2-3x^2)+\effswim(5x^2-1+\effquad(3x^4-x^2-2))^3 = 0,
\end{multline}
where $x = \cos \theta$ and the definition for effective swimming speed, $\effswim = u\quadavg^2/\forceavg^3$, is maintained from the analysis for trivial steady states. Therefore, as \cref{eq: FD+Q+swim poly} depends on three parameters only, $\effforce,\, \effquad,\, \effswim$, the possible behaviours predicted by the model can be determined from these effective body shape and swimming speed parameters as well as if the swimmer is an effective pusher, $\forceavg>0$, or puller, $\forceavg<0$. By exploring the steady states and associated linear stabilities numerically, we find the same possible behaviours: hovering, escaping and crashing. Following the parameter and behaviour sweep detailed in \cref{subsec: FD+SD emergent behaviour}, we consider the possible behaviours for a range of effective swimming speeds and effective body shape parameters as in \cref{subsec: FD+SD+U}

\begin{figure}[hbtp]
\vspace{0.1cm}
\centering
\begin{subfigure}[hbtp]{0.24\textwidth}
  \centering
  \begin{overpic}[width = \textwidth, grid = false]{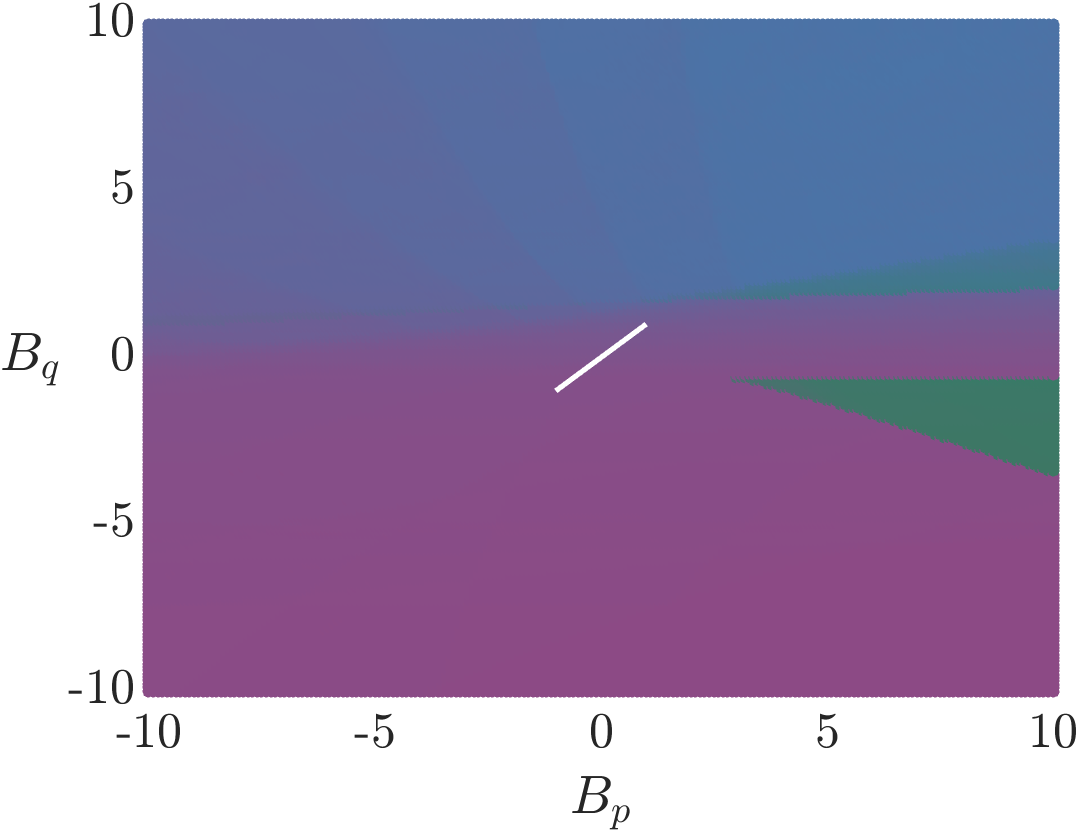}
  \put(-4,72){(a)}\put(203,72){(c)} \put(308,72){(d)} \put(-29,45){Pusher}\put(-27,35){$\forceavg>0$} \put(40,78){$U = -1$} \put(138,78){$U = -0.05$} \put(245,78){$U = 0.05$}
      \put(355,78){$U = 1$}
\end{overpic}

\end{subfigure}
\hfill
\begin{subfigure}[hbtp]{0.24\textwidth}
  \centering
    \begin{overpic}[width=0.935\textwidth, grid = false]{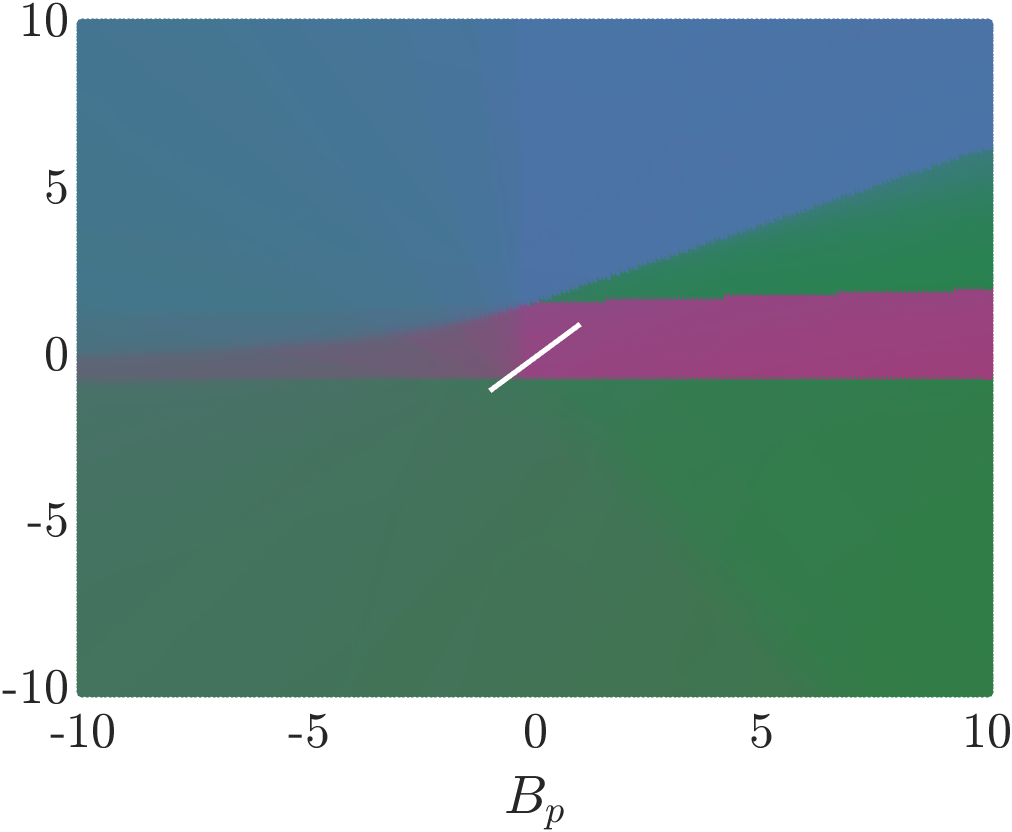}
  \put(-10, 76.75){(b)} 
  \end{overpic}
\end{subfigure}
\hfill
\begin{subfigure}[hbtp]{0.24\textwidth}
  \centering
  \includegraphics[width=0.935\textwidth]{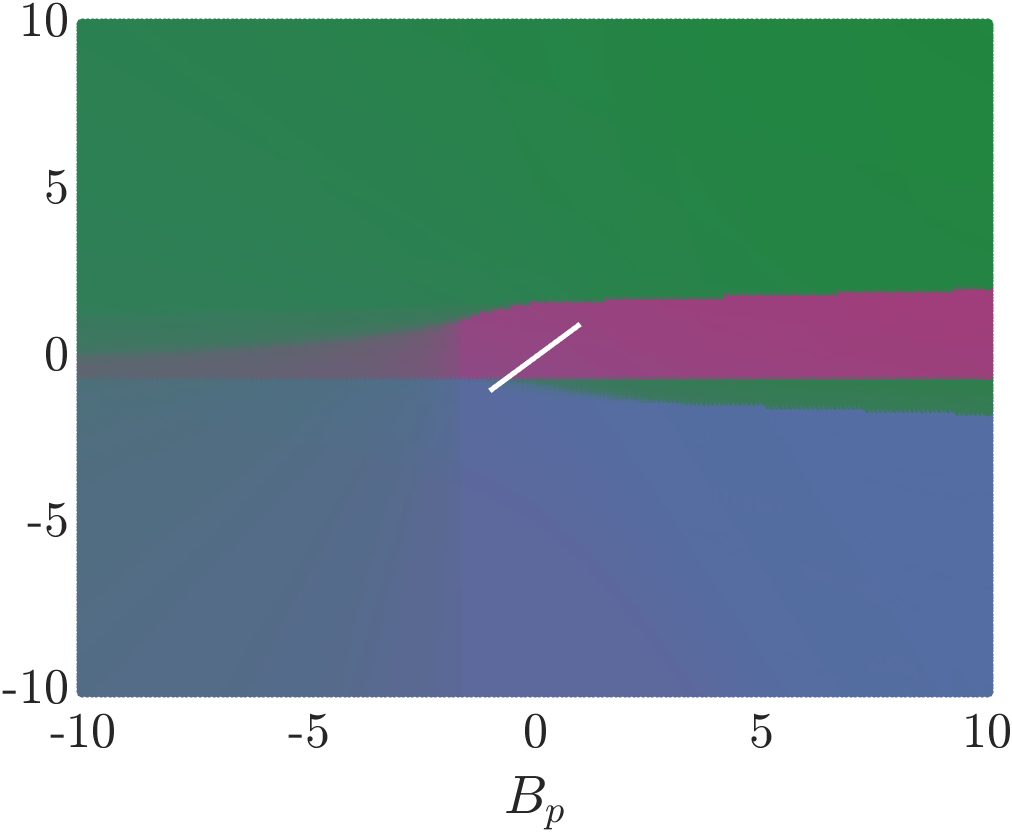}

\end{subfigure}
\hfill
\begin{subfigure}[hbtp]{0.24\textwidth}
  \centering
  \includegraphics[width=0.935\textwidth]{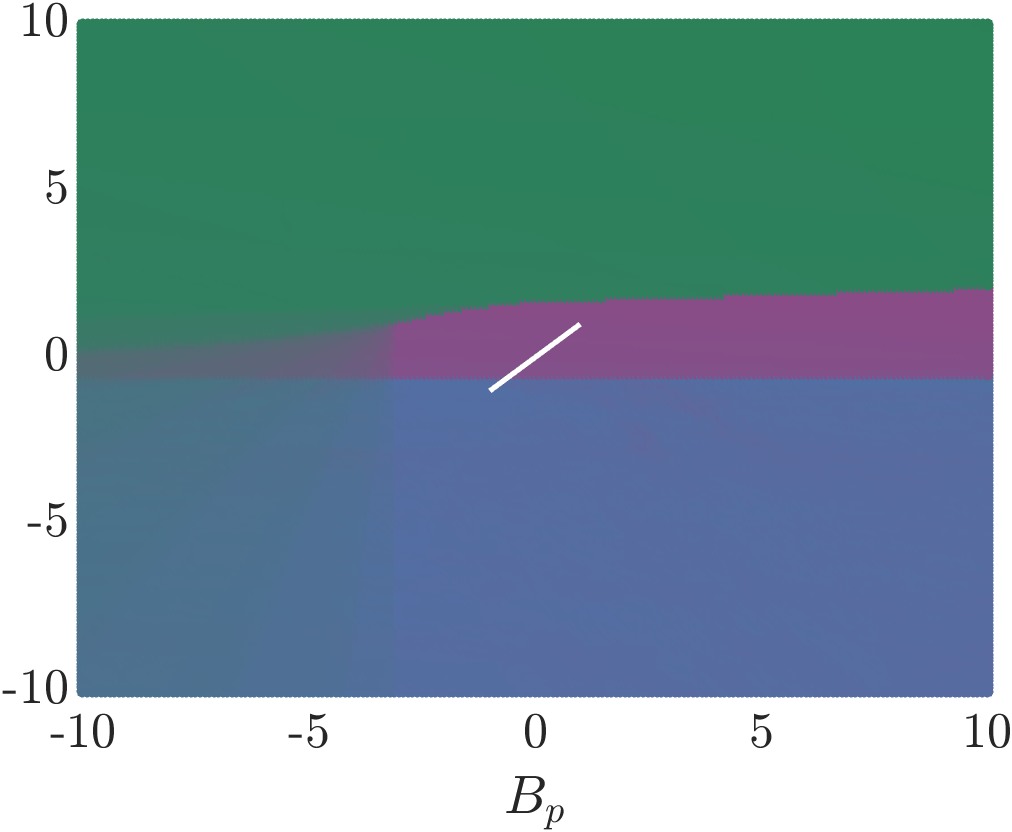}

\end{subfigure}
\vspace{0.4cm}

\begin{subfigure}[hbtp]{0.24\textwidth}
  \centering
    \begin{overpic}[width=\textwidth]{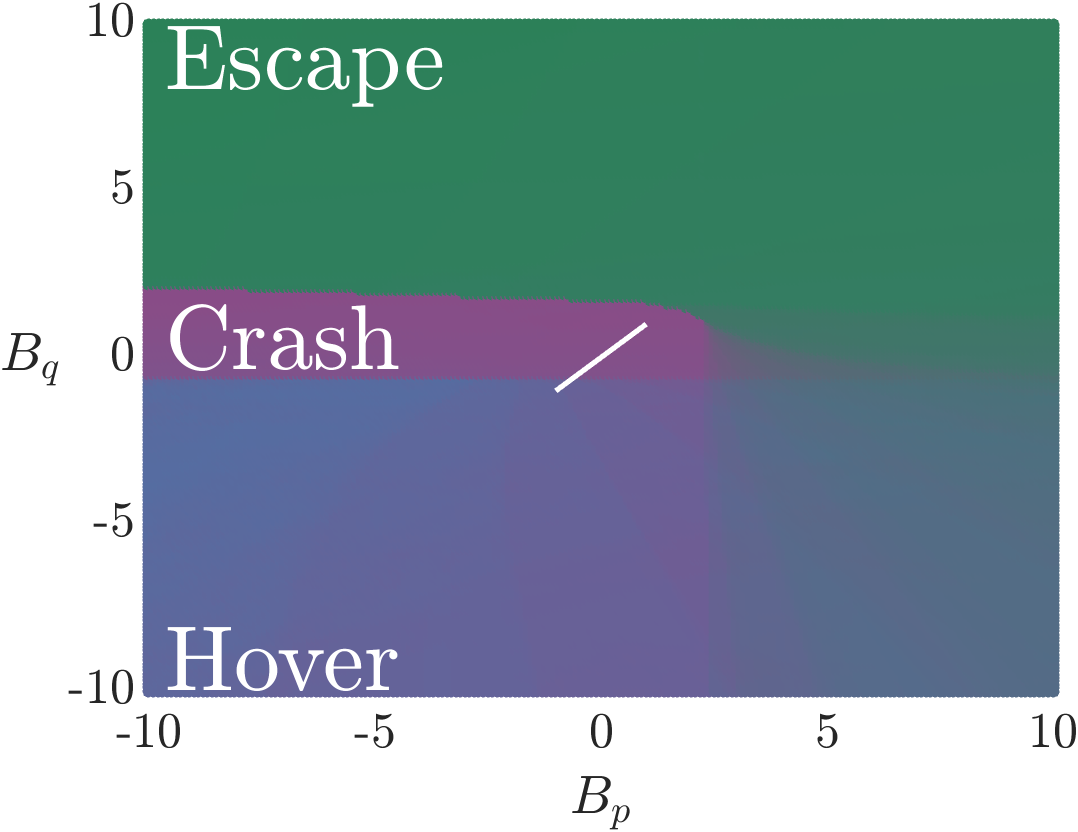}
      \put(-4,72){(e)} \put(203,72){(g)} \put(308,72){(h)} \put(-29,45){Puller} \put(-27,35){$\forceavg<0$}
  \end{overpic}
 
\end{subfigure}
\hfill
\begin{subfigure}[hbtp]{0.24\textwidth}
  \centering
  \begin{overpic}[width=0.935\textwidth, grid = false]{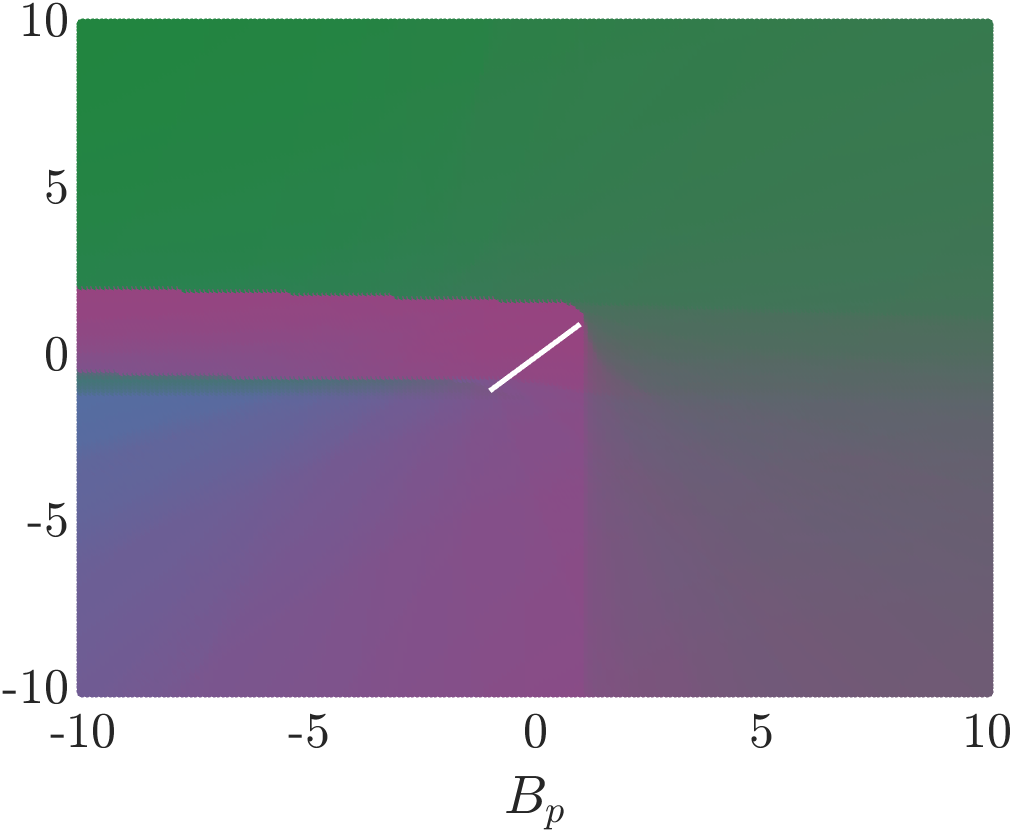}
  \put(-10, 76.75){(f)} 
  \end{overpic}
  
\end{subfigure}
\hfill
\begin{subfigure}[hbtp]{0.24\textwidth}
  \centering
  \includegraphics[width=0.935\textwidth]{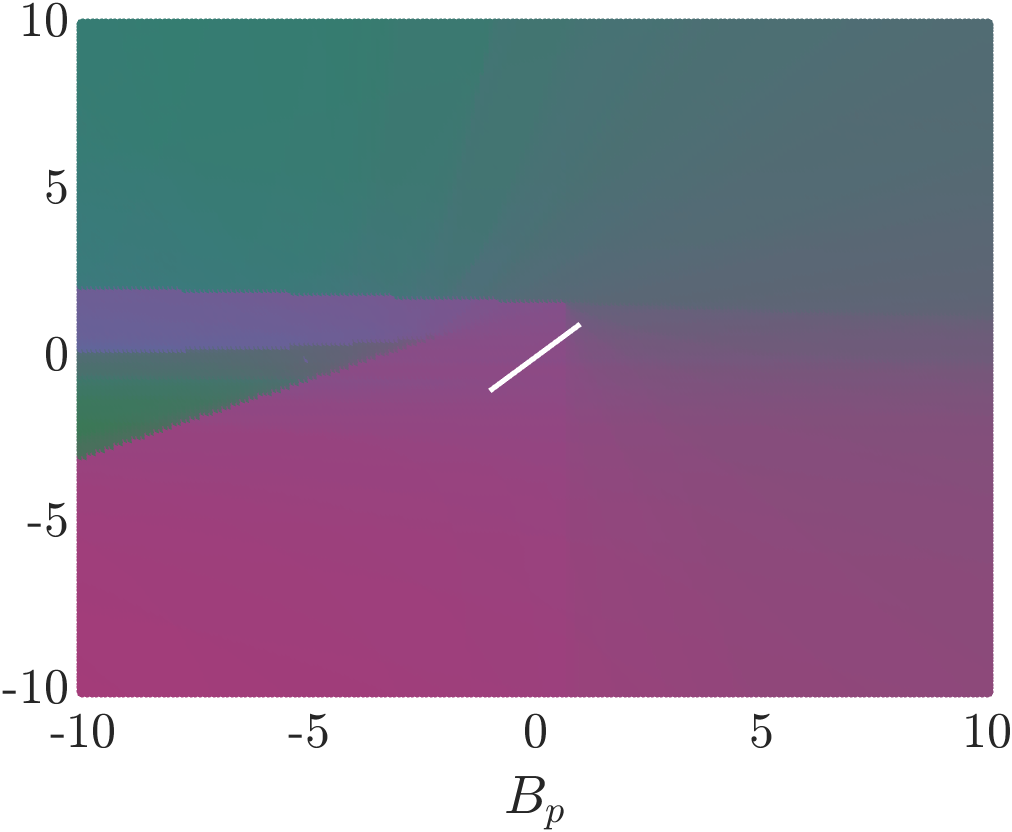}

\end{subfigure}
\hfill
\begin{subfigure}[hbtp]{0.24\textwidth}
  \centering
  \includegraphics[width=0.935\textwidth]{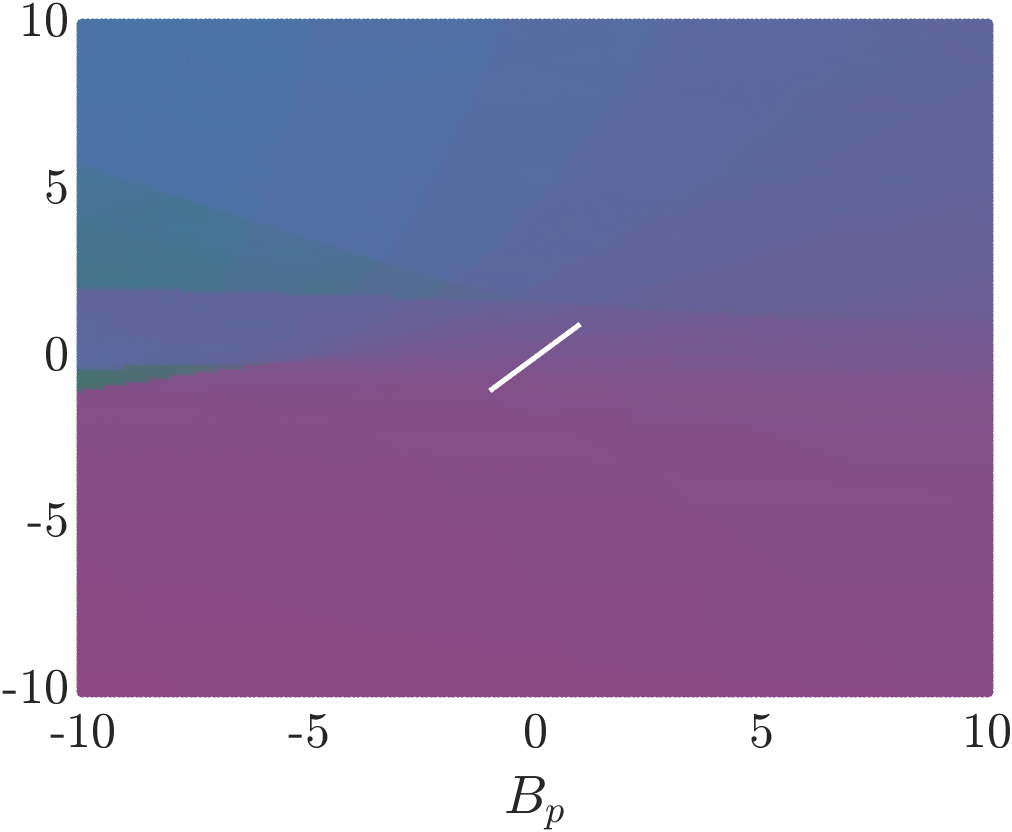}

\end{subfigure}
\vspace{0.1cm}
\caption{ The effect of a changing effective swimming speed on the prevalent  behaviours for a single swimmer modelled as a force dipole and quadrupole above a stress-free surface.  In panels (a)-(d), the swimmer is an effective pusher, $\forceavg>0$; in panels (e)-(h), the swimmer is an effective puller $\forceavg<0$.  We vary the effective swimming speed to show the four distinct regions of main behaviour. Therefore $\effswim =$ (a),(e) $-1$, (b),(f) $-0.05$, (c),(g) $0.05$, and (d)(h) $1$.  The possible dynamics of the \textit{a priori}-averaged system correspond to the white line segments which do not generally reflect the proportions observed in the wider $(\effforce,\effsource)$ space of the systematically averaged system.  We take $\quadavg = 1$, have $\forceavg = \pm 1$ and vary $\effforce$, and $\effsource$ between $-10$ and $10$ beyond which the trends continue.}
\label{fig: FD+Q +swim}

\end{figure}
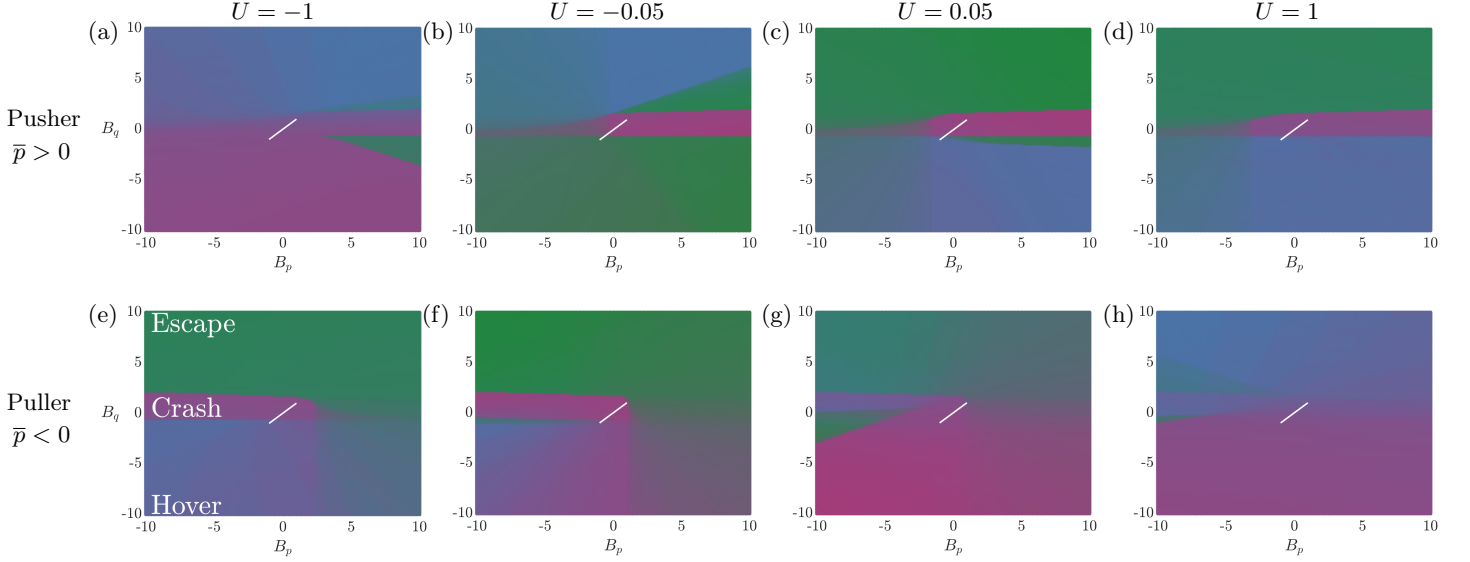

As the effective swimming speed, $\effswim$, varies, we consider the prevalent qualitative behaviour across the whole of the effective-body-shape parameter space for effective pushers, $\forceavg>0$, and pullers, $\forceavg<0$ separately. As for the source dipole case, we find four regions of the effective swimming speed, $\effswim$, within which the numerically found prevalent behaviour in the $\effforce,\,\effquad$ space is qualitatively similar  (\cref{fig: FD+Q +swim}). These regions are bounded by the critical effective swimming speed $\effswim = \pm 4/54$ for the trivial steady states, $U = 0$ and $U \approx \mp 0.5$ for pushers $\forceavg>0$ and pullers $\forceavg< 0$, respectively. 

\cref{fig: FD+Q +swim} shows that trends from the source dipole case persist despite the change in additional singularity. In particular, the predicted behaviour from the \textit{a priori}-averaged system is not necessarily the same as the wider system, and the possible dynamics vary significantly across  swimming speed regimes.  However, for sufficiently large swimming speeds ($U>4/54$ for pushers and $U<4/54$ for pullers), we now have hovering behaviour in a large proportion of the parameter space. It is therefore important to account for the effects of multi-timescale swimming for an appropriate additional singularity through a systematic multiscale analysis. 

\section{Including both higher-order singularities}\label{sec: FD+SD+Q analysis}
Up to this point, we have considered the effect of a single additional, higher-order singularity to the force dipole model. As the source dipole accounts for the effect of the finite cell body and the quadrupole for the fore-aft asymmetry, we considered how the individual singularities changed the dynamics. We therefore consider any potential effect on the dynamics due to the interaction between the higher-order singularities. Due to the complexity and high-dimensionality of these interactions, general statements about stability are difficult, but in this section we highlight what can be said about the dynamics when we include all three singularities. For convenience, we limit the analysis to a zero swimming speed, $u = 0$, and a stress-free boundary so that \cref{eq: FD+SD+Q averaged equations freeslip} becomes
\begin{subequations}
\label{eq: FD+SD+Q averaged freeslip u 0}
\begin{align}
    \diff{\theta}{t}  =& \frac{3 \forceavg \sin 2\theta}{32h^3}\left[2+\effforce(1+\cos 2\theta)\right] - \frac{3\effsource \sourceavg \sin\theta}{32h^4}\left(3+\cos2 \theta\right)  -\frac{3\quadavg\sin \theta}{32 h^4}\left[3+5\cos 2\theta  +\frac{\effquad}{4}(3\cos 4\theta + 8 \cos 2\theta -11)\right] \label{eq: FD+SD+Q avg theta final}\\
    \diff{h}{t} =& \frac{\forceavg}{8h^2}\left(1+3\cos 2\theta\right)  -\frac{\sourceavg\cos\theta}{4h^3}+ \frac{\quadavg \cos \theta}{4h^3}(1-3\cos2\theta) \label{eq: FD+SD+Q avg freeslip h}.
\end{align}
\end{subequations}
As a result of systematic averaging, we now have a six parameter system: $\forceavg$, $\sourceavg$, $\quadavg$, $\effforce$, $\effsource$, $\effquad$, compared to four in earlier \cref{sec: FD+SD,sec: FD+Q,sec: no slip} and five in \cref{sec: nonzero swim}. We can reduce the systematically averaged system \cref{eq: FD+SD+Q averaged freeslip u 0} to the \textit{a priori}-averaged system by setting $\forceavg = \force,\,\sourceavg = \source,\, \quadavg = \quadrupole$ and $\effforce = \effsource = \effquad = B \in (-1, 1)$.  
\subsection{Emergent dynamics}
From the $\theta$ equation \cref{eq: FD+SD+Q avg theta final}, there are several potential steady states. We first consider $\theta = 0$, $\theta =  \pi$ and substitute them into the $h$ equation \cref{eq: FD+SD+Q avg freeslip h}. We find 
\begin{equation}\label{eq: FD+SD+Q trivial ss}
    \theta = 0, \; h = \dfrac{\sourceavg+2\quadavg}{2\forceavg} \quad \text{and}\quad \theta =  \pi,\; h = -\dfrac{\sourceavg+2\quadavg}{2\forceavg},
\end{equation}
are steady states of \cref{eq: FD+SD+Q averaged freeslip u 0} and we continue to refer to them as trivial steady states. Calculating the linear stability, we find the corresponding eigenvalues to be
\begin{equation}
  \lambda_0 = \left\{  \frac{3\forceavg^4}{(2\quadavg+\sourceavg)^4}\left(2\quadavg(\effforce-1)+\sourceavg(\effforce-2\effsource+1)\right), \frac{8\forceavg^4}{(\quadavg+\sourceavg)^4}(2\quadavg+5\sourceavg) \right\} \; \text{and} \; \lambda_{\pm \pi} = -\lambda_0.
\end{equation}
 The linear stability of the trivial steady states depends on four parameters, $\sourceavg,\quadavg,\effforce,\effsource$, corresponding to the strength of the source dipole and quadrupole as well as the effective shape parameters for the force dipole and source dipole. Compared to the single higher order singularity case, where the linear stability depends at most on the two effective body shape parameters and the sign of the force dipole, the conditions for linearly stability are more cumbersome to classify. 

For non-trivial steady states, the steady height as a function of $\theta$ can be determined from \cref{eq: FD+SD+Q avg freeslip h}:
\begin{equation}\label{eq: height FD+SD+Q steady}
h = 2\frac{\sourceavg\cos\theta-\quadavg\cos \theta \left(1-3\cos2\theta\right)}{\forceavg(1+3\cos\theta)}.
\end{equation}

Continuing as we did for a single higher order singularity, we factor out $3\sin \theta /(32h^4)$ from \cref{eq: FD+SD+Q avg theta final}, roots of which correspond to the trivial steady states. After substituting \cref{eq: height FD+SD+Q steady} into \cref{eq: FD+SD+Q avg theta final} and setting $x = \cos ^2 \theta$, the remaining  non-trivial steady states are therefore solutions to the equation
\begin{equation}
2 x \frac{\sourceavg-\quadavg(2-3 x)}{3 x-1}\left(1+\effforce x\right) -\sourceavg\effsource(1+x)-\quadavg\left( -1+5 x+\effquad(3 x^2-x-2)\right)=0.
\end{equation}
 Now the existence of any non-trivial steady states depends on five parameters: the three effective body shape parameters $\effforce,\, \effsource,\, \effquad$ and the strengths of the  next-order singularities $\sourceavg,\,\quadavg$. Although the dynamics continue to depend on whether the swimmer is an effective pusher or puller, the dimension of the remaining parameter space means we can no longer compactly describe the prevalent dynamics.

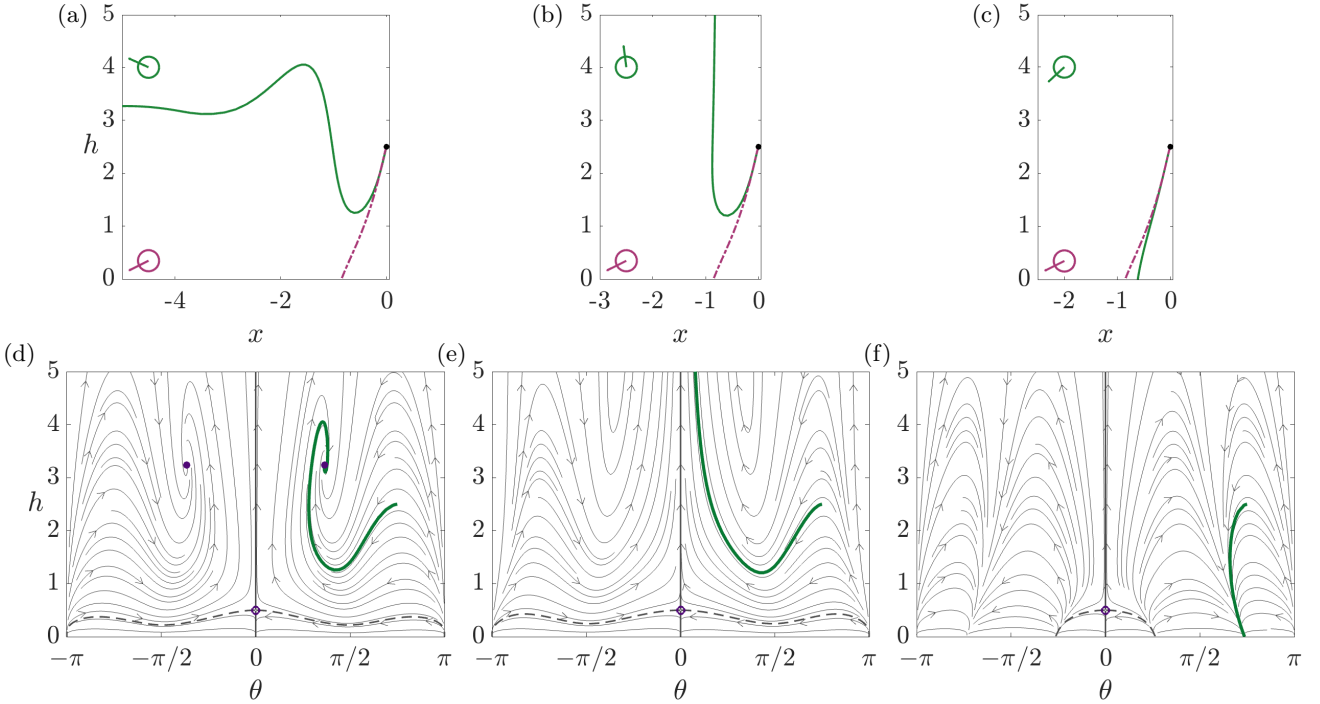
\begin{figure}[hbtp]
\centering
\begin{overpic}[width=\textwidth, grid = false]{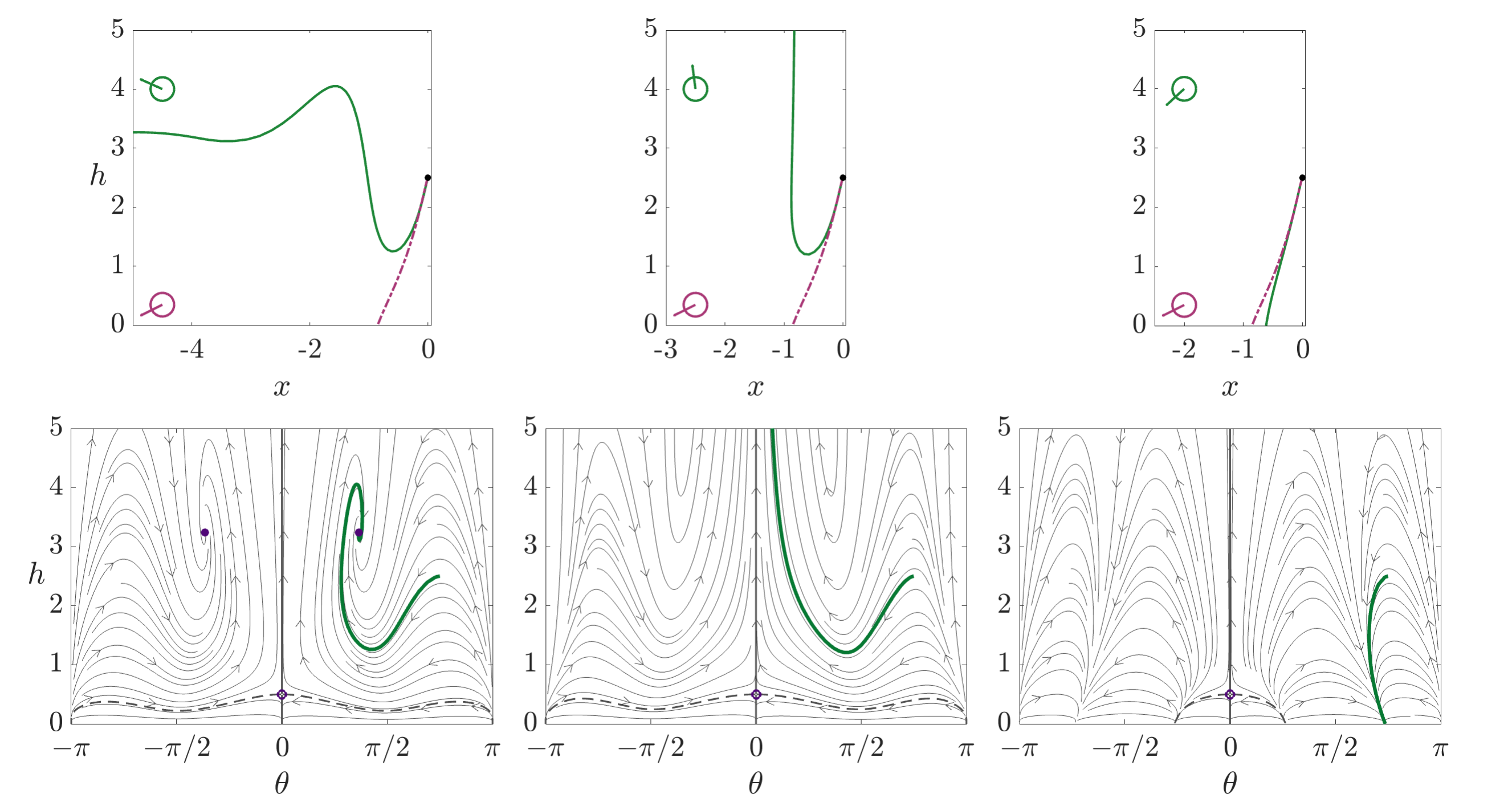}
\put(4,51){(a)} \put(39,51){(b)} \put(71.5,51){(c)}
\put(0,26){(d)} \put(32,26){(e)} \put(63.5,26){(f)}
\end{overpic}
\caption{ Dynamics of a single swimmer modelled as a combination of a force dipole, a source dipole, and a quadrupole near a  stress-free surface. (a)-(c) Hovering, escaping and crashing trajectories, respectively, in the $(x,h)$ plane and illustrative long-time configurations.  The systematically averaged system (green) can exhibit different behaviour to the\textit{a priori}-averaged system (pink, dashed-dot) which crashes into the boundary despite the same initial condition. (d)-(f) The systematically averaged, $(\theta,h)$, phase portrait with the same trajectory from (a)-(c). We take $s(T) = -1-4\sin T$, $B(T) = (\sin T)/2$, (a),(d) $p(T) = 1+4\sin T$, $q(T) =1-4\sin T $, (b),(e) $p(T) = 1-6\sin T$, $q(T) =1-4\sin T $, (c),(f) $p(T) = 1+4\sin T$, $q(T) =1+6\sin T $. We therefore have $ \forceavg = 1$, $\sourceavg = -1$ $\quadavg = 1$, $\effsource = -1$, $\avg{B} = 0$ and (a),(d) $\effforce = 1$, $\effquad = -1$, (b),(e) $\effforce = -1.5$, $\effquad = -1$, (c),(f) $\effforce = 1$, $\effquad = 1.5$ .}
\label{fig: FD+SD+Q +trajectories}
\end{figure}

To generate some understanding of these results, we numerically explore the  $(\theta, h)$ phase planes for a range of different parameter values. We find instances of the swimmer following a trajectory to a stable hovering state for both effective pushers, $\forceavg>0$, \cref{fig: FD+SD+Q +trajectories}(a) and pullers, $\forceavg<0$. We also find examples of the swimmer escaping from the boundary and crashing, highlighted for pushers in \cref{fig: FD+SD+Q +trajectories} (b) and (c).

Whilst examples of hovering, escaping and crashing can be found in the \textit{a priori}-averaged system, the systematically averaged predictions are not generally the same. In particular,  Figures \ref{fig: FD+SD+Q +trajectories}(a)-(c) show the disagreement between systematically averaged and \textit{a priori}-averaged predictions, demonstrating the importance of accounting for multi-timescale effects. 

\section{Discussion and conclusions}\label{sec: conclusion}

In this study, we have derived and explored systematically averaged minimal models for a microswimmer near a boundary. Extending beyond the (leading-order) force-dipole model, we studied the effects of including next-order terms from the multipole expansion of fundamental singular solutions to Stokes flow. We considered the addition of a source dipole and a quadrupole,  which account for the finite size of the cell body and any fore-aft asymmetry. We further assumed that the model parameters oscillated rapidly, modelling a change in body shape of the swimmer used to induce swimming. Through a systematic multiscale analysis,  we derived effective governing equations for swimming trajectories.  Whereas the \textit{a priori}-averaged system only quantifies the shape of a swimmer through the Bretherton parameter alone, the systematically averaged system highlights that it is necessary to consider the interactions between the change in body shape and singularity strengths during rapid oscillations. This leads to the emergence of new effective body shape parameters for both stress-free and no-slip boundary conditions, and for all appropriate next-order singularities. When there is only one next-order singularity included, three main behaviours occur in the systematically averaged system: the swimmer reaches a stable steady state and hovers above the boundary, the swimmer is deflected away from the boundary and ultimately escapes, or the swimmer crashes into the boundary. Whereas crashing dominates the predictions of the \textit{a priori}-averaged system, the variation in possible dynamics of the systematically averaged system demonstrates the importance of accounting for multi-timescale effects in making reliable predictions for microscale swimming. 

This variation in different possible behaviours is a direct result of new parameters emerging from the multiscale averaging. In particular, incorporating one additional singularity expands the parameter space from four dimensions to five; including both singularities increases the dimensionality from five to seven. In either scenario, the increase in dimensionality,  due to the interaction between the fast-timescale oscillations in body shape and singularity strengths, qualitatively changes the predictions of the systematically averaged model when compared to the \textit{a priori}-averaged model. The resulting dynamics are controlled by the new effective shape parameters, allowing these different possible behaviours to be more readily accessible when multi-timescale effects are accounted for.

Whilst for some choices of parameters the prevalent behaviour predicted by the systematically averaged system is the same as the \textit{a priori}-averaged system, this is not true in general and the predicted behaviour for similar swimmers depends on the choice of model. Therefore, the choice of which parameters and subsequent model to consider is important. Throughout this study, we have considered a wide range of effective body parameters to understand the possible system behaviours for any values. However, to enable direct comparison with experiments, estimates of these time-dependent quantities for specific swimmers would be required.  Although this is theoretically possible, we are unfortunately not aware of time-dependent measurements that report the quantities identified in this study as being key for understanding behaviour. Therefore, this study provides further motivation for new experiments that record and report temporally varying data on the smallest timescales of microswimmer motion. 

Though we have included singularities beyond the leading-order force dipole considered in the most minimal models, the foundational far-field expansion nevertheless loses validity as a swimmer draws close to a boundary. Thus, our conclusions should have appropriate caveats, and warrant the consideration of more detailed models, such as those including steric effects or more refined representations of swimmer geometry. Whilst the additional singularities considered in this study account for geometric attributes of the swimmer, the interaction with  the multi-timescale nature of microscale swimming is key for the increase in diversity of possible dynamics. It is therefore important to consider the effect of any rapid oscillations within more detailed models. Furthermore, by only considering a single swimmer in a plane, the model does not capture three-dimensional effects such as circular swimming exhibited above a surface \citep{lauga_swimming_2006,bianchi_3d_2019}. However, by including a rotlet dipole, the remaining next-order singularity, the approach could be extended to three-dimensions to include rotation above the surface. Even though the averaging procedure employed in this study is not particularly involved for this class of models, it is nevertheless a key component of the analysis. In particular, the predictions of the systematically averaged system are qualitatively different from those of the \textit{a priori}-averaged model. As such, we may conclude that, if present, the inclusion and subsequent consequences of rapid oscillations must be treated carefully. In particular, this suggests that the techniques used could have similar results for many other swimmer models where the far-field assumption breaks down, such as two nearby swimmers. For swimmer-swimmer interactions such as oscillatory microorganisms driving fluid through channels, the predictions of hydrodynamic effects on possible synchronisation could be affected \citep{golestanian_hydrodynamic_2011,elgeti_emergence_2013,friedrich_hydrodynamic_2016}.

To conclude, in this study we have extended the minimal force-dipole model for the swimming trajectories of a microswimmer near a boundary to include next-order terms from the far-field expansion of the flow field generated by the swimmer. Through a systematic multiscale analysis, we have accounted for the interaction between the fast-timescale oscillations of the swimmer and the terms from the far-field expansion, thereby expanding the parameter space and causing different behaviours to emerge. In particular, the predictions of the \textit{a priori}-averaged system are not necessarily representative of the full range of possible behaviours in the systematically averaged system, highlighting the importance of considering both the next-order terms and the effect of oscillatory, fast-timescale changes during swimming even in minimal swimmer models.

\begin{acknowledgments}
    We are grateful to Prof. Eric Keaveny for motivating discussions. For the purpose of open access, the authors have applied a Creative Commons Attribution (CC BY) licence to any Author Accepted Manuscript version arising from this submission.
\end{acknowledgments}

\section*{Data availability}
The code used to generate the figures in this manuscript is freely available at \url{https://github.com/sara-curtis/multi-timescale_swimmer_near_boundary}.
\bibliography{references.bib}

@article{spagnolie_hydrodynamics_2012,
    title = {Hydrodynamics of self-propulsion near a boundary: predictions and accuracy of far-field approximations},
    volume = {700},
    copyright = {https://www.cambridge.org/core/terms},
    issn = {0022-1120, 1469-7645},
    shorttitle = {Hydrodynamics of self-propulsion near a boundary},
    url = {https://www.cambridge.org/core/product/identifier/S0022112012001012/type/journal_article},
    doi = {10.1017/jfm.2012.101},
    urldate = {2025-05-19},
    journal = {J. Fluid Mech.},
    author = {Spagnolie, Saverio E. and Lauga, Eric},
    month = jun,
    year = {2012},
    pages = {105--147},
}

@book{holmespertubation,
  title={Introduction to perturbation methods},
  author={Holmes, Mark H},
  volume={20},
  year={2012},
publisher={Springer Science \& Business Media},

}

@book{lauga_fluid_2020,
    place={Cambridge}, series={Cambridge Texts in Applied Mathematics}, title={The fluid dynamics of cell motility}, publisher={Cambridge University Press}, author={Lauga, Eric}, year={2020}}

@article{rothschild_non-random_1963,
    title = {Non-random {distribution} of {bull} {spermatozoa} in a {drop} of {sperm} {suspension}},
    volume = {198},
    copyright = {http://www.springer.com/tdm},
    issn = {0028-0836, 1476-4687},
    url = {https://www.nature.com/articles/1981221a0},
    doi = {10.1038/1981221a0},
    
    number = {4886},
    urldate = {2025-05-20},
    journal = {Nature},
    author = {{Rothschild}},
    month = jun,
    year = {1963},
    pages = {1221--1222},
}

@article{frymier_three-dimensional_1995,
    title = {Three-dimensional tracking of motile bacteria near a solid planar surface.},
    volume = {92},
    issn = {0027-8424, 1091-6490},
    url = {https://pnas.org/doi/full/10.1073/pnas.92.13.6195},
    doi = {10.1073/pnas.92.13.6195},
    
    number = {13},
    urldate = {2025-06-05},
    journal = {PNAS},
    author = {Frymier, P D and Ford, R M and Berg, H C and Cummings, P T},
    month = jun,
    year = {1995},
    pages = {6195--6199},
}

@article{di_leonardo_swimming_2011,
    title = {Swimming with an image},
    volume = {106},
    copyright = {http://link.aps.org/licenses/aps-default-license},
    issn = {0031-9007, 1079-7114},
    url = {https://link.aps.org/doi/10.1103/PhysRevLett.106.038101},
    doi = {10.1103/PhysRevLett.106.038101},
    
    number = {3},
    urldate = {2025-05-20},
    journal = {Phys. Rev. Lett.},
    author = {Di Leonardo, R. and Dell’Arciprete, D. and Angelani, L. and Iebba, V.},
    month = jan,
    year = {2011},
    pages = {038101},
}

@article{bianchi_3d_2019,
    title = {{3D} dynamics of bacteria wall entrapment at a water–air interface},
    volume = {15},
    issn = {1744-683X, 1744-6848},
    url = {https://xlink.rsc.org/?DOI=C9SM00077A},
    doi = {10.1039/C9SM00077A},
    
    number = {16},
    urldate = {2025-06-12},
    journal = {Soft Matter},
    author = {Bianchi, Silvio and Saglimbeni, Filippo and Frangipane, Giacomo and Dell'Arciprete, Dario and Di Leonardo, Roberto},
    year = {2019},
    pages = {3397--3406},
}

@article{berg_chemotaxis_1990,
    title = {Chemotaxis of bacteria in glass capillary arrays. {Escherichia} coli, motility, microchannel plate, and light scattering},
    volume = {58},
    copyright = {https://www.elsevier.com/tdm/userlicense/1.0/},
    issn = {00063495},
    url = {https://linkinghub.elsevier.com/retrieve/pii/S000634959082436X},
    doi = {10.1016/S0006-3495(90)82436-X},
    
    number = {4},
    urldate = {2025-09-02},
    journal = {Biophys. J.},
    author = {Berg, H.C. and Turner, L.},
    month = oct,
    year = {1990},
    pages = {919--930},
}

@article{berke_hydrodynamic_2008,
    title = {Hydrodynamic {attraction} of {swimming} {microorganisms} by {surfaces}},
    volume = {101},
    copyright = {http://link.aps.org/licenses/aps-default-license},
    issn = {0031-9007, 1079-7114},
    url = {https://link.aps.org/doi/10.1103/PhysRevLett.101.038102},
    doi = {10.1103/PhysRevLett.101.038102},
    
    number = {3},
    urldate = {2025-06-05},
    journal = {Phys. Rev. Lett.},
    author = {Berke, Allison P. and Turner, Linda and Berg, Howard C. and Lauga, Eric},
    month = jul,
    year = {2008},
    pages = {038102},
}

@article{giacche_hydrodynamic_2010,
    title = {Hydrodynamic entrapment of bacteria swimming near a solid surface},
    volume = {82},
    copyright = {http://link.aps.org/licenses/aps-default-license},
    issn = {1539-3755, 1550-2376},
    url = {https://link.aps.org/doi/10.1103/PhysRevE.82.056309},
    doi = {10.1103/PhysRevE.82.056309},
    
    number = {5},
    urldate = {2025-05-20},
    journal = {Phys. Rev. E},
    author = {Giacché, Davide and Ishikawa, Takuji and Yamaguchi, Takami},
    month = nov,
    year = {2010},
    pages = {056309},
}

@article{smith_human_2009,
    title = {Human sperm accumulation near surfaces: a simulation study},
    volume = {621},
    copyright = {https://www.cambridge.org/core/terms},
    issn = {0022-1120, 1469-7645},
    shorttitle = {Human sperm accumulation near surfaces},
    url = {https://www.cambridge.org/core/product/identifier/S0022112008004953/type/journal_article},
    doi = {10.1017/S0022112008004953},
    
    urldate = {2025-05-20},
    journal = {J. Fluid Mech.},
    author = {Smith, D. J. and Gaffney, E. A. and Blake, J. R. and Kirkman-Brown, J. C.},
    month = feb,
    year = {2009},
    pages = {289--320},
}

@article{ishimoto_squirmer_2013,
    title = {Squirmer dynamics near a boundary},
    volume = {88},
    copyright = {http://link.aps.org/licenses/aps-default-license},
    issn = {1539-3755, 1550-2376},
    url = {https://link.aps.org/doi/10.1103/PhysRevE.88.062702},
    doi = {10.1103/PhysRevE.88.062702},
    
    number = {6},
    urldate = {2025-05-22},
    journal = {Phys. Rev. E},
    author = {Ishimoto, Kenta and Gaffney, Eamonn A.},
    month = dec,
    year = {2013},
    pages = {062702},
}

@article{lauga_swimming_2006,
    title = {Swimming in {circles}: {motion} of {bacteria} near {solid} {boundaries}},
    volume = {90},
    issn = {00063495},
    shorttitle = {Swimming in {Circles}},
    url = {https://linkinghub.elsevier.com/retrieve/pii/S0006349506722214},
    doi = {10.1529/biophysj.105.069401},
    
    number = {2},
    urldate = {2025-05-20},
    journal = {Biophys. J.},
    author = {Lauga, Eric and DiLuzio, Willow R. and Whitesides, George M. and Stone, Howard A.},
    month = jan,
    year = {2006},
    pages = {400--412},
}

@article{manabe_shape_nodate,
     title={Shape matters: entrapment of a model ciliate at interfaces}, volume={892}, DOI={10.1017/jfm.2020.160}, journal={J. Fluid Mech.}, author={Manabe, Junichi and Omori, Toshihiro and Ishikawa, Takuji}, year={2020}, pages={A15}}

@article{or_dynamics_2009,
    title={Dynamics and stability of a class of low Reynolds number swimmers near a wall},
  author={Or, Yizhar and Murray, Richard M},
  journal={Phys. Rev. E: Stat. Nonlinear Soft Matter Phys.},
  volume={79},
  number={4},
  pages={045302},
  year={2009},
}

@article{shum_modelling_2010,
    title = {Modelling bacterial behaviour close to a no-slip plane boundary: the influence of bacterial geometry},
    volume = {466},
    issn = {1364-5021, 1471-2946},
    shorttitle = {Modelling bacterial behaviour close to a no-slip plane boundary},
    url = {https://royalsocietypublishing.org/doi/10.1098/rspa.2009.0520},
    doi = {10.1098/rspa.2009.0520},
    
    number = {2118},
    urldate = {2025-05-20},
    journal = {Proc. R. Soc. A},
    author = {Shum, H. and Gaffney, E. A. and Smith, D. J.},
    month = jun,
    year = {2010},
    pages = {1725--1748},
}

@article{htet_hydrodynamic_2024,
    title = {Hydrodynamic hovering of swimming bacteria above surfaces},
    volume = {6},
    issn = {2643-1564},
    url = {https://link.aps.org/doi/10.1103/PhysRevResearch.6.L032070},
    doi = {10.1103/PhysRevResearch.6.L032070},
    
    number = {3},
    urldate = {2025-06-12},
    journal = {Phys. Rev. Res.},
    author = {Htet, Pyae Hein and Das, Debasish and Lauga, Eric},
    month = sep,
    year = {2024},
    pages = {L032070},
}

@article{walker_systematic_2023,
    title = {Systematic parameterizations of minimal models of microswimming},
    volume = {8},
    issn = {2469-990X},
    url = {https://link.aps.org/doi/10.1103/PhysRevFluids.8.034102},
    doi = {10.1103/PhysRevFluids.8.034102},
    
    number = {3},
    urldate = {2025-05-20},
    journal = {Phys. Rev. Fluids},
    author = {Walker, Benjamin J. and Ishimoto, Kenta and Gaffney, Eamonn A.},
    month = mar,
    year = {2023},
    pages = {034102},
}

@article{walker_emergent_2022,
    title = {Emergent rheotaxis of shape-changing swimmers in {Poiseuille} flow},
    volume = {944},
    issn = {0022-1120, 1469-7645},
    url = {https://www.cambridge.org/core/product/identifier/S0022112022004748/type/journal_article},
    doi = {10.1017/jfm.2022.474},
    
    urldate = {2025-09-02},
    journal = {J. Fluid Mech.},
    author = {Walker, Benjamin J. and Ishimoto, K. and Moreau, C. and Gaffney, Eamonn A. and Dalwadi, Mohit P.},
    month = aug,
    year = {2022},
    pages = {R2},
}

@article{omori_rheotaxis_2022,
    title = {Rheotaxis and migration of an unsteady microswimmer},
    volume = {930},
    issn = {0022-1120, 1469-7645},
    url = {https://www.cambridge.org/core/product/identifier/S0022112021009216/type/journal_article},
    doi = {10.1017/jfm.2021.921},
    
    urldate = {2025-09-02},
    journal = {J. Fluid Mech.},
    author = {Omori, T. and Kikuchi, K. and Schmitz, M. and Pavlovic, M. and Chuang, C.-H. and Ishikawa, T.},
    month = jan,
    year = {2022},
    pages = {A30},
}

@article{elgeti_microswimmers_2016,
    title = {Microswimmers near surfaces},
    volume = {225},
    issn = {1951-6355, 1951-6401},
    url = {http://link.springer.com/10.1140/epjst/e2016-60070-6},
    doi = {10.1140/epjst/e2016-60070-6},
    
    number = {11-12},
    urldate = {2025-09-01},
    journal = {Eur. Phys. J. Spec. Top.},
    author = {Elgeti, Jens and Gompper, Gerhard},
    month = nov,
    year = {2016},
    pages = {2333--2352},
}

@article{kantsler_ciliary_2013,
    title = {Ciliary contact interactions dominate surface scattering of swimming eukaryotes},
    volume = {110},
    issn = {0027-8424, 1091-6490},
    url = {https://pnas.org/doi/full/10.1073/pnas.1210548110},
    doi = {10.1073/pnas.1210548110},
    
    number = {4},
    urldate = {2025-07-07},
    journal = {PNAS},
    author = {Kantsler, Vasily and Dunkel, Jörn and Polin, Marco and Goldstein, Raymond E.},
    month = jan,
    year = {2013},
    pages = {1187--1192},
}

@article{bianchi_holographic_2017,
    title = {Holographic {imaging} {reveals} the {mechanism} of {wall} {entrapment} in {swimming} {bacteria}},
    volume = {7},
    copyright = {https://creativecommons.org/licenses/by/4.0/},
    issn = {2160-3308},
    url = {https://link.aps.org/doi/10.1103/PhysRevX.7.011010},
    doi = {10.1103/physrevx.7.011010},
    
    number = {1},
    urldate = {2025-07-07},
    journal = {Phys. Rev. X},
    author = {Bianchi, Silvio and Saglimbeni, Filippo and Di Leonardo, Roberto},
    month = jan,
    year = {2017},
}

@article{suarez_sperm_2006,
    title = {Sperm transport in the female reproductive tract},
    volume = {12},
    issn = {1460-2369, 1355-4786},
    url = {http://academic.oup.com/humupd/article/12/1/23/607817/Sperm-transport-in-the-female-reproductive-tract},
    doi = {10.1093/humupd/dmi047},
    
    number = {1},
    urldate = {2025-09-03},
    journal = {Hum. Reprod. Update},
    author = {Suarez, S.S. and Pacey, A. A.},
    month = jan,
    year = {2006},
    pages = {23--37},
}

@article{elgeti_hydrodynamics_2010,
    title = {Hydrodynamics of {sperm} {cells} near {surfaces}},
    volume = {99},
    copyright = {https://www.elsevier.com/tdm/userlicense/1.0/},
    issn = {00063495},
    url = {https://linkinghub.elsevier.com/retrieve/pii/S0006349510006156},
    doi = {10.1016/j.bpj.2010.05.015},
    
    number = {4},
    urldate = {2025-05-20},
    journal = {Biophys. J.},
    author = {Elgeti, Jens and Kaupp, U. Benjamin and Gompper, Gerhard},
    month = aug,
    year = {2010},
    pages = {1018--1026},
}

@article{klapper_mathematical_2010,
    title = {Mathematical {description} of {microbial} {biofilms}},
    volume = {52},
    issn = {0036-1445, 1095-7200},
    url = {http://epubs.siam.org/doi/10.1137/080739720},
    doi = {10.1137/080739720},
    
    number = {2},
    urldate = {2025-09-03},
    journal = {SIAM Rev.},
    author = {Klapper, Isaac and Dockery, Jack},
    month = jan,
    year = {2010},
    pages = {221--265},
}

@article{guttenplan_regulation_2013,
    title = {regulation of flagellar motility during biofilm formation},
    volume = {37},
    issn = {1574-6976},
    url = {https://academic.oup.com/femsre/article-lookup/doi/10.1111/1574-6976.12018},
    doi = {10.1111/1574-6976.12018},
    
    number = {6},
    urldate = {2025-09-03},
    journal = {FEMS Microbiol. Rev.},
    author = {Guttenplan, Sarah B. and Kearns, Daniel B.},
    month = nov,
    year = {2013},
    pages = {849--871},
}

@article{bretherton_motion_1962,
    title = {The motion of rigid particles in a shear flow at low {Reynolds} number},
    volume = {14},
    copyright = {https://www.cambridge.org/core/terms},
    issn = {0022-1120, 1469-7645},
    url = {https://www.cambridge.org/core/product/identifier/S002211206200124X/type/journal_article},
    doi = {10.1017/S002211206200124X},
    
    number = {2},
    urldate = {2025-10-08},
    journal = {J. Fluid Mech.},
    author = {Bretherton, F. P.},
    month = oct,
    year = {1962},
    pages = {284--304},
}

@article{jeffery_motion_1922,
    title = {The motion of ellipsoidal particles immersed in a viscous fluid},
    volume = {102},
    copyright = {https://royalsociety.org/journals/ethics-policies/data-sharing-mining/},
    issn = {0950-1207, 2053-9150},
    url = {https://royalsocietypublishing.org/doi/10.1098/rspa.1922.0078},
    doi = {10.1098/rspa.1922.0078},
    
    number = {715},
    urldate = {2025-09-22},
    journal = {Proc. R. Soc. Lond. A},
    author={Jeffery, George Barker},
    month = nov,
    year = {1922},
    pages = {161--179},
}

@misc{MATLAB,
author = {Shampine, Lawrence F. and Reichelt, Mark W.},
title = {The MATLAB ODE Suite},
journal = {SIAM Journal on Scientific Computing},
volume = {18},
number = {1},
pages = {1-22},
year = {1997},
doi = {10.1137/S1064827594276424},

URL = { 
    
        https://doi.org/10.1137/S1064827594276424
    
    

},
eprint = { 
    
        https://doi.org/10.1137/S1064827594276424
},
}

@article{friedrich_hydrodynamic_2016,
    title = {Hydrodynamic synchronization of flagellar oscillators},
    volume = {225},
    issn = {1951-6355, 1951-6401},
    url = {http://link.springer.com/10.1140/epjst/e2016-60056-4},
    doi = {10.1140/epjst/e2016-60056-4},
    
    number = {11-12},
    urldate = {2025-10-27},
    journal = {Eur. Phys. J. Spec. Top.},
    author = {Friedrich, Benjamin},
    month = nov,
    year = {2016},
    pages = {2353--2368},
}

@article{golestanian_hydrodynamic_2011,
    title = {Hydrodynamic synchronization at low {Reynolds} number},
    volume = {7},
    issn = {1744-683X, 1744-6848},
    url = {https://xlink.rsc.org/?DOI=c0sm01121e},
    doi = {10.1039/c0sm01121e},
    
    number = {7},
    urldate = {2025-10-27},
    journal = {Soft Matter},
    author = {Golestanian, Ramin and Yeomans, Julia M. and Uchida, Nariya},
    year = {2011},
    pages = {3074},
}

@article{elgeti_emergence_2013,
    title = {Emergence of metachronal waves in cilia arrays},
    volume = {110},
    issn = {0027-8424, 1091-6490},
    url = {https://pnas.org/doi/full/10.1073/pnas.1218869110},
    doi = {10.1073/pnas.1218869110},
    
    number = {12},
    urldate = {2025-10-27},
    journal = {PNAS},
    author = {Elgeti, Jens and Gompper, Gerhard},
    month = mar,
    year = {2013},
    pages = {4470--4475},
}

@book{bender_advanced_1999,
    address = {New York, NY},
    title = {Advanced {mathematical} {methods} for {scientists} and {engineers} {I}},
    copyright = {http://www.springer.com/tdm},
    isbn = {978-1-4419-3187-0 978-1-4757-3069-2},
    url = {http://link.springer.com/10.1007/978-1-4757-3069-2},
    
    urldate = {2025-11-13},
    author = {Bender, Carl M. and Orszag, Steven A.},
    year = {1999},
    doi = {10.1007/978-1-4757-3069-2},
}

@article{walker_effects_2022,
    title = {Effects of rapid yawing on simple swimmer models and planar {Jeffery}'s orbits},
    volume = {7},
    issn = {2469-990X},
    url = {https://link.aps.org/doi/10.1103/PhysRevFluids.7.023101},
    doi = {10.1103/PhysRevFluids.7.023101},
    
    number = {2},
    urldate = {2025-11-17},
    journal = {Phys. Rev. Fluids},
    author = {Walker, Benjamin J. and Ishimoto, Kenta and Gaffney, Eamonn A. and Moreau, Clément and Dalwadi, Mohit P.},
    month = feb,
    year = {2022},
    pages = {023101},
}

@article{gaffney_canonical_2022,
    title = {Canonical orbits for rapidly deforming planar microswimmers in shear flow},
    volume = {7},
    issn = {2469-990X},
    url = {https://link.aps.org/doi/10.1103/PhysRevFluids.7.L022101},
    doi = {10.1103/PhysRevFluids.7.L022101},
    
    number = {2},
    urldate = {2025-11-17},
    journal = {Phys. Rev. Fluids},
    author = {Gaffney, Eamonn A. and Dalwadi, Mohit P. and Moreau, Clément and Ishimoto, Kenta and Walker, Benjamin J.},
    month = feb,
    year = {2022},
    pages = {L022101},
}

@article{dalwadi_generalised_2024,
    title = {Generalised {Jeffery}'s equations for rapidly spinning particles. {Part} 1. {Spheroids}},
    volume = {979},
    issn = {0022-1120, 1469-7645},
    url = {https://www.cambridge.org/core/product/identifier/S0022112023009230/type/journal_article},
    doi = {10.1017/jfm.2023.923},
    
    urldate = {2025-11-17},
    journal = {J. Fluid Mech.},
    author = {Dalwadi, Mohit P. and Moreau, C. and Gaffney, Eamonn A. and Ishimoto, K. and Walker, Benjamin J.},
    month = jan,
    year = {2024},
    pages = {A1},
}

@article{dalwadi_rapidly_2025,
    title = {Rapidly yawing spheroids in viscous shear flow: emergent loss of symmetry},
    volume = {1009},
    issn = {0022-1120, 1469-7645},
    shorttitle = {Rapidly yawing spheroids in viscous shear flow},
    url = {https://www.cambridge.org/core/product/identifier/S0022112025002174/type/journal_article},
    doi = {10.1017/jfm.2025.217},
    
    urldate = {2025-11-17},
    journal = {J. Fluid Mech.},
    author = {Dalwadi, Mohit P.},
    month = apr,
    year = {2025},
    pages = {A27},
}

@article{dalwadi_generalised_2024_b,
	title = {Generalised {Jeffery}'s equations for rapidly spinning particles. {Part} 2. {Helicoidal} objects with chirality},
	volume = {979},
	issn = {0022-1120, 1469-7645},
	url = {https://www.cambridge.org/core/product/identifier/S0022112023009242/type/journal_article},
	doi = {10.1017/jfm.2023.924},
	
	urldate = {2025-11-17},
	journal = {J. Fluid Mech.},
	author = {Dalwadi, Mohit P. and Moreau, C. and Gaffney, Eamonn A. and Walker, Benjamin J. and Ishimoto, K.},
	month = jan,
	year = {2024},
	pages = {A2},
}

@article{chwang_hydromechanics_1975,
    title = {Hydromechanics of low-{Reynolds}-number flow. {Part} 2. {Singularity} method for {Stokes} flows},
    volume = {67},
    copyright = {https://www.cambridge.org/core/terms},
    issn = {0022-1120, 1469-7645},
    url = {https://www.cambridge.org/core/product/identifier/S0022112075000614/type/journal_article},
    doi = {10.1017/S0022112075000614},
    
    number = {4},
    urldate = {2025-12-03},
    journal = {J. Fluid Mech.},
    author = {Chwang, Allen T. and Wu, T. Yao-Tsu},
    month = feb,
    year = {1975},
    pages = {787--815},
}

@book{kim2013microhydrodynamics,
  title={Microhydrodynamics: principles and selected applications},
  author={Kim, Sangtae and Karrila, Seppo J},
  year={2013},
  publisher={Butterworth-Heinemann}
}

@article{yeo_shear-induced_2025,
  title={A shear-induced limit on bacterial surface adhesion in fluid flow},
  author={Yeo, Edwina F and Walker, Benjamin J and Pearce, Philip and Dalwadi, Mohit P},
  journal={PNAS},
  volume={123},
  number={4},
  pages={e2516069123},
  year={2026},
}

@article{drescher_direct_2010,
    title = {Direct {Measurement} of the {Flow} {Field} around {Swimming} {Microorganisms}},
    volume = {105},
    copyright = {http://link.aps.org/licenses/aps-default-license},
    issn = {0031-9007, 1079-7114},
    url = {https://link.aps.org/doi/10.1103/PhysRevLett.105.168101},
    doi = {10.1103/PhysRevLett.105.168101},
    
    number = {16},
    urldate = {2025-06-12},
    journal = {Phys. Rev. Lett.},
    author = {Drescher, Knut and Goldstein, Raymond E. and Michel, Nicolas and Polin, Marco and Tuval, Idan},
    month = oct,
    year = {2010},
    pages = {168101},
}

@article{drescher_fluid_2011,
    title = {Fluid dynamics and noise in bacterial cell–cell and cell–surface scattering},
    volume = {108},
    issn = {0027-8424, 1091-6490},
    url = {https://pnas.org/doi/full/10.1073/pnas.1019079108},
    doi = {10.1073/pnas.1019079108},
    
    number = {27},
    urldate = {2026-02-24},
    journal = {PNAS},
    author = {Drescher, Knut and Dunkel, Jörn and Cisneros, Luis H. and Ganguly, Sujoy and Goldstein, Raymond E.},
    month = jul,
    year = {2011},
    pages = {10940--10945},
}

@article{guasto_oscillatory_2010,
    title = {Oscillatory {flows} {induced} by {microorganisms} {swimming} in {two} {dimensions}},
    volume = {105},
    copyright = {http://link.aps.org/licenses/aps-default-license},
    issn = {0031-9007, 1079-7114},
    url = {https://link.aps.org/doi/10.1103/PhysRevLett.105.168102},
    doi = {10.1103/PhysRevLett.105.168102},
    
    number = {16},
    urldate = {2026-02-24},
    journal = {Phys. Rev. Lett.},
    author = {Guasto, Jeffrey S. and Johnson, Karl A. and Gollub, J. P.},
    month = oct,
    year = {2010},
    pages = {168102},
}

@article{ishimoto_coarse-graining_2017,
    title = {Coarse-{graining} the {fluid} {flow} around a {human} {sperm}},
    volume = {118},
    copyright = {http://link.aps.org/licenses/aps-default-license},
    issn = {0031-9007, 1079-7114},
    url = {https://link.aps.org/doi/10.1103/PhysRevLett.118.124501},
    doi = {10.1103/PhysRevLett.118.124501},
    
    number = {12},
    urldate = {2025-05-21},
    journal = {Phys. Rev. Lett.},
    author = {Ishimoto, Kenta and Gadêlha, Hermes and Gaffney, Eamonn A. and Smith, David J. and Kirkman-Brown, Jackson},
    month = mar,
    year = {2017},
    pages = {124501},
}

@article{schoeller_flagellar_2018,
    title = {From flagellar undulations to collective motion: predicting the dynamics of sperm suspensions},
    volume = {15},
    issn = {1742-5689, 1742-5662},
    shorttitle = {From flagellar undulations to collective motion},
    url = {https://royalsocietypublishing.org/doi/10.1098/rsif.2017.0834},
    doi = {10.1098/rsif.2017.0834},
    number = {140},
    urldate = {2026-06-05},
    journal = {Journal of The Royal Society Interface},
    author = {Schoeller, Simon F. and Keaveny, Eric E.},
    month = mar,
    year = {2018},
    pages = {20170834},
}

\end{document}